# Topological Magnetism in 2D and 3D: Skyrmions and Hopfions, Classical and quantum approaches.

A.S. T. Pires

Departamento de Fisica, Universidade Federal de Minas Gerais, Belo Horizonte, MG, CP 702, 30123-970, Brazil.
e-mail: antpires@fisica.ufmg.br

## Abstract

Magnetic skyrmions and related topological spin textures have emerged as important subjects in modern condensed-matter physics because of their nontrivial topology, rich underlying physics, and potential relevance to future magnetic technologies. This review provides a unified theoretical overview of skyrmions and related topological structures, covering their description from classical continuum models to quantum regimes. We first introduce the mathematical concepts required to characterize magnetic textures and formulate the continuum theory of ferromagnets. We then discuss the properties of ferromagnetic skyrmions in relation to the Dzyaloshinskii–Moriya interaction, nonlinear sigma models, merons, vortices, frustration, and dipole–dipole interactions. Particular attention is given to skyrmion–skyrmion interactions and to the dynamical equations governing magnetic textures, including the Landau–Lifshitz–Gilbert, Thiele, and Landau–Lifshitz–Bloch equations. The review subsequently extends the discussion to antiferromagnetic systems and to topological structures in three dimensions, with emphasis on Hopfions and their topological characterization. We then examine quantum skyrmions, including approaches that go beyond the semiclassical description, and discuss the emergence of quantum Hopfions. The theoretical framework is complemented by appendices presenting analytical solutions and ansatzes for skyrmions, spin coherent states and path integrals, quantum antiferromagnets, spin torques, fiber-bundle concepts and the Hopf map, and the Euler–Lagrange formalism in three dimensions. By connecting continuum field theory, magnetic dynamics, topology, and quantum mechanics, this review aims to provide a coherent framework for understanding the formation, stability, dynamics, interactions, and quantum properties of topological magnetic textures across different dimensions and magnetic systems.

# Contents.

# 1. Introduction.

Topological concepts have become central to modern condensed-matter physics, providing a unifying language for understanding robust, defect-protected structures in electronic, photonic, and magnetic systems. Among these topological magnetic textures, skyrmions, antiskyrmions, and hopfions are nanoscale spin configurations whose stability arises not from conventional energetic minima but from topological invariants associated with their spin-field mappings. Their discovery in chiral magnets and multilayer heterostructures has transformed the spintronics landscape, offering particle-like excitations that can be created, manipulated, and transported at remarkably low energy cost. The stability of magnetic skyrmions makes them candidates for future information carriers.

Magnetic skyrmions were first predicted theoretically in chiral magnets, where competition between symmetric exchange and antisymmetric Dzyaloshinskii–Moriya interaction (DMI) stabilizes swirling spin textures with quantized topological charge. Subsequent real-space imaging experiments confirmed their existence and revealed their ability to form ordered lattices, move under ultralow current densities, and interact with magnons, defects, and thermal fluctuations in fundamentally distinct ways from conventional domain walls [1]. These properties have positioned skyrmions as promising candidates for high-density memory, logic architectures, and neuromorphic computing.

Beyond skyrmions, the field has expanded to include antiskyrmions, stabilized by lower crystal symmetries, and hopfions, three-dimensional knotted spin textures characterized by a nontrivial Hopf invariant. These emergent structures highlight the richness of topological magnetism and underscore the importance of symmetry, dimensionality, and confinement in determining the stability and dynamics of magnetic textures. Recent advances in ultrafast optics, multilayer engineering, and frustrated magnetism have further broadened the range of accessible topological states, enabling controlled nucleation, annihilation, and transformation between distinct topological classes.

This review provides a theoretical overview of topological magnetic textures in two and three dimensions. Because excellent review papers already cover the physics and technological applications of these excitations [1-12], I aim to briefly outline the underlying mathematics. This text contains little original material. I obtained most of the material from research literature. However, I have tried to present the subject pedagogically by developing nearly all the mathematical passages.

## 2. Some mathematical concepts.

A soliton is a stable, localized nonlinear wave that travels long distances without changing its shape, speed, or energy. The defining feature is that even after colliding with another soliton, it re-emerges unchanged, behaving almost like a particle rather than a typical wave. A typical one-dimensional example is the solution of the sine-Gordon equation $\phi_{tt} - \phi_{xx} + \sin\phi = 0$. A soliton in this model is a localized, stable solution that connects two neighboring minima of the periodic potential $V(\phi) = 1 - \cos\phi$.

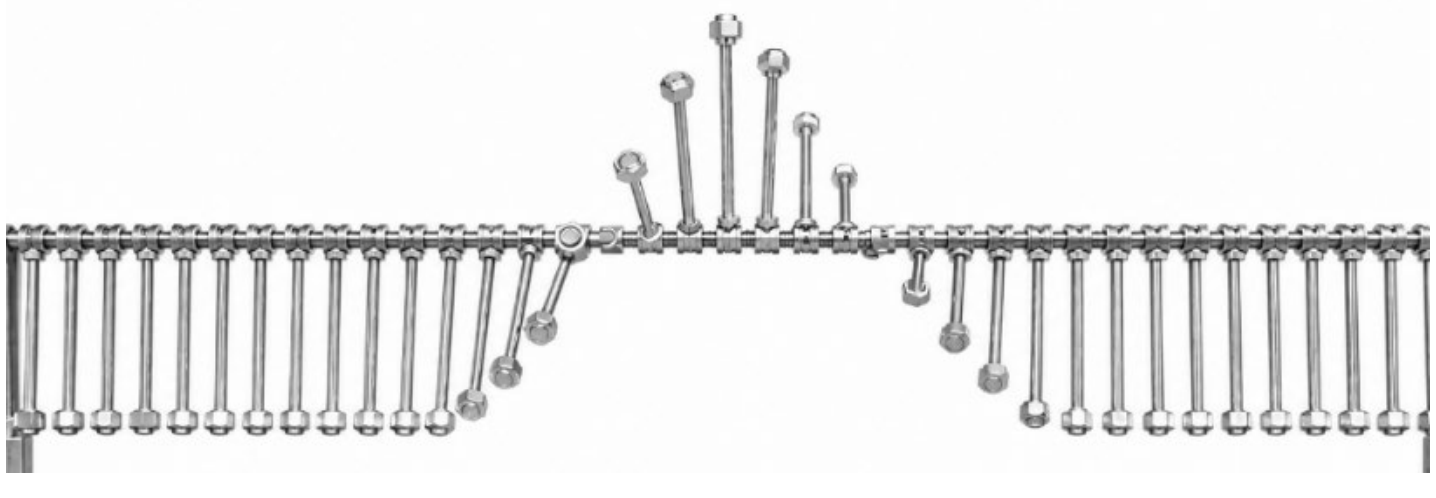

Figure 1. A sine-Gordon soliton.

A soliton connects neighboring minima as $x$ goes from $x\ -\infty$ to $+\infty$. We have $\phi(-\infty) = 0,\ \phi(+\infty) = 2\pi$. An antisoliton does the opposite: $\phi(-\infty) = 2\pi,\ \phi(+\infty) = 0$.

These solitons behave like particles: they propagate without changing shape and survive collisions with only a phase shift. A classic physical realization of the sine-Gordon equation is a chain of pendulums coupled by torsion springs (Fig.1). Imagine many identical pendulums, equally spaced along a line, each able to swing in a vertical plane. Neighboring pendulums are connected by torsion springs so that if one rotates, it tends to twist its neighbors. We keep the pendulum at the far left fixed and then rotate a pendulum at the far right by $2\pi$ . The resulting localized twist propagating along the chain is a soliton. Another example is a ferromagnetic chain along the *z*-direction, with easy-plane anisotropy that forces the spins to lie in the *xy*-plane, and an applied magnetic field along the *x*-direction. Skyrmions and hopfions, studied here, are other soliton examples. Topology matters in soliton studies because it explains why certain localized field configurations cannot disappear, unwind, or decay without crossing an infinite-energy barrier. It classifies field configurations into distinct sectors that cannot be continuously deformed into one another, making some solitons "topologically protected."

Topology studies the properties of objects that remain unchanged under continuous deformations such as stretching, twisting, crumpling, or bending. These operations do not create or destroy holes, tear the object, or glue parts together. A classic example is that a mug can be continuously deformed into a donut because both have one hole; topologically, they are equivalent. In physics, this idea translates into physical properties that remain robust even in the presence of impurities, defects, or temperature variations. These robust features are characterized by **topological invariants**, integer-valued quantities (such as the number of holes in a surface) that remain unchanged under continuous transformations. In this section, I will introduce the mathematical concepts used later. For a more detailed discussion of topology in condensed matter physics, see Ref. [13]. I will start by defining some concepts.

A topological space is a set equipped with a collection of subsets called *open sets* that formalize the idea of *closeness* or *continuity* without relying on numerical distance. A **manifold** is a topological space with extra structure. It looks locally like Euclidean space. Even if space may be curved, twisted, or globally complicated, every small neighborhood around any point behaves like $\mathbb{R}^n$ of some dimension $n$. In other words, a manifold is a topological space in which each point has a neighborhood homeomorphic to an open subset of $\mathbb{R}^n$. The number $n$ is the manifold's dimension. Example: a sphere is a 2-dimensional manifold because every point on it has a neighborhood that resembles a patch of the plane.

A map $(U, \phi)$ of a manifold $M$ is an open set $U$ in $M$, called the map domain, along with a homeomorphism $\phi: U \to V$ of $U$ in the open set $V$ in $\mathbb{R}^n$. The coordinates $(x_1, x_2, ..., x_n)$ of the image $\phi(x)$ of the point $x \in M$ are called coordinates of $x$ on the map $(U, \phi)$. The sphere can be given a differential structure by two stereographic projections from the north and south poles (Fig. 2). For a point $P(x, y, z)$ on the sphere $x^2 + y^2 + z^2 = 1$ , its stereographic projection $p(u, v)$ on the plane $z = 0$ is

$$u = \frac{x}{1-z}, \quad v = \frac{y}{1-z}.$$

The geometry of the sphere will be important for our study of skyrmions and hopfions.

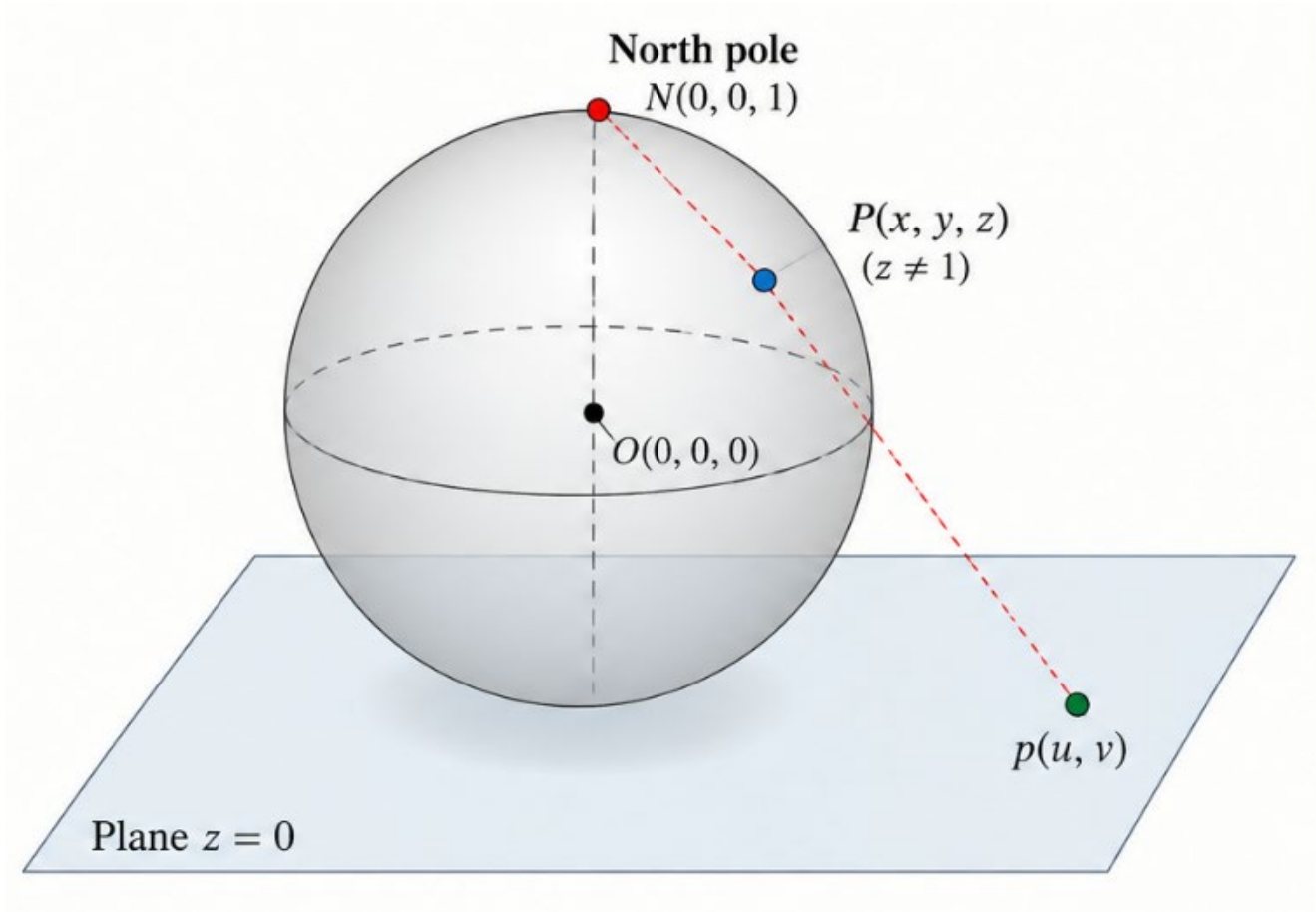

Figure 2. Stereographic projection of a sphere.

Homotopy theory provides mathematical language for classifying topological defects and solitons. Different homotopy classes correspond to configurations that cannot be continuously transformed into one another without creating singularities. Therefore, I will briefly discuss this subject.

Let $p_1$ and $p_2$ be points in a manifold $X$. If there is a curve $C$ in $X$ that goes from $p_1$ to $p_2$, we say that $X$ is path-connected. If, in addition, $X$ is such that given any two curves $C_1$ and $C_2$ going from any point $p_1$ to any point $p_2$, $C_1$ can be smoothly transformed into $C_2$ in $X$; that is, if there exists a continuous function $p(t, s)$ such that for each $s$ in the range [0, 1], $p(t, s)$ describes a curve from $p_1$ to $p_2$, when $t$ varies and if this curve coincides with $C_1$ for $s = 0$, and with $C_2$ for $s = 1$, then $X$ is called simply connected. Two curves $C_0$ and $C_1$ in a manifold $X$, both having the same starting point $p_0$ and the same final point $p_1$ are called homotopic if one of them can be deformed into the other by continuous deformations in $X$. Thus, a manifold is connected if any two curves in it, having the same initial and final points, are homotopic. Given the points $p_0$ and $p_1$, the class of all curves that are homotopic to a given curve from $p_0$ to $p_1$ form an equivalence class, or homotopy class, at $X$.

Let $X$ and $Y$ be two manifolds. A map $\psi_0 : X \to Y$, with $\psi_0(x_0) = y_0$, where $x_0 \in X$ and $y_0 \in Y$ is said to be homotopic to another such map $\psi_1$, if $\psi_0$ can be continuously deformed into $\psi_1$( $x_0$ and $y_0$ are fixed points called base points). We can also say that $\psi_0$ is homotopic to $\psi_1$ if there is a continuous map

$$\tilde{\psi} : X \times [0,1] \to Y \,, \qquad (1)$$

with $t \in [0,1]$ such that $\tilde{\psi}(x_0, t) = y_0$ for all $t$ and $\tilde{\psi}\big|_{t=0} = \psi_0$, $\tilde{\psi}\big|_{t=1} = \psi_1$. The maps $\psi$ are symmetric, transitive, and reflexive; thus, they can be classified into homotopy classes.

Let us consider the case where $X$ is an $n$-sphere $S^n$ (that is, the set of points in $\Re^{n+1}$ at unit distance from the origin). The set of homotopy classes of maps $\psi : S^n \to Y$ is called the $n$th homotopy group and denoted by $\pi_n(Y)=[S^n,Y]$. For $n\geq 1$, the set $\pi_n(Y)$ forms a group, called the $n$th homotopy group of $Y$. A map $S^1\to Y$ is called a loop. The class of the constant map $S^1\to y_0$ is the identity element of the group $\pi_1(Y)$, called the fundamental group of $Y$. The group for $n > 1$ is generally non-Abelian. If $Y$ is connected and $\pi_1(Y)=I$, where $I$ denotes the trivial group with just the identity element, the space $Y$ is said to be simply connected. In this case, every loop is contractible (can be continuously deformed to a point), i. e., homotopic to the trivial loop. If $\Re^d$ has the origin as base point, any loop $\psi_0:S^1\to\Re^d$ is contractible (parameterize $S^1$ by $\theta\in[0,2\pi]$, and define $\widetilde{\psi}(\theta,t)=(1-t)\psi_0(\theta)$, therefore $\pi_1(\Re^d)=I$). If there is a hole in the space $\Re^n$, the loops can be divided into classes, each characterized by the number of times the loop winds around the hole (Fig.3).

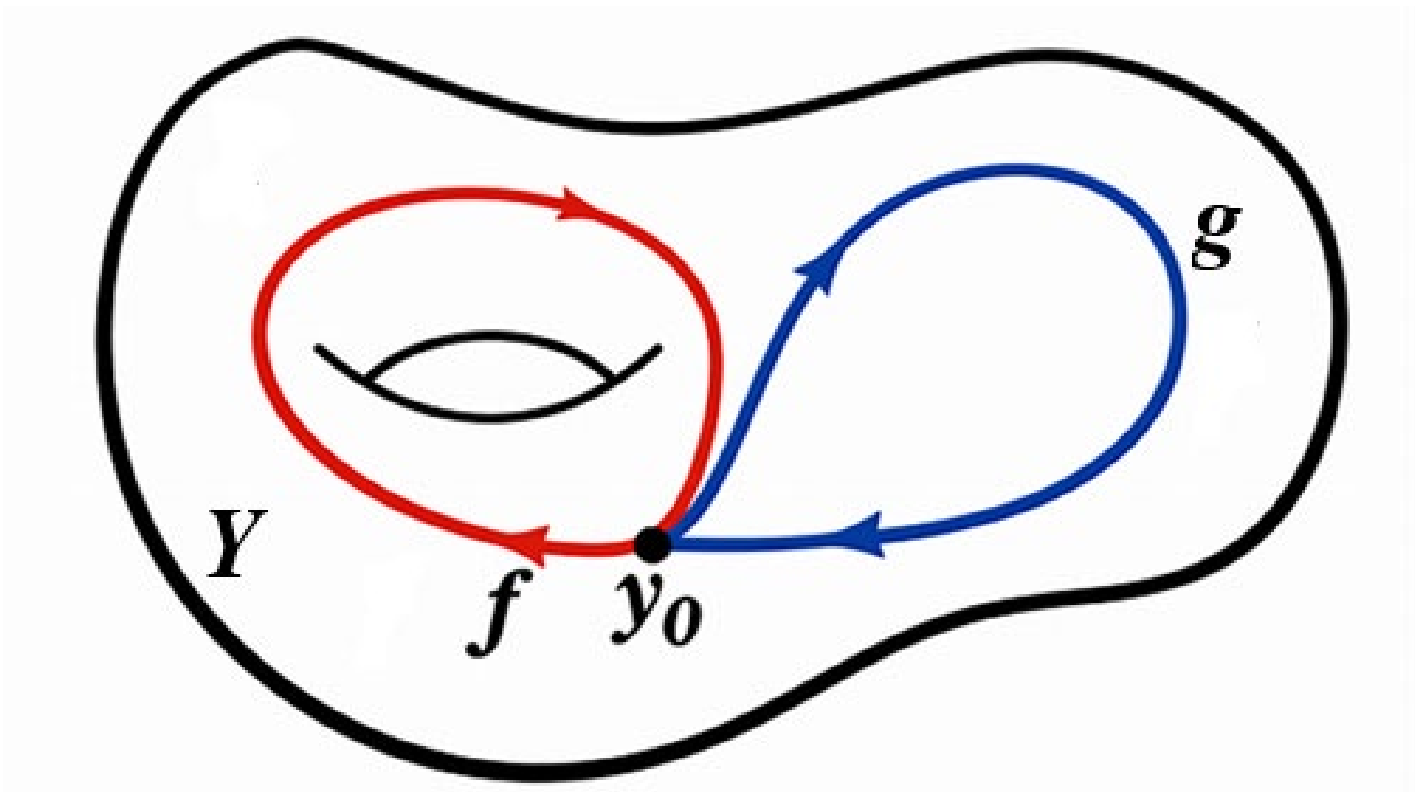

Figure 3. We can view the mapping $\pi_1(Y), f:S^1\to Y$, as a mapping from the interval [0, 1], such that $f(0)=f(1)=y_0$. These maps are associated with a path in $Y$ that begins and ends at the point $y_0$.

A simple case is the mapping of circles into circles, i. e. $S^1\to S^1$. We can parameterize the circle using an angle $\theta$ defined modulo $2\pi$. A mapping can be defined by a continuous function $\Lambda(\theta)$ modulo $2\pi$. As an example, let us consider two such mappings as

$$\Lambda_0(\theta)=0 \quad \text{for all } \theta \tag{2}$$

And

$$\widetilde{\Lambda}_0(\theta)=\begin{cases} t\theta & \text{for } 0\leq\theta<\pi \\ t(2\pi-\theta) & \text{for } \pi\leq\theta<2\pi \end{cases}, \tag{3}$$

where $t\in[0,1]$. By varying $t$ continuously down to zero, we can continuously deform the second mapping into the first. These two mappings, therefore, belong to the same homotopy class.

Let us now consider the mapping

$$\Lambda_1(\theta)=\theta \quad \text{for all } \theta. \tag{4}$$

As $\theta$ completes a full circle, so does $\Lambda_1$. But now, it cannot be continuously deformed into (2) or (3), since in equation (4) the second circle is wound once around the first circle, whereas in (2) and (3) it is effectively wound zero times. Thus, (4) belongs to a different homotopy class from (2) or (3). The integer distinguishing the two classes is the winding number defined by

$$W = \frac{1}{2\pi}\int_0^{2\pi} \frac{d\Lambda}{d\theta} d\theta. \tag{5}$$

Therefore, $\Lambda_n(\theta) = n\theta$ is the prototype mapping belonging to the $W = n$ class (negative values of $W$ are obtained by doing the winding in the opposite sense). The winding number $W$ is the net number of times that the image $\Lambda(\theta)$ winds around the target as $\theta$ goes once around the domain. The product $\gamma_1\gamma_2$ of path $\gamma_1$ and $\gamma_2$ characterized by winding numbers $W_1$ and $W_2$, respectively, has winding number $W = W_1 + W_2$.

Let us now consider the non-singular mappings of a sphere $S^2$ into another sphere $S^2$. As was said above, these mappings can be classified into homotopy sectors. A mapping in one sector can be continuously deformed into another, whereas mappings from two different sectors cannot. There is a denumerable infinity of such homotopy sectors, or classes, which can be characterized by integers (positive, negative, and zero). That is, these homotopy classes form a group isomorphic to the integers. We can write in a compact form

$$\pi_2(S^2) = \mathbb{Z}, \tag{6}$$

where $\pi_n(S^m)$ means the homotopy group associated with the mappings $S^n \to S^m$ and $\mathbb{Z}$ is the group of integers. The integer characterizing the homotopy classes of $S^2 \to S^2$ is the number of times one sphere has been wrapped around the other. We also have:

$$\pi_n(S^n) = \mathbb{Z}, \quad \pi_n(S^m) = 0 \text{ for } n < m, \quad \pi_n(S^1) = 0 \text{ for } n > 1. \tag{7}$$

As an example, we have $\pi_1(S^2) = 0$, since any loop in a sphere can be deformed to a point.

A one-dimensional ferromagnetic chain with periodic boundary conditions can be compactified into a circle $S^1$. Topological classes are classified by the first homotopy group $\pi_1(M)$. If the spins are restricted to the $xy$-plane, we can take the order parameter as an angle $\phi$. Suppose as $x$ goes around the chain, we have $\phi(L) - \phi(0) = 2\pi W$, where $W \in \mathbb{Z}$. As we saw, $W$ counts how many times the target circle is wrapped. Mathematically $\pi_1(S^1) = \mathbb{Z}$. Therefore, there are infinitely many disconnected topological sectors labeled by the winding number $W$. Now let the spin be a three-component unit vector. The map is $S^2 \to S^2$. But now $\pi_1(S^2) = 0$. Every closed loop on a sphere can continuously shrink to a point. There is no topological invariant.

Calculating homotopy groups for $n > m$ is very difficult. One highly nontrivial result is $\pi_3(S^2) = \mathbb{Z}$. The integer number labeling homotopy classes, in this case, is called the Hopf invariant. The Hopf invariant is a mathematical tool that classify maps $S^3 \to S^2$ by measuring the linking number of the preimages of two regular values.

In physics and materials science, a 3D texture often means a continuous field: $\vec{n}(\vec{r}) : \mathbb{R}^3 \to S^2$, such as the magnetization in a ferromagnet. Such a field assigns a unit vector to every point in space. If the field is well-behaved at infinity and points to the same direction (then we can compactify space to a 3-sphere), the map is then: $S^3 \to S^2$.

That is exactly the setting where the Hopf invariant lives. Two textures can look similar locally but have different Hopf invariants, meaning they belong to different topological classes and cannot be smoothly morphed into each other.

The homotopy group $\pi_3(S^2)$ means: all continuous maps from the 3-sphere to the 2-sphere, modulo smooth deformations. Take any continuous map: $f : S^3 \to S^2$ . Pick a point *p* on the target sphere. The preimage $f^{-1}(p)$ is (generically) a closed loop in $S^3$ . Pick another point *q*. Its preimage $f^{-1}(q)$ is another loop. So, a 3D texture (map to $S^2$) naturally produces families of loops in 3-space.

A **knot** in topology is a closed loop that may be tangled but has no loose ends. The key idea is that two knots are considered the *same* if one can be smoothly deformed into the other without cutting the loop or letting it pass through itself. A loop may be embedded in three-dimensional space in arbitrary complicated ways.. Two knots are equivalent if one can be transformed into the other by a continuous deformation. We use the crossing number (a topological invariant) to distinguish knots. In Fig. 4, I show the trivial knot and the simplest nontrivial knot, the trefoil knot. The trefoil knot has three crossings. Because it is permanently tangled, no amount of wiggling or stretching can untie it into a standard, untangled circle.

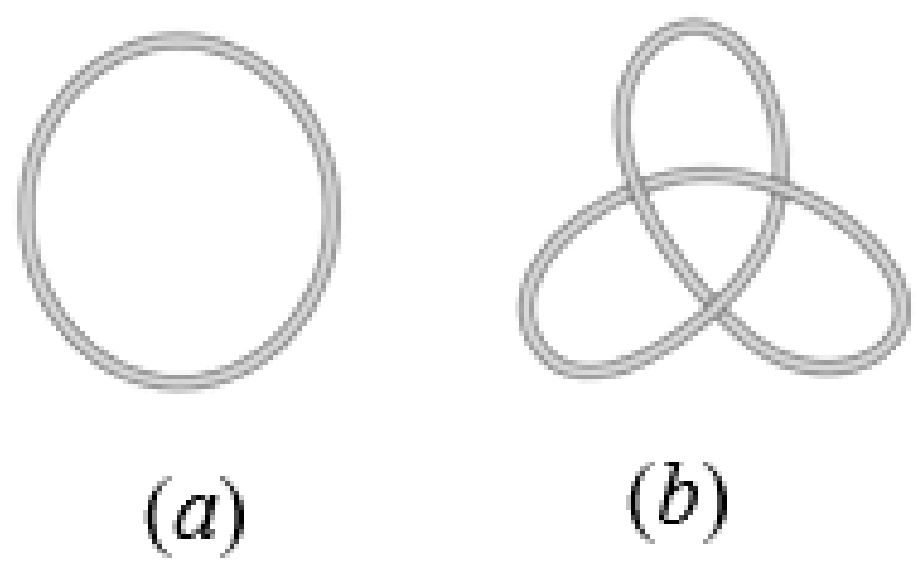

Figure 4. (a) the unknot (trivial knot) and (b) the trefoil.

The *Hopf invariant* relates to knot theory primarily through the *linking number* and maps between spheres ($\pi_3(S^2)$). Mathematically, it measures how many times the preimages of two points in a mapped space intertwine, which translates directly to the topological interlocking of closed loops (knots and links). In 3D physical systems, a **hopfion** is a stable, knot-like twisting of fields (e.g., in magnetic materials, liquid crystals, or quantum matter). The Hopf invariant determines the mathematical stability and topological "charge" of these knotted defects. I will discuss hopfions in section 14.

Two circles linked exactly once, like a simple chain, are called a Hopf link. It is the simplest nontrivial link. It has two components, each an unknot (Fig. 5).

In the following, I will show how to calculate the topological invariant for the mapping $S^2 \to S^2$ . Let $\vec{r}(u,v)$ be a point on a surface at (*u*, v). If we shift *u* by a small amount *du*, the position changes along the surface, approximately by

$$d\vec{r}_u \approx \frac{\partial \vec{r}}{\partial u} du \, . \tag{8}$$

Similarly, shifting *v* by *dv* leads to the displacement

$$d\vec{r}_v \approx \frac{\partial \vec{r}}{\partial v} dv . \tag{9}$$

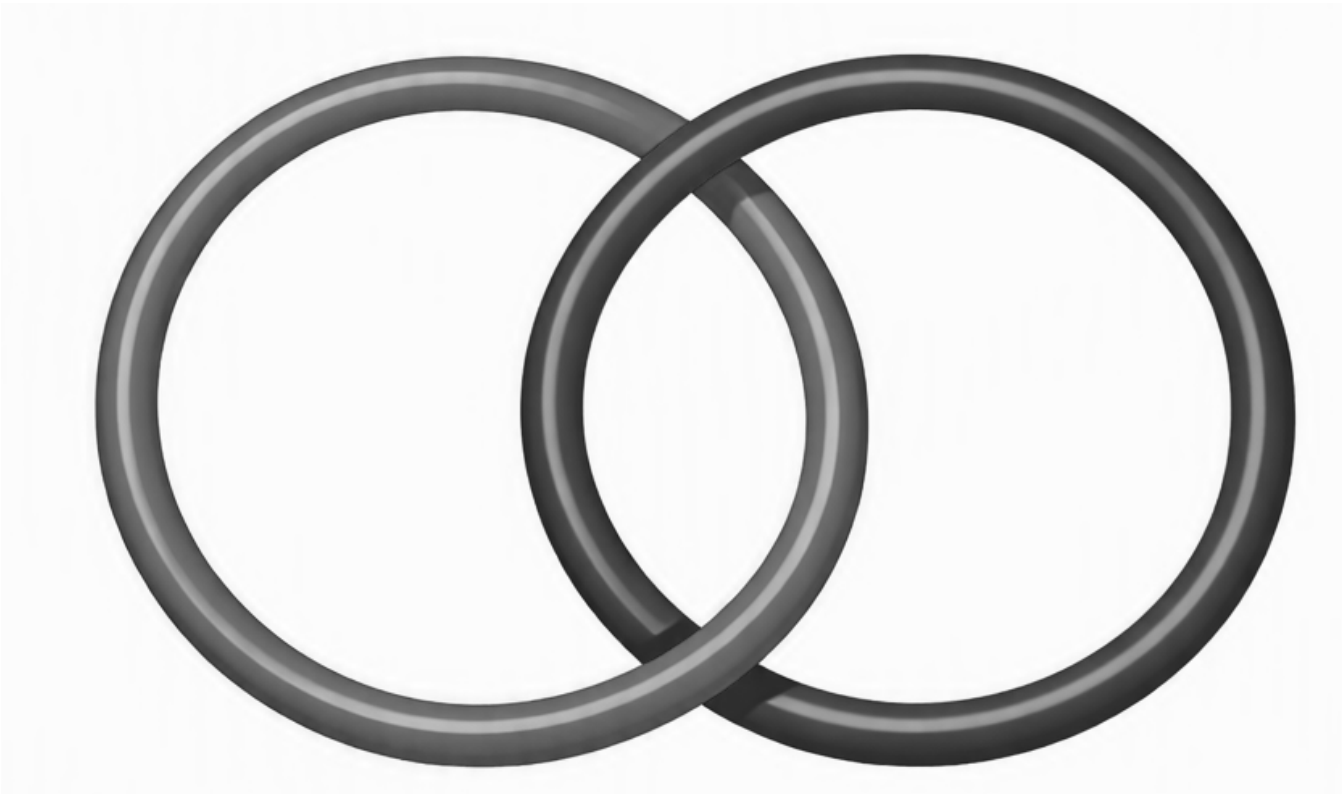

Figure 5. A Hopf link.

These two displacements form two adjacent sides of a parallelogram on the surface. The cross product $d\vec{A} = d\vec{r}_u \times d\vec{r}_v$ is a vector that points perpendicular (normal) to the plane formed by the two displacement vectors, and its magnitude is the area of the parallelogram spanned by the two vectors. The oriented area element $d\vec{A}$ can then be written as

$$d\vec{A} = \left( \frac{\partial \vec{r}}{\partial u} \times \frac{\partial \vec{r}}{\partial v} \right) dudv . \tag{10}$$

If $\vec{n}(x, y)$ is a surface on the unit sphere, we have the map

$$\vec{n} : \mathbb{R}^2 \to S^2 , \qquad |\vec{n}(x, y)| = 1 . \tag{11}$$

Every point $(x, y)$ in the plane is mapped to a point on the unit sphere. So, the mapping defines a (possibly self-overlapping) surface on the sphere parameterized by $(x, y)$. The normal to the sphere is just $\vec{n}$. Thus, the total oriented area on the unit sphere is

$$A_{oriented} = \int \vec{n}.(\partial_x \vec{n} \times \partial_y \vec{n}) dxdy . \tag{12}$$

The total area of the unit sphere is $4\pi$. So, to obtain how many times (with orientation) the mapping covers the sphere, we write

$$N = \frac{1}{4\pi} \int \vec{n}. \left( \frac{\partial \vec{n}}{\partial x} \times \frac{\partial \vec{n}}{\partial y} \right) dxdy . \tag{13}$$

Assuming that the vector field $\vec{n}$ points in the same direction far away in the plane, we add a single point at infinity to the plane $\mathbb{R}^2$, turning it into a compact space that is topologically equivalent to the sphere $S^2$, in which case the "coverage number" is an integer.

Skyrmions in two-dimensional magnets correspond to mapping $S^2 \to S^2$ classified by $\pi_2(S^2) = \mathbb{Z}$. The study of skyrmions and other topological textures in condensed-matter systems relies heavily on homotopy theory. In magnetic systems, the relevant mappings often involve assigning a magnetization direction to every point in space. Because the magnetization vector has a fixed magnitude, each point corresponds to a point on the sphere $S^2$.

## 3. Ferromagnet in the continuum limit.

For a two-dimensional ferromagnetic square lattice with lattice constant $a = 1$, and nearest-neighbor interaction, a typical Hamiltonian is [14]

$$H = -J\sum_{<i,j>}\vec{S}_i.\vec{S}_j + \sum_{<i,j>}\vec{D}.\vec{S}_i\times\vec{S}_j - B\sum_i S_i^z \,. \tag{14}$$

The second term is the Dzyaloshinskii-Moriya interaction (DMI), an antisymmetric exchange interaction that arises in magnetic systems with spin–orbit coupling (SOC) and broken inversion symmetry. In ultrathin ferromagnetic films interfaced with heavy metals, this interaction becomes purely interfacial and stabilizes a variety of chiral spin textures, including Néel-type domain walls and skyrmions. The isotropic two-dimensional Heisenberg ferromagnet exhibits long-range magnetic order only at zero temperature, in accordance with the Mermin-Wagner theorem. Long-range order at finite temperature can be stabilized by perturbations that break the continuous spin-rotation symmetry, such as magnetic anisotropy or an external magnetic field.

The quantum treatment of Hamiltonian (14) is typically carried out using the spin-wave approach, in which spin operators are expressed in terms of bosonic operators through representations such as Holstein–Primakoff or Schwinger bosons [14]. For a spin-½ system, for example, the spin has only two eigenstates, $|+\rangle$ and $|-\rangle$. A quantum spin is described by a quantum state rather than by a classical vector pointing in a definite direction. The classical picture emerges only as an approximation, for example through spin-coherent states. . To investigate topological excitations, however, we adopt a semiclassical description and represent a quantum spin using a coherent spin state $\vec{S} = \hbar S\vec{n}$, as discussed in Appendix C. For $S$ = ½, this state can be written as $|\vec{n}\rangle = [e^{-i\phi}\cos(\theta/2), \sin(\theta/2)]$,, where $\theta$ and $\phi$ are the spherical angles defining the classical spin direction of $\vec{n}$. The coherent spin state thus provides a bridge between the quantum spin operators and their classical interpretation, allowing us to treat spin textures and topological excitations in a semiclassical framework.

Introducing the unit vector $\vec{n} = \vec{S}/S$, with $|\vec{n}| = 1$ we can rewrite Hamiltonian (14) as

$$H = -JS^2\sum_{<i,j>}\vec{n}_i.\vec{n}_j + S^2\sum_{<i,j>}\vec{D}.\vec{n}_i\times\vec{n}_j \,. \tag{15}$$

If D = 0, the zero-temperature ground state consists of uniformly aligned spins, which we choose to point along the z-axis. For sufficiently large D, the ground state is a helix. In the experimentally relevant regime D/J ≪ 1, the magnetic texture varies slowly on the scale of the lattice spacing, and the continuum approximation becomes appropriate. We have

$$-\vec{n}_i.\vec{n}_j = \frac{1}{2}[(\vec{n}_i - \vec{n}_j)^2 - \vec{n}_i^2 - \vec{n}_j^2] \,. \tag{16}$$

In the continuum limit, we can approximate the spin difference using a spatial gradient. We assume the continuous magnetization field varies smoothly on the scale of the lattice spacing. Let $\vec{r}_i$ be the position of site *i*, and define a smooth field $\vec{n}(\vec{r})$ such that $\vec{n}_i = \vec{n}(\vec{r}_i)$. For a nearest neighbor in the *x*-direction, we have

$$\vec{n}_{i+\hat{x}} = \vec{n}(\vec{r}_i + \hat{x}) \approx \vec{n}(\vec{r}_i) + \partial_x \vec{n}(\vec{r}_i), \tag{17}$$

(since $a = 1$). Similarly, for the *y*-direction. We can write

$$\vec{n}_i.\vec{n}_{i+\hat{x}} = \vec{n}.(\vec{n} + \partial_x \vec{n}) = \vec{n}^2 + \vec{n}.\partial_x \vec{n} . \tag{18}$$

We have $\vec{n}^2 = 1$. Also

$$\vec{n}.\partial_x \vec{n} = \frac{1}{2}\partial_x(\vec{n}^2) = 0 . \tag{19}$$

The first-order contribution vanishes; therefore, the leading nontrivial contribution arises from the second-order term.

$$\vec{n}_{i+\hat{x}} = \vec{n} + \partial_x \vec{n} + \frac{1}{2}\partial_x^2 \vec{n} + ... \tag{20}$$

Then

$$\vec{n}_i.\vec{n}_{i+\hat{x}} = 1 + \vec{n}.\partial_x \vec{n} + \frac{1}{2}\vec{n}.\partial_x^2 \vec{n} + ... \tag{21}$$

Using again $\vec{n}.\partial_x \vec{n} = 0$, and differentiating $\vec{n}^2 = 1$ twice, we get

$$\partial_x^2(\vec{n}^2) = 2(\partial_x \vec{n})^2 + 2\vec{n}.\partial_x^2 \vec{n} = 0 \quad \rightarrow \quad \vec{n}.\partial_x^2 \vec{n} = -(\partial_x \vec{n})^2 . \tag{22}$$

Thus

$$\vec{n}_i.\vec{n}_{i+\hat{x}} = 1 - \frac{1}{2}(\partial_x \vec{n})^2 . \tag{23}$$

The exchange Hamiltonian becomes

$$H_J = -JS^2 \sum_{<i,j>} \vec{n}_i.\vec{n}_j \approx -JS^2 \sum_i \left\{ 1 - \frac{1}{2}\left[ (\partial_x \vec{n})^2 + (\partial_y \vec{n})^2 \right] \right\} . \tag{24}$$

Dropping the constant and converting the sum to an integral, we find.

$$H_J = \frac{JS^2}{2}\int d^2 r (\nabla \vec{n})^2 . \tag{25}$$

Following the same procedure as before, we obtain

$$\vec{n}_i \times \vec{n}_j = \vec{n}_i \times (\vec{n}_i + \nabla \vec{n}) = \vec{n}_i \times \nabla \vec{n} , \tag{26}$$

For concreteness, we conder a Bloch-type DM interaction for which the vector $\vec{D}$ is perpendicular to the bond direction. Thus

$$\vec{D}.\vec{n}_i \times \vec{n}_j \rightarrow D\vec{n}.(\vec{\nabla} \times \vec{n}) . \tag{27}$$

In the continuum limit, the full Hamiltonian becomes

$$H = \int d^2 r \left[ \frac{JS^2}{2}(\nabla \vec{n})^2 + DS^2 \vec{n}.(\vec{\nabla} \times \vec{n}) - BSn_z \right] . \tag{28}$$

For a Néel-type DM interaction, the DM term is given by

$$H_D = DS^2 \int d^2r[n_z(\vec{\nabla}.\vec{n}) - (\vec{n}.\vec{\nabla})n_z]. \tag{29}$$

To obtain the equation of motion, a simple procedure is to use the quantum expression $\dot{\vec{S}} = i[H, \vec{S}]$, and then take the continuum limit. For the exchange term, we get

$$\frac{\partial \vec{n}}{\partial t} = JS\vec{n} \times \nabla^2 \vec{n}. \tag{30}$$

We can also obtain the equation of motion from the classical ferromagnetic Lagrangian, as discussed in Appendix C. Some authors use the difference between the skyrmion energy density and that of the saturated state and write $(n_z - 1)$ for the magnetic field term.

Before discussing skyrmions, it is useful to briefly review two conventional magnetic textures: domain walls and magnetic bubbles. Domain walls are specific configurations in ferromagnetic materials (Fig. 6). Under fixed boundary conditions, domain walls can be regarded as one-dimensional topological solitons connecting distinct magnetic ground states. The magnetization orientation can be either in-plane or out-of-plane, depending on the thickness of the magnetic layer and the surrounding layers. Domain walls are topologically protected insofar as they separate distinct magnetic ground states and cannot be removed by local continuous deformation. However, their protection depends on the boundary conditions, and they can often annihilate at sample edges or with other domain walls. Their topological protection is generally less robust than that of skyrmions. In a domain wall, spins at the far left and far right are in opposite directions. Because the magnetization approaches different orientations at the two ends of the system, the order parameter does not approach a unique value at infinity. Consequently, the spatial domain cannot be compactified to a sphere in the usual way, and one cannot define a skyrmion topological charge.

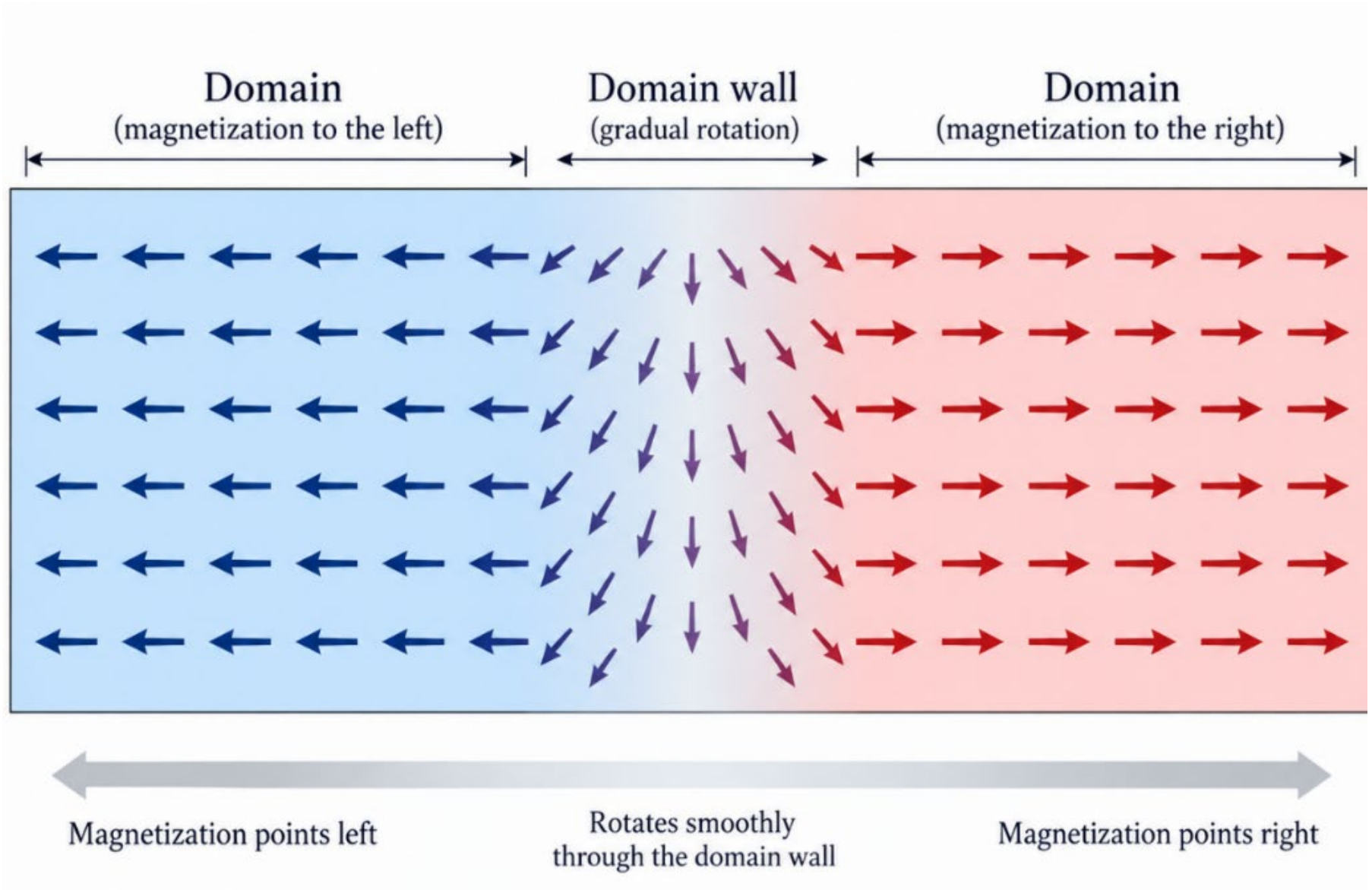

Figure 6. A domain wall.

A magnetic bubble (Fig.7) is a closed magnetic domain embedded within a thin ferromagnetic film and surrounded by a domain wall. It consists of a small cylindrical region whose magnetization is oriented opposite to that of the surrounding material. For example, the magnetization inside the bubble may point upward, while the surrounding film's magnetization points downward. The transition between these oppositely magnetized regions occurs through a domain wall, where the spins rotate continuously between the two orientations. Magnetic bubbles typically arise in materials exhibiting strong perpendicular magnetic anisotropy when an external magnetic field is applied near a critical value. Under these conditions, the system minimizes its total energy by forming isolated reversed domains rather than extended stripe domains. A magnetic bubble's stability results from a balance among competing energy contributions, including domain-wall energy, magnetostatic interactions, magnetic anisotropy, and the applied magnetic field. In addition, material defects and inhomogeneities can act as pinning sites that further stabilize the bubble. Consequently, creating or annihilating a bubble requires overcoming an energy barrier, allowing these structures to exist as metastable magnetic configurations. Magnetic bubbles can be manipulated by magnetic field gradients and may expand, contract, merge, split, and collapse. Despite their stability, conventional magnetic bubbles are not generally protected by topology. A conventional magnetic bubble generally carries zero skyrmion number and therefore lacks topological protection. Although the magnetization within the domain wall may rotate clockwise or counterclockwise, the overall texture usually carries zero topological charge. As a result, the bubble can continuously shrink until its radius vanishes, leading to complete annihilation without encountering a fundamental topological obstruction. Therefore, a conventional magnetic bubble's stability is energetic rather than topological. This behavior contrasts with magnetic skyrmions, which possess a nonzero topological charge. Their annihilation requires a change of topological sector, typically involving a singular configuration or a substantial energy barrier, making them considerably more robust than conventional magnetic bubbles.

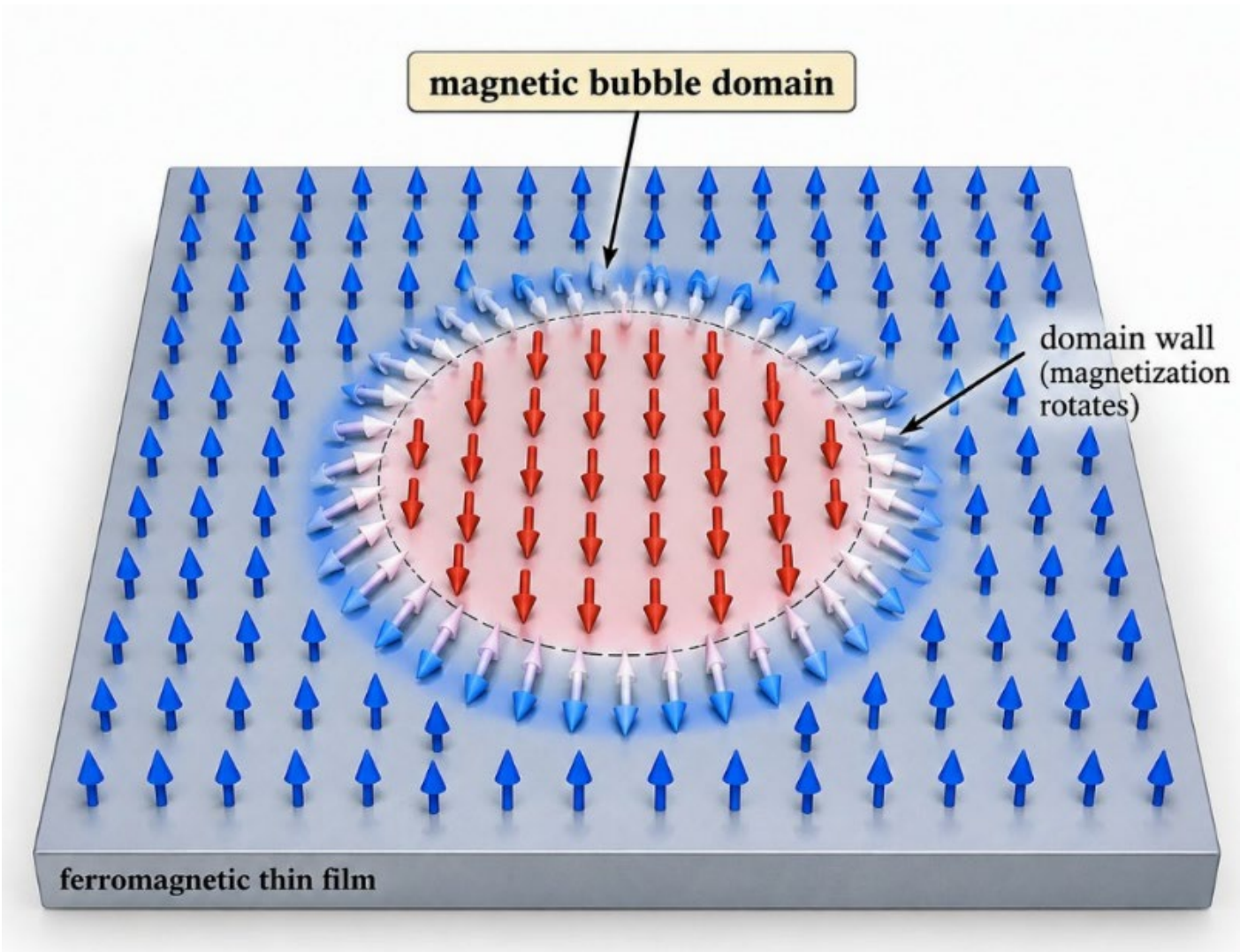

Figure 7. A magnetic bubble.

# 4. Ferromagnetic Skyrmions.

Skyrmions were originally proposed by Tony Skyrme in 1960 in the context of particle physics [15]. In magnetic systems, they are noncollinear, spatially inhomogeneous spin textures in which the spin orientation changes continuously over characteristic length scales ranging from a few nanometers to several micrometers. They can appear as individual quasiparticles or as ordered crystal lattices in bulk materials and thin film heterostructures. Mathematically, a skyrmion is a finite-energy field configuration characterized by a nontrivial topological charge. Its stability may arise from topology, energetic barriers, or both. This distinguishes skyrmions from certain other topological textures, such as isolated vortices or hedgehog defects, whose energies may diverge with system size in simple continuum models. Magnetic skyrmions have attracted considerable attention due to their potential applications in spintronics, high-density data storage, and quantum computing. Their stability, nanoscale size, and ability to be manipulated by extremely low current densities make them promising candidates for future information technologies. Skyrmion formation requires a microscopic mechanism that continuously twists the spin orientation. One of the most important mechanisms is the Dzyaloshinskii–Moriya interaction, mentioned before. In addition, competing exchange interactions, magnetic dipolar (long-range) interactions, magnetic anisotropy, or geometrical confinement can stabilize skyrmions.

Using $\Theta$ and $\Phi$ to denote field variables, and $\theta$, and $\varphi$ for angular coordinates, we write

$$\vec{n}(\vec{r}) = (\cos\Phi(\varphi)\sin\Theta(\vec{r}), \sin\Phi(\varphi)\sin\Theta(\vec{r}), \cos\Theta(\vec{r})) . \quad (31)$$

and

$$\vec{r} = (x, y) = (r\cos\varphi, r\sin\varphi) , \qquad r = |\vec{r}\,| . \quad (32)$$

To define a skyrmion, we usually impose a boundary condition such as $\vec{n}(\vec{r} \to \infty) =$ constant. For example, all spins pointing upward far away from the skyrmion. Because the magnetization approaches a unique value at infinity, we can identify all points at infinity. The plane is therefore compactified to a sphere: $\mathbb{R}^2 \bigcup \{\infty\} \simeq S^2$, and we can write the map $S^2 \to S^2$. We take $\Theta(\vec{r}) = \Theta(r)$.

Thus, the exchange term is given by

$$H_J = \frac{JS^2}{2}\int d^2r[(\nabla\Theta)^2 + \sin^2\Theta(\nabla\Phi)^2] . \quad (33).$$

The equation of motion in terms of $\Theta$, and $\Phi$ is more involved, but for static solutions, we can use $\vec{n} \times \nabla^2\vec{n} = 0$, and obtain

$$(\partial_x^2 + \partial_y^2)\Theta - \sin\Theta\cos\Theta[(\partial_x\Phi)^2 + (\partial_y\Phi)^2] = 0 . \quad (34)$$

$$2\cos\Theta(\partial_x\Theta\partial_x\Phi + \partial_y\Theta\partial_y\Phi) + +\sin\Theta(\partial_x^2 + \partial_y^2)\Phi = 0 . \quad (35)$$

Or

$$\nabla^2\Theta = \sin\theta\cos\theta\left(\vec{\nabla}\Phi\right)^2 , \qquad \nabla^2\Phi = -2\cot\Theta\vec{\nabla}\Theta.\vec{\nabla}\Phi . \quad (36)$$

A trivial solution is $\Theta = 0$. It represents the ferromagnetic configuration with all spins aligned along the *z*–direction.

Using polar coordinates $(r,\varphi)$ and the ansatz $\Theta = \Theta(r)$, $\Phi = m\varphi + \gamma$, where $m$ is the vorticity and $\gamma$ the helicity (the helicity is a phase factor that describes the azimuthal angle between $\vec{n}$ and the radial vector), we get

$$(\nabla\Phi)^2 = \frac{m^2}{r^2}, \qquad \nabla^2\Theta = \frac{d^2\Theta}{dr^2} + \frac{1}{r}\frac{d\Theta}{dr}. \tag{37}$$

Then, we arrive at the equation

$$\frac{d^2\Theta}{dr^2} + \frac{1}{r}\frac{d\Theta}{dr} - \frac{m^2}{r^2}\sin\Theta\cos\Theta = 0 \ . \tag{38}$$

One solution for this equation, as shown in Appendix A, is given by (Fig. 8)

$$\Theta(r) = 2\arctan(R/r), \tag{39}$$

where $R$ is the skyrmion radius, with the origin at the skyrmion center. We see that the spins are oriented in the down direction at $r = 0$, and in the up direction at $r \to \infty$. The solution with $R \leftrightarrow r$ corresponds to the same texture shape but with core and background magnetization swapped. In a skyrmion, the core magnetic moments are antiparallel to the external magnetic field (oriented in the $z$-direction). Using Eq. (39) we get

$$n_z(r) = \cos\Theta(r) = \frac{r^2 - R^2}{r^2 + R^2}. \tag{40}$$

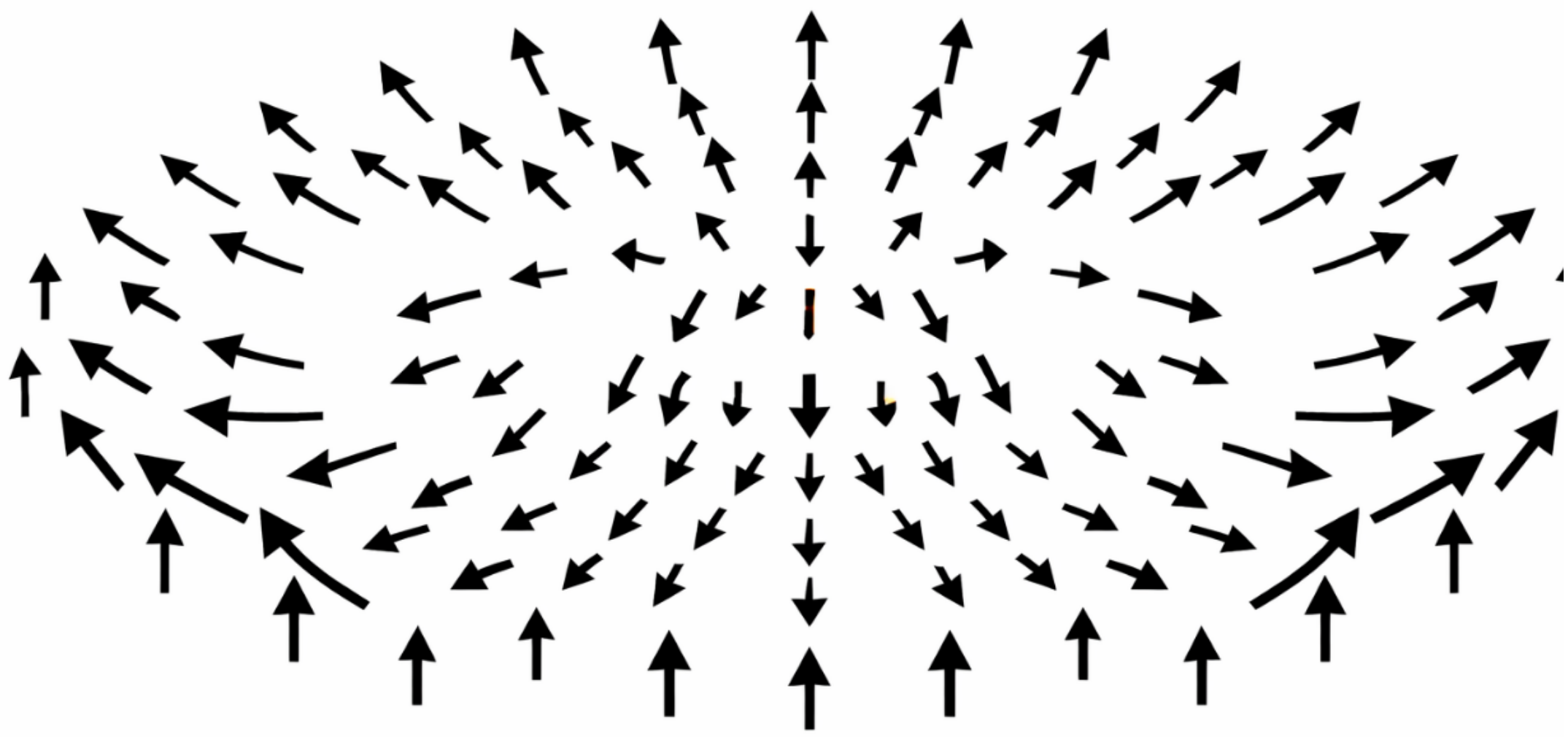

Figure 8: In this example of a skyrmion, the field at infinity points in the positive $z$-direction, while near the origin, it points in the $-z$-direction.

In Figure 9, I show the skyrmion profile $\theta(r)$. As noted earlier, a skyrmion can be considered as a mapping $S^2 \to S^2$, and therefore, as discussed in section 2, it can be characterized by a topological charge, also called skyrmion number, given by the expression

$$Q = \frac{1}{4\pi}\int d^2r\vec{n}.\left(\frac{\partial\vec{n}}{\partial x}\times\frac{\partial\vec{n}}{\partial y}\right). \tag{41}$$

As mentioned before, $Q$ counts how many times $\vec{n}(x,y)$ wraps around the unit sphere as the coordinates $x$ and $y$ span the entire plane. It corresponds to the degree of the map $S^2 \to S^2$. If we smoothly deform the skyrmion (for instance, stretching it, squeezing it, moving the core, etc.), as long as the spin field remains continuous and the boundary spins remain fixed, the mapping $S^2 \to S^2$ stays in the same homotopy class. The integrand in Eq. (41) may change, but the total integral remains the same. Thus, skyrmions can have different shapes but the same $Q$. Note that this holds for an idealized mathematical theory in which the spin field remains smooth and continuous. However, in real materials, several mechanisms can break this continuity. The most common mechanism is that the skyrmion shrinks until its radius reaches the atomic scale. At that point, the continuum description fails, and neighboring spins can no longer vary smoothly. For instance, the topological charge can change from $Q = 1$ to $Q = 0$. We can also say that topology is exact only in the continuum model, where $\vec{n}(x, y)$ is continuous everywhere. But a crystal is a discrete lattice of spins, and the topological invariant is not mathematically exact. However, an energy barrier may still be very high, which is why skyrmions can be long-lived. Skyrmions can be metastable with lifetimes ranging from microseconds to years, depending on the barrier height. At finite temperature, fluctuations can overcome the energy barrier, leading to spontaneous annihilation or creation events. Note also that a skyrmion can escape through a boundary (in a finite system) without requiring a singularity in the bulk.

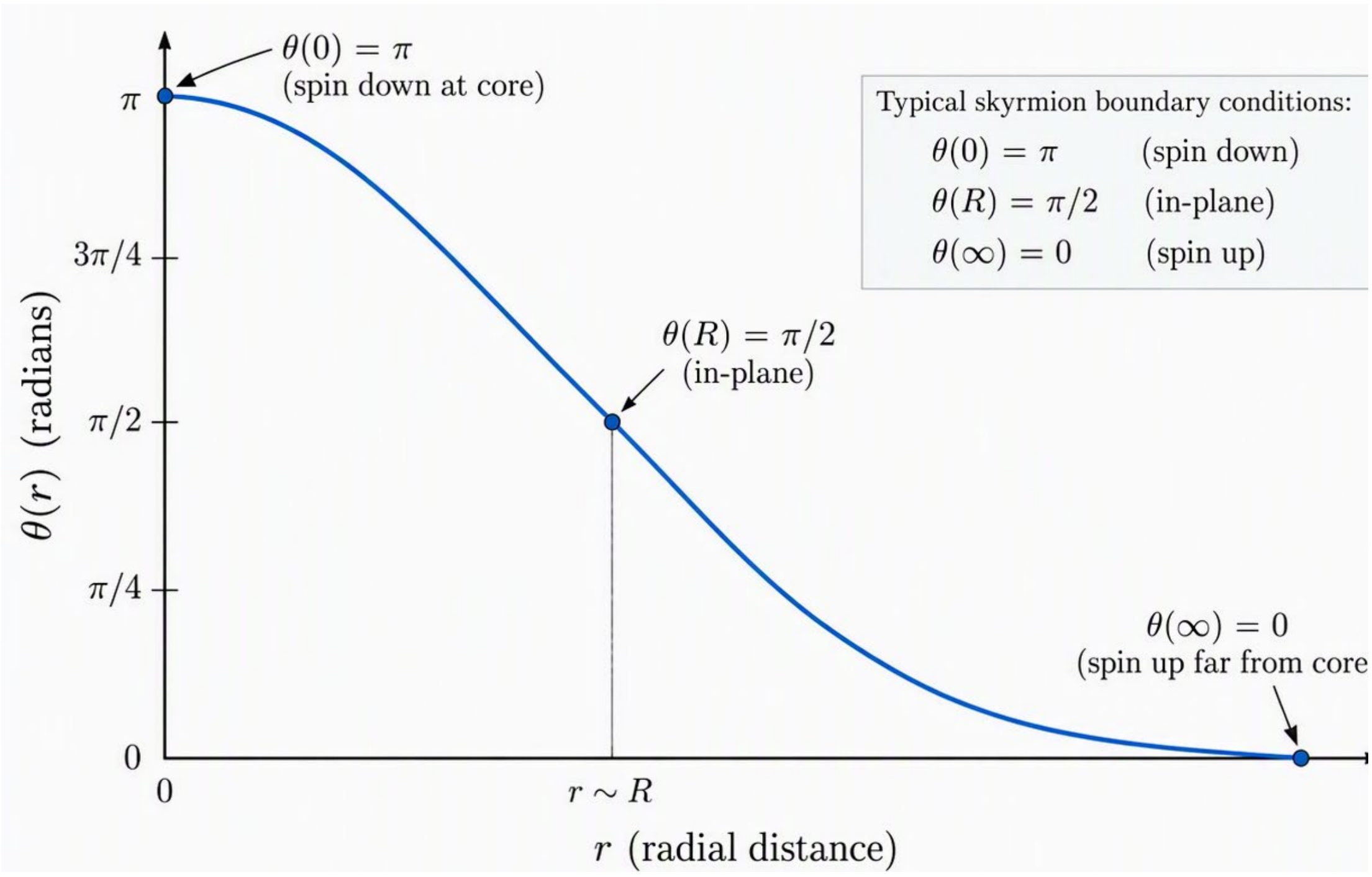

Figure 9. The skyrmion profile.

The term

$$b_z = \vec{n}.\left(\frac{\partial \vec{n}}{\partial x} \times \frac{\partial \vec{n}}{\partial y}\right), \tag{42}$$

can be interpreted as the $z$-component of a fictitious magnetic field. Using the ansatz

$$\vec{n}(\vec{r}) = (\cos\Phi(m\varphi+\gamma)\sin\Theta(r), \sin\Phi(m\varphi+\gamma)\sin\Theta(r), \cos\Theta(r)), \tag{43}$$

a straightforward calculation gives, in polar coordinates

$$\vec{n}.(\partial_r \vec{n} \times \partial_\varphi \vec{n}) = m \frac{d\Theta}{dr} \sin\Theta \, .$$

Thus

$$Q = \frac{m}{4\pi} \int_0^\infty dr \int_0^{2\pi} d\varphi \frac{d\Theta(r)}{dr} \sin\Theta(r) = m \frac{\cos\Theta(0) - \cos\Theta(\infty)}{2} . \tag{44}$$

If $\Theta(0) = \pi$, $\Theta(\infty) = 0$, we have $Q$ = - $m$. As discussed before, we define the integer number $m = [\phi(\varphi)]_{\varphi=0}^{\varphi=2\pi} / 2\pi$ as the vorticity of the skyrmion. The helicity is defined by the phase $\gamma$ in $\Phi(\varphi) = m\varphi + \gamma$. It follows that the topological charge is determined by the product of the vorticity (which describes how the in-plane spins rotate around the center) and the polarity $p$ (which describes whether the core magnetization points up or down). Several possible skyrmion types correspond to different values of vorticity and helicity. They can be classified into two types: Bloch and Néel, depending on the type of DM interaction that stabilizes them. Bloch and Néel skyrmions have the same vorticity, $m$ = 1, but differ in their helicity. Bloch skyrmions correspond to helicities $\gamma = \pm\pi / 2$, leading to tangential in-plane magnetization. Néel skyrmions correspond to helicities $\gamma =$ 0 or $\pi$, leading to radial in-plane magnetization (Fig. 10). Antiskyrmions have opposite vorticity, which results in a topological charge of opposite sign for a fixed polarity,

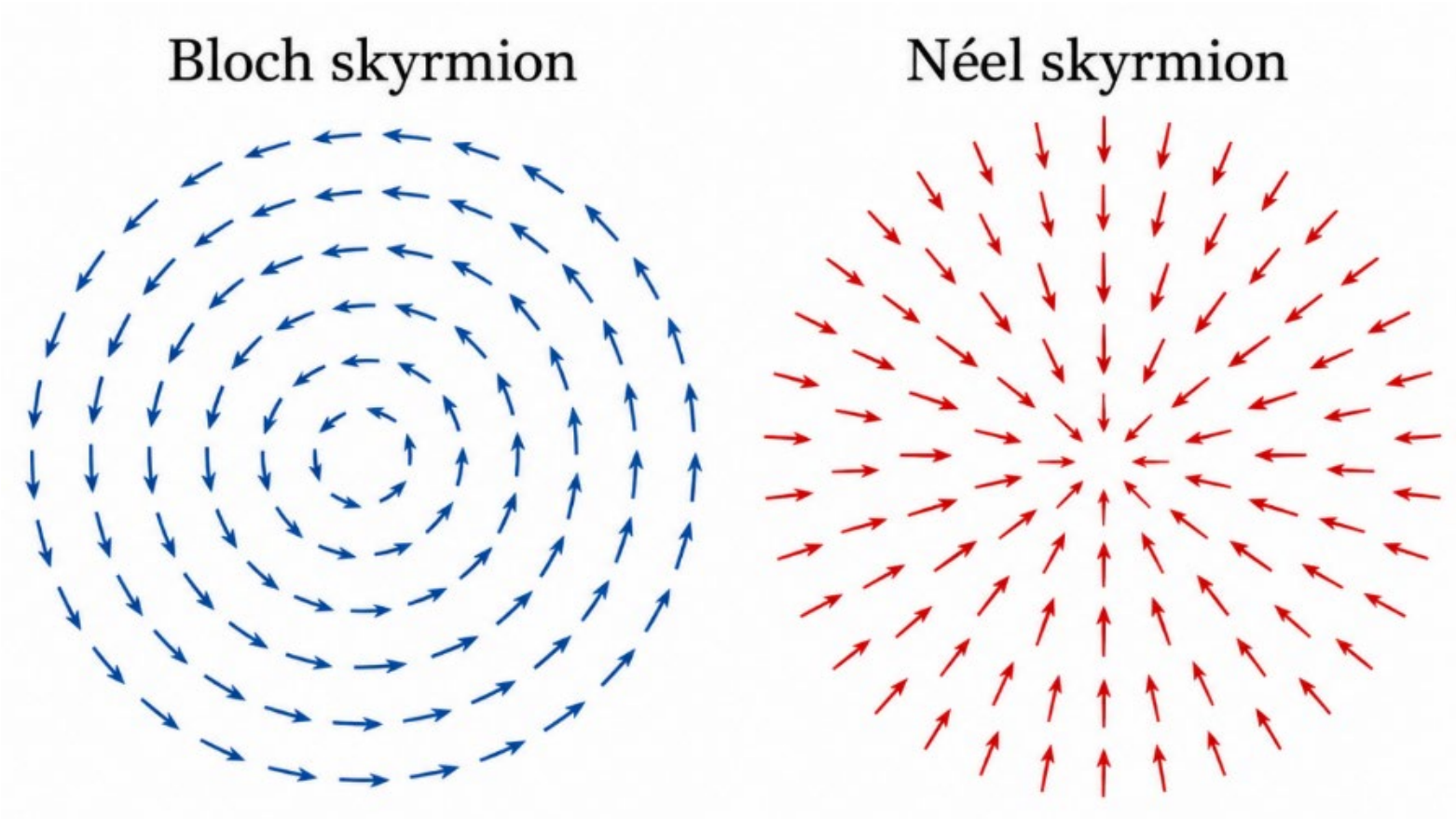

Figure 10. In-plane spin configurations for Bloch and Néel skyrmions.

Skyrmion stability results from energy barriers (the energy required to transition between states with different topological charges). However, thermal fluctuations and discrete magnetic moments can overcome these barriers. The spins in a magnetic skyrmion rotate continuously varying spin texture with nonzero topological charge. from the center to the outer edge. The essential point is that the magnetization field wraps around the unit sphere, giving the skyrmion its non-zero topological charge. At the center, the spins point fully up or down (out of the plane). Toward the edge, the spins gradually rotate into the in-plane direction. At the boundary, the spins point fully opposite to the center (down if the center is up, or vice versa). This continuous rotation forms a *vortex-* *like* or *hedgehog-like* pattern depending on the skyrmion type. In a Néel skyrmion, spins point radially outward or inward in the plane, forming a "hedgehog" pattern. It is common in thin films with strong interfacial Dzyaloshinskii–Moriya interaction. The rotation is *radial*: center → outward twist. For the Bloch skyrmion, spins rotate tangentially, forming a swirling vortex. This is usual in bulk chiral magnets like MnSi. The rotation is *azimuthal*: center → circular twist.

Using the relation $\vec{a}.(\vec{b}\times\vec{c}) = \vec{b}.(\vec{c}\times\vec{a})$, we have the identity

$$(\nabla\vec{n})^2 = (\partial_x\vec{n} \mp \vec{n}\times\partial_y\vec{n})^2 \pm 2\vec{n}.(\partial_x\vec{n}\times\partial_y\vec{n}) . \tag{45}$$

Then, for the pure exchange model, the energy can be written as

$$E = \frac{JS^2}{2}\int d^2r(\nabla\vec{n})^2 = \frac{JS^2}{2}\int d^2r(\partial_x\vec{n} \mp \vec{n}\times\partial_y\vec{n})^2 \pm JS^2\int d^2r\vec{n}.(\partial_x\vec{n}\times\partial_y\vec{n}) . \tag{46}$$

Since the square term is positive, we have

$$E \geq \pm JS^2\int d^2r\vec{n}.(\partial_x\vec{n}\times\partial_y\vec{n}) = \pm 4\pi JS^2 Q . \tag{47}$$

We choose the sign so that the right-hand side is positive, For the isotropic exchange model without anisotropy, magnetic field, or Dzyaloshinskii-Moriya interaction, the skyrmion energy satisfies the Belavin-Polyakov bound

$$E \geq 4\pi JS^2 |Q| . \tag{48}$$

For the radial solution with vorticity *m*, we have *Q* = - *m*, so

$$E \geq 4\pi JS^2 |m| . \tag{49}$$

Thus, for fixed boundary conditions, the minimal energy is obtained for $|m| = 1$. Higher $|m|$ configurations cost more energy and tend to split into $|m|$ unit skyrmions. Skyrmions have been experimentally observed in bulk chiral magnets, ultrathin ferromagnetic films with interfacial Dzyaloshinskii–Moriya interaction, multilayer hetereostructures, multiferroics, and synthetic antiferromagnets. Advanced imaging techniques such as Lorentz transmission electron microscopy, spin-polarized scanning tunneling microscopy, and magnetic force microscopy have enabled direct visualization of skyrmion structures.

## 5. Dzyaloshinskii- Moriya interaction.

The Dzyaloshinskii–Moriya interaction (DMI) appears in several distinct mathematical forms, depending on the symmetry of the magnetic system [1, 16,17,18]. These forms determine whether skyrmions are Néel-type, Bloch-type, or possess more exotic spin textures. There are three major forms of the Dzyaloshinskii- Moriya interaction: i) General Lifshitz-invariant form, ii) Bloch-type DMI, iii) Interfacial Néel-type DMI. The different forms emerge purely from symmetry. We begin with the Hamiltonian

$$H_{DM} = \sum_{<i,j>} \vec{D}_{ij}.(\vec{S}_i\times\vec{S}_j) , \tag{50}$$

where $\vec{D}_{ij}$ depends on the crystal symmetry and the bond connecting sites $i$ and $j$. We have $\vec{D}_{ij} = -\vec{D}_{ji}$. Passing to the continuum limit, we write

$$\vec{n}(\vec{r}+\vec{\delta}) = \vec{n}(\vec{r}) + (\vec{\delta}.\vec{\nabla})\vec{n}(\vec{r}). \tag{51}$$

And obtain

$$H_{DM} = S^2 \sum_{<i,j>} \vec{D}_{ij}.[\vec{n}\times(\vec{\delta}.\vec{\nabla})\vec{n}]. \tag{52}$$

In component form, this becomes

$$H_{DM} = S^2 \sum_{<i,j>} D_{ij,k}\varepsilon_{klm} n_l \delta_p \partial_p n_m. \tag{53}$$

The antisymmetric Levi-Civita tensor $\varepsilon_{klm}$ equals +1 for an even permutation of the sequence (1, 2, 3), -1 for an odd permutation, and vanishes whenever two indices are identical. It is convenient to define the Lifshitz invariants [1]:

$$L_{ij}^{(k)} = n_i \frac{\partial n_j}{\partial x_k} - n_j \frac{\partial n_i}{\partial x_k}, \tag{54}$$

which are the only antisymmetric combinations of the magnetization components that are linear in spatial derivatives. The most general DMI Hamiltonian can then be written as

$$H_{DM} = S^2 \sum_{i,j,k} D_{ij,k} L_{ij}^{(k)}. \tag{55}$$

*Bloch-type skyrmion*

Bulk DMI arises in non-centrosymmetric chiral crystals. These systems possess broken inversion symmetry, while retaining cubic rotation symmetry. The only allowed term, in this case, is the isotropic DMI term $\vec{n}.(\vec{\nabla}\times\vec{n})$, which favors Bloch-type skyrmions.

*Néel -type skyrmions*

Interfacial DMI occurs in ultrathin ferromagnetic films adjacent to a heavy metal. In this case, inversion symmetry is broken along the interface normal $\hat{z}$ while rotational symmetry around $\hat{z}$ is preserved. Mirror symmetries further constrain the allowed Lifshitz invariant. Then only in-plane derivatives $\partial_x$, $\partial_y$ survive. As a consequence, the only allowed combination is

$$n_z(\partial_x n_x + \partial_y n_y) - (n_x \partial_x n_z + n_y \partial_y n_z). \tag{56}$$

And the corresponding Hamiltonian can be written as

$$H_{DM} = S^2 D \int dr^2 [n_z(\vec{\nabla}.\vec{n}) - (\vec{n}.\vec{\nabla})n_z]. \tag{57}$$

This form stabilizes Néel-type skyrmions.

For Bloch Skyrmions in two-dimensional (no $z$ dependence, $\partial_z \vec{n} = 0$) polar coordinates, we have

$$\vec{n}.(\vec{\nabla}\times\vec{n}) = n_r \frac{1}{r}\frac{\partial n_z}{\partial \Phi} - n_\Phi \frac{\partial n_z}{\partial r} + n_z \frac{1}{r}\left(\frac{\partial(n_\Phi r)}{\partial r} - \frac{\partial n_r}{\partial \Phi}\right), \tag{58}$$

where

$$\vec{n} = n_r \hat{e}_r + n_\Phi \hat{e}_\Phi + n_z \hat{e}_z.$$

With

$$n_r = \sin\Theta\cos\Phi\,, \quad n_\Phi = \sin\Theta\sin\Phi\,, \quad n_z = \cos\Theta\,. \tag{59}$$

For a skyrmion $\Theta = \Theta(r)$, and $\Phi = \phi + \gamma$. Then

$$\vec{n}.(\vec{\nabla}\times\vec{n}) = \sin\gamma\left[\frac{d\Theta}{dr} + \frac{\sin 2\Theta}{2r}\right]. \tag{60}$$

And the Hamiltonian becomes

$$H_{DM} = DS^2\int dr^2 \sin\gamma\left[\frac{d\Theta}{dr} + \frac{\sin\Theta\cos\Theta}{r}\right]. \tag{61}$$

For the Bloch skyrmion: $\gamma = \pm\pi/2 \;\rightarrow\; \sin\gamma = \mp 1$ and the DMI contribution is maximal in magnitude. In contrast, for the Néel skyrmion $\gamma = 0$, and the contribution vanishes.

Repeating the same procedure for the Néel-type skyrmions, one finds

$$H_{DM} = DS^2\int dr^2 \cos\gamma\left[\frac{d\Theta}{dr} + \frac{\sin\Theta\cos\gamma}{r}\right]. \tag{62}$$

For a Néel skyrmion ($\gamma$ = 0 or $\pi$), and the term reduces to the standard radial DMI form. Eq (62) vanishes for a Bloch skyrmion. Each type of DMI selectively stabilizes the spin texture compatible with the underlying crystal symmetry.

An anisotropy term

$$H_A = -A\sum_i (S_i^z)^2\,, \tag{63}$$

may also be present in the Hamiltonian. It leads to a term $-A(\cos\Theta)^2$ in the continuous Hamiltonian. Magnetic anisotropy determines the preferred magnetization direction and strongly influences skyrmion stability. Two important cases are: perpendicular magnetic anisotropy (PMA, A > 0), which favors out-of-plane magnetization and stabilizes compact skyrmions, and easy-plane anisotropy (A < 0), which favors in-plane alignment and can stabilize vortex-like textures. The anisotropy energy strength determines skyrmion size, domain-wall width, the stability of multi-skyrmion configurations, and the shape of hopfion tubes in 3D. In thin films, PMA is often essential for stabilizing isolated skyrmions at room temperature. An external magnetic field also plays an important role. By aligning the background magnetization, it stabilizes isolated skyrmions, modifies their radius, and drives transitions between helical, skyrmion-lattice, and ferromagnetic phases.

The role of the DMI can now be understood more clearly. As discussed previously, topology protects a skyrmion against smooth deformations into the uniform state. However, topology alone does not prevent the skyrmion from shrinking to zero size and disappearing through a singular configuration. Exchange interactions favor parallel spins and thus tend to homogenize the magnetization. Magnetic anisotropy and Zeeman coupling similarly favor uniform alignment along preferred directions. Consequently, without DMI, a skyrmion generally has no preferred finite radius and tends to collapse. The DMI counteracts this tendency by introducing an energetic preference for a finite twist of the magnetization. It selects a preferred chirality, lowers the energy of chiral spin textures, and competes with exchange, anisotropy, and Zeeman energies to stabilize a finite-size skyrmion. The magnitude and sign of $D$ determine important properties such as skyrmion radius, chirality, and stability [9].

In some chiral magnets, including MnSi and FeGe, the DMI is sufficiently strong that the thermodynamic ground state is a skyrmion lattice rather than a simple ferromagnetic state. In simplified models, increasing $D$ generally reduces the equilibrium skyrmion radius. Chiral point groups $T$, $O$, and $D_n$ favor Bloch skyrmions with helicity $\gamma = \pm\pi/2$, whereas polar point groups $C_{nv}$ favor Néel skyrmions with helicity $\gamma = 0$ or $\pi$. Point groups $C_n$, which are simultaneously chiral and polar, may stabilize skyrmions with intermediate helicities [9].

The asymmetric skyrmion energy density (with $m$ = 1) including the DMI and magnetic anisotropy terms is given by

$$\varepsilon = JS^2\left[\left(\frac{\partial\Theta}{\partial r}\right)^2 + \frac{\sin^2\Theta}{r^2}\right] - DS^2\sin\gamma\left[\frac{d\Theta}{dr} + \frac{\sin 2\Theta}{2r}\right] - AS^2\cos^2\Theta\,. \tag{64}$$

For a radial functional

$$E = 2\pi\int r dr \varepsilon(r,\Theta,\Theta_r)\,. \tag{65}$$

The Euler-Lagrange equation is

$$\frac{1}{r}\frac{d}{dr}\left(r\frac{\partial\varepsilon}{\partial\Theta_r}\right) - \frac{\partial\varepsilon}{\partial\Theta} = 0\,, \tag{66}$$

which leads to

$$2J\left[\frac{\partial^2\Theta}{\partial r^2} + \frac{1}{r}\frac{\partial\Theta}{\partial r} - \frac{\sin\Theta\cos\Theta}{r^2}\right] + D\frac{\sin\gamma}{r}(\cos 2\Theta - 1) - 2A\sin\Theta\cos\Theta = 0\,. \tag{67}$$

Since Eq. (67) cannot be solved analytically, it must be treated numerically. Nevertheless, an approximate analytic solution can be obtained for large values of anisotropy A, following the approach of Ref.[18]. Physically, we can view a skyrmion with a core $\Theta = \pi$, surrounded by a uniform background $\Theta \approx 0$, with a narrow transition region around some radius $R$. This transition is essentially a circular domain wall. Near the wall, we use the following approximation: inside the wall $\Theta(r < R) \approx \pi$, outside $\Theta(r > R) \approx 0$. The angle $\Theta$ varies rapidly in the region near $R$. In that narrow region, we treat the wall locally as a one-dimensional wall along a coordinate $x = r - R$. Curvature terms (like 1 / $r$) are small corrections and can be neglected. Thus, we can write the equation

$$2J\frac{d^2\Theta}{dx^2} = A\sin(2\Theta)\,, \tag{68}$$

which is the well-known sine-Gordon equation with solution

$$\Theta(x) = 2\arctan[\exp(x/\Delta)], \tag{69}$$

where $\Delta = \sqrt{J/A}$, up to a shift and sign conventions. This profile interpolates smoothly from $\Theta = 0$ to $\Theta = \pi$, over a characteristic width $\Delta$. Then, we can write

$$\Theta(r) = 2\arctan[\exp((R-r)/\Delta)]. \tag{70}$$

Where $\Theta(R) = \pi/2$. In a more refined treatment, the DMI modifies the effective wall width $\Delta$. Substituting this ansatz into the equation of the full energy, treating $R$ and $\Delta$ as variational parameters, allows us to determine the skyrmion radius and wall width by energy minimization. This approximation satisfies the expected boundary conditions: $r << R : \Theta \approx \pi$, $r >> R : \Theta \approx 0$. For $r \sim R$ there is a rapid change over scale $\Delta$. We can think of this solution as a circular wall at radius $R$. Taking this ansatz in the Hamiltonian, one finds that the contributions of each term scale as follows.

The exchange energy is $E_{ex} \sim 4\pi J + c_1 JR/\Delta$, where $4\pi J$ is the topological core energy, and the second term comes from the domain wall. The anisotropic term gives $E_A \sim c_2 AR\Delta$, and for the DMI, we find $E_D \sim c_3 DR$. Here $c_i$ are order-one numerical constants from the detailed integrals. If a magnetic field $B$ is present, the Zeeman energy is

$$E_z = -\mu_0 M_S B \int (1-n_z) d^2 r,$$

per unit area. The core area scales as $\pi R^2$. Hence

$$E_z \sim \gamma B R^2$$

Thus, the total energy is approximately given by

$$E(R) \sim 4\pi J + c_1 JR/\Delta + c_2 AR\Delta - c_3 DR + \gamma BR^2. \tag{71a}$$

First, we minimize with respect to $\Delta$.

$$\frac{\partial E}{\partial \Delta} = -c_1 J \frac{R}{\Delta^2} + c_2 AR = 0 \quad \rightarrow \quad \Delta_* \sim \sqrt{\frac{J}{A}}. \tag{71b}$$

Hence

$$E(R) \sim 4\pi J + c_4 \sqrt{JA} R - c_3 DR + \gamma BR^2, \tag{72a}$$

with $c_4 \sim 2\sqrt{c_1 c_2}$. Then we minimize with respect to $R$:

$$\frac{dE}{dR} \sim c_4 \sqrt{JA} - c_3 D + 2\gamma BR = 0, \quad , \tag{72b}$$

and obtain the critical radius

$$R_* = \frac{c_3 D - c_4 \sqrt{J} A}{2\gamma B}. \tag{72c}$$

Otherwise, for $B = 0$ and a circular wall, the exchange energy density contains a term $J \sin^2\theta / r^2$. Integrating near $r = R$. we get

$$\int drr \left( \frac{J \sin^2\theta}{r^2} \right) \sim \frac{J\Delta}{R}. \tag{73a}$$

Therefore the exchange energy contains a correction

$$E_{ex} = 4\pi J + c_1 J \frac{R}{\Delta} + c_5 J \frac{\Delta}{R}. \tag{73b}$$

After minimizing over $\Delta$, we get

$$E(R) = 4\pi J + c_4\sqrt{JA}R - c_3DR + \frac{\eta J}{R}. \tag{73c}$$

Now

$$\frac{dE}{dR} = c_4\sqrt{JA} - c_3D - \frac{\eta J}{R} = 0, \tag{73d}$$

gives

$$R_* = \sqrt{\frac{\eta J}{c_3D - c_4\sqrt{JA}}}. \tag{74}$$

. The approximate solution

$$\Theta(r) = 2\arctan[\exp((R_* - r)/\Delta_*)], \tag{75}$$

is a good analytic skyrmion shape in both ferromagnets and antiferromagnets.

## 6. Nonlinear sigma model.

Below, I derive the nonlinear sigma model that describes the classical Heisenberg antiferromagnet in the continuum limit. Antiferromagnetic skyrmions are skyrmions in the topological sense (the static equation is the same as for the ferromagnet), but they exhibit different dynamics than ferromagnetic skyrmions. For clarity, I focus on the one-dimensional case along the *z*-direction; the extension to higher dimensions proceeds analogously. Appendix C presents the derivation for the quantum model using coherent spin states. The Hamiltonian for a one-dimensional antiferromagnet is [14,19,20]:

$$H = J\sum_i \vec{S}_i.\vec{S}_{i+1} \tag{76}$$

In this section, I will write the spin variables using four angles in spherical coordinates [21,22] :

$$\vec{S}_i = (-1)^i S[\sin(\theta_i + (-1)^i v_i)\cos(\phi_i + (-1)^i \alpha_i),$$
$$\sin(\theta_i + (-1)^i v_i)\sin(\phi_i + (-1)^i \alpha_i), \cos(\theta_i + (-1)^i v_i)], \tag{77}$$

I label by $i$ any site of the lattice and parametrize the local spin orientation in terms of angles $\theta$ and $\phi$ that describe the smooth Néel phase. The fields $v$ and $\alpha$ represent staggered fluctuations, describing deviations from perfect antiferromagnetic alignment. At low temperatures, these fluctuations are small, so we retain only terms up to second order in $v$ and $\alpha$. In the continuum limit, we further assume that spatial variations of the fields are slow, so we focus on the system's long-wavelength behavior. Using the equation of motion

$$\frac{d\vec{S}_i}{dt} = \{\vec{S}_i, H\}_{PB} = J\vec{S}_i \times (\vec{S}_{i+1} + \vec{S}_{i-1}), \tag{78}$$

where *PB* means the Poisson bracket, taking $z = ia$, where $a$ is the lattice spacing, which we set to unity, we obtain

$$\dot{\theta}/c = 2\alpha \sin\theta, \qquad \dot{\phi}/c = -2v/\sin\theta.$$

$$\dot{v}/c = -2\alpha v \cos\theta - \theta_z \phi_z \cos\theta - \phi_{zz} \sin\theta, \tag{79}$$

$$\dot{\alpha}\frac{\sin\theta}{c} = \frac{1}{2}\theta_{zz} + (\phi_z)^2 - \left(\frac{2v}{\sin\theta}\right)^2 + 4\alpha^2 . \tag{80}$$

Where $c = 2JS$ is the spin wave velocity. Inserting Eq. (77) into Eq. (76), using Eqs. (79) and (80) to eliminate the small angles, and taking the continuum limit, we obtain the classical O (3) nonlinear sigma model:

$$H = \frac{JS^2}{2}\int dz \left\{ \left(\frac{\partial\theta}{\partial z}\right)^2 + \sin^2\theta \left(\frac{\partial\phi}{\partial z}\right)^2 + \frac{1}{c^2}\left[ \left(\frac{\partial\theta}{\partial t}\right)^2 + \sin^2\theta\left(\frac{\partial\phi}{\partial t}\right)^2 \right] \right\}. \tag{81}$$

Note that the essential nonlinearities related to the Hamiltonian are included fully since no restrictions on the values of $\theta$ and $\phi$ are made.

To obtain the equations of motion from Eq.(76) using the Hamilton equations, we need to define the momenta

$$p_\theta = \frac{1}{2J}\frac{\partial\theta}{\partial t}, \qquad p_\phi = \frac{1}{2J}\sin^2\theta\frac{\partial\phi}{\partial t}, \tag{82}$$

and write

$$H = \frac{JS^2}{2}\int dz \left\{ \left(\frac{\partial\theta}{\partial z}\right)^2 + \sin^2\theta\left(\frac{\partial\phi}{\partial z}\right)^2 + \frac{4}{S^2}p_\theta^2 + \frac{4}{S^2}\frac{p_\phi^2}{\sin^2\theta} \right\}. \tag{83}$$

From the equation of motion,

$$\dot{\theta} = \frac{\partial\mathscr{H}}{\partial p_\theta}, \qquad \dot{p}_\theta = -\frac{\partial\mathscr{H}}{\partial\theta} + \frac{\partial}{\partial z}\left(\frac{\partial\mathscr{H}}{\partial\theta_z}\right), \tag{84}$$

where $\mathscr{H}$ is the Hamiltonian density, we obtain

$$\frac{\partial^2\theta}{\partial z^2} - \frac{1}{c^2}\frac{\partial^2\theta}{\partial t^2} = \sin\theta\cos\theta\left[ \left(\frac{\partial\phi}{\partial z}\right)^2 - \frac{1}{c^2}\left(\frac{\partial\phi}{\partial t}\right)^2 \right]. \tag{85}$$

Using a similar equation for the $\phi$ variable, we get

$$\frac{\partial^2\phi}{\partial z^2} - \frac{1}{c^2}\frac{\partial^2\phi}{\partial t^2} = -2\cot\theta\left[ \left(\frac{\partial\theta}{\partial z}\right)\left(\frac{\partial\phi}{\partial z}\right) - \frac{1}{c^2}\left(\frac{\partial\theta}{\partial t}\right)\left(\frac{\partial\phi}{\partial t}\right) \right]. \tag{86}$$

The static Euler-Lagrange equations of the antiferromagnetic nonlinear sigma model coincide with those of the ferromagnetic exchange model when expressed in terms of the Néel vector. Therefore, we can use the formalism developed below to obtain static solutions for the ferromagnet.

The Lagrangian density is obtained using the transformation

$$\mathscr{L} = -\mathscr{H} + p_\theta\dot{\theta} + p_\phi\dot{\phi}, \tag{87}$$

which leads to

$$L = \frac{JS^2}{2}\int dz\left\{-\left[\left(\frac{\partial\theta}{\partial z}\right)^2 + \sin^2\theta\left(\frac{\partial\phi}{\partial z}\right)^2\right] + \frac{1}{c^2}\left[\left(\frac{\partial\theta}{\partial t}\right)^2 + \sin^2\theta\left(\frac{\partial\phi}{\partial t}\right)^2\right]\right\}. \quad (88)$$

As before, we define the unitary vector

$$\vec{n} = (\sin\theta\cos\phi, \sin\theta\sin\phi, \cos\theta). \quad (89)$$

We can write Eq. (81) for any number of dimensions as

$$H = \frac{JS^2}{2}\int dV\left[\frac{1}{c^2}\left(\frac{\partial\vec{n}}{\partial t}\right)^2 + (\vec{\nabla}\vec{n})^2\right], \quad (90)$$

The canonical momentum associated with the Hamiltonian (90) is given by

$$\vec{p} = \frac{1}{c}\frac{\partial\vec{n}}{\partial t}. \quad (91)$$

Using Eqs. (89) and (91), we obtain Eq. (82). From the equation of motion for the Hamiltonian (90), we find (in one dimension, for simplicity)

$$\frac{\partial\vec{n}}{\partial t} = \frac{\partial\vec{n}}{\partial z}\times\vec{n}. \quad (92)$$

Hence, the Lagrangian density is

$$\mathscr{L} = -\mathscr{H} + \vec{p}.\frac{\partial\vec{n}}{\partial t} = -\mathscr{H} + \frac{1}{c}\frac{\partial\vec{n}}{\partial t}.\frac{\partial\vec{n}}{\partial z}\times\vec{n}. \quad (93)$$

Which leads to

$$L = \frac{JS^2}{2}\int dV\left[-\frac{1}{c^2}\left(\frac{\partial\vec{n}}{\partial t}\right)^2 + (\vec{\nabla}\vec{n})^2\right]. \quad (94)$$

In the presence of an external magnetic field $\vec{B}$, the Lagrangian becomes

$$\mathscr{L}(x,\tau) = \frac{JS^2}{2}\left[-\frac{1}{c^2}\left(\partial_t\vec{n} - i\vec{B}\times\vec{n}\right)^2 + (\vec{\nabla}\vec{n})^2\right]. \quad (95)$$

To enforce the constraint $|\vec{n}|=1$, we introduce a Lagrange multiplier $\lambda$ and add the term $\lambda\delta(\vec{n}^2-1)$ in the integrand of Eqs. (90), (94), and (95).

The problem is more complicated in one dimension because it includes a topological term that distinguishes between integer and half-integer spins. This term vanishes in two dimensions (see Appendix C), the case of interest to us.

Let us write the Lagrangian density in two dimensions *in* the following form

$$\mathscr{L}(x,y,\tau) = \frac{1}{2g}(\partial_\mu\vec{n}).(\partial^\mu\vec{n}), \quad (96)$$

with $\mu = x, y, it$, $g = 1/JS^2$ and subject to the constraint $\vec{n}.\vec{n} = 1$. The resulting Euler-Lagrange equation after eliminating the constraint in the term $\lambda(1-\vec{n}.\vec{n})$ added to Eq. (96) is [23]

$$\partial_\mu\partial^\mu\vec{n} + (\partial_\mu\vec{n}.\partial^\mu\vec{n})\vec{n} = 0. \quad (97)$$

Writing $\vec{n} = (n_1, n_2, n_3)$ and considering that the allowed values of $\vec{n}$ form the surface of a sphere $S^2_{\text{int}}$, the stereographic projection of this sphere from the north pole onto the complex plane $(w_1, w_2)$, is given by

$$w_1 = \frac{n_1}{1-n_3}, \qquad w_2 = \frac{n_2}{1-n_3}. \tag{98}$$

That is the projection $S^2 \setminus \{N\} \to \mathbb{C}$. Introducing the complex quantity $w = w_1 + iw_2$, the inverse map is

$$n_1 = \frac{2w_1}{1+|w|^2}, \quad n_2 = \frac{2w_2}{1+|w|^2}, \quad n_3 = \frac{|w|^2 - 1}{1+|w|^2}. \tag{99}$$

Some authors use the projection from the south pole and in this case the sign of $n_3$ is reversed. The plane $\{w_1, w_2\}$ is parallel to the plane $\{n_1, n_2\}$. So in the non-linear sigma model, instead of working with $\vec{n}(x, y)$, we work with the complex field $w(x, y)$. We see that $w = 0$ corresponds to $n_3 = -1$, while $w = \infty$ corresponds to $n_3 = +1$. Note that $x_1 = x$, $x_2 = y$ are cartesian coordinates of the original two-dimensional physical space; $w_1$ and $w_2$, describe the plane in the internal space on which $S^2$ has been projected stereographically.

To derive an expression for the Lagrangian, I will derive some relations. We have

$$n_3^2 = 1 - n_1^2 - n_2^2 \quad \to \quad n_3 \partial_\mu n_3 = -n_1 \partial_\mu n_1 - n_2 \partial_\mu n_2, \tag{100}$$

$$n_3 \partial_\mu n_3 = -n_a \partial_\mu n_a, \quad \text{with} \quad a = 1,2 \tag{101}$$

(Note that we sum over repeated indices),

$$\partial_\mu n_3 \partial_\mu n_3 = \frac{(n_3 \partial_\mu n_3)(n_3 \partial_\mu n_3)}{n_3^2} = \frac{1}{n_3^2}(n_a \partial_\mu n_a)(n_b \partial_\mu n_b). \tag{102}$$

Using these expressions, we can write the Lagrangian as

$$\mathscr{L}(x, y, \tau) = \frac{1}{2g}\left( \partial_\mu n_a \partial_\mu n_a + \frac{n_a n_b}{n_3^2} \partial_\mu n_a \partial_\mu n_b \right). \tag{103}$$

We also have

$$\partial_\mu n_a = (1+n_3)\partial_\mu w_a + w_a \partial_\mu n_3, \tag{104}$$

$$w\bar{w} = w_1^2 + w_2^2 = \frac{n_1^2 + n_2^2}{(1+n_3)^2} = \frac{1-n_3^2}{(1+n_3)^2} = \frac{1-n_3}{1+n_3}, \qquad , \tag{105}$$

$$w_a \partial_\mu w_a = -\frac{\partial_\mu n_3}{(1+n_3)^2}.,\tag{106}$$

Taking (106) into (104), we obtain

$$\partial_\mu n_a = (1+n_3)[\partial_\mu w_a + w_a (1+n_3) w_b \partial_\mu w_b]. \tag{107}$$

Using the above results, we get, after some calculations

$$\mathscr{L} = \frac{1}{2g}(1+n_3)^2 \partial_\mu w_a \partial_\mu w_a. \tag{108}$$

Using now

$$(1+n_3)^2 = \frac{4}{(1+w\overline{w})^2}, \tag{109}$$

we can write the Lagrangian density as

$$\mathscr{L}(x,\tau) = \frac{2}{g}\frac{\partial_\mu w \partial^\mu \overline{w}}{(1+|w|^2)^2}. \tag{110}$$

Using $z = x + iy$, $\partial_z = (\partial_x - i\partial_y)/2$, $\partial_{\overline{z}} = (\partial_x + i\partial_y)/2$, we have

$$(\partial_i \vec{n})^2 = \frac{4(|\partial_z w|^2 + |\partial_{\overline{z}} w|^2)}{(1+|w|^2)^2}. \tag{111}$$

Then, the energy of a static field configuration, as obtained from Lagrangian (110), is given by

$$E = 4JS^2 \int d^2x \frac{(|\partial_z w|^2 + |\partial_{\overline{z}} w|^2)}{(1+|w|^2)^2}. \tag{112}$$

In terms of $w$ the topological charge has the following expression

$$Q = \frac{1}{\pi}\int d^2x \frac{(|\partial_z w|^2 - |\partial_{\overline{z}} w|^2)}{(1+|w|^2)^2}. \tag{113}$$

We can split the integral for the energy into two parts, and write

$$E = 4JS^2 \int d^2x \frac{(|\partial_z w|^2 - |\partial_{\overline{z}} w|^2)}{(1+|w|^2)^2} + 8JS^2 \int d^2x \frac{|\partial_{\overline{z}} w|^2}{(1+|w|^2)^2}. \tag{114}$$

Using the equation for $Q$ we get

$$E = 4JS^2 \pi Q + 8JS^2 \int d^2x \frac{|\partial_{\overline{z}} w|^2}{(1+|w|^2)^2}. \tag{115}$$

Since the second term is nonnegative, we have $E \geq 4\pi JS^2 Q$ ( $Q > 0$). For $Q <$ ), we group with $|\partial_z w|^2$. Thus

$$E \geq 4\pi JS^2 |Q|. \tag{116}$$

This result is called the Bogomolny bound equation. The bound is saturated if the nonnegative term vanishes. For $Q > 0$, we have

$$\int d^2x \frac{|\partial_{\overline{z}} w|^2}{(1+|w|^2)^2} = 0 \quad \rightarrow \quad \partial_{\overline{z}} = 0. \tag{117}$$

Thus, the function $w(z,\overline{z})$ must be holomorphic. If $Q < 0$, we obtain $\partial_z = 0$, and the function $w(z,\overline{z})$ is antiholomorphic.

The general static finite-energy solutions of the nonlinear sigma model are given by

$$w(z) = \frac{P(z)}{Q(z)}. \tag{118}$$

where $P(z)$ and $Q(z)$are polynomials with no common zeros. For a topological charge $Q = 1$, we have $w(z) = \frac{\lambda}{z - z_0}$, (a translated and rescaled skyrmion). For $Q = 2$, we get

$$w(z) = \frac{\lambda}{(z - z_1)(z - z_2)}, \tag{119}$$

The parameter $\lambda$ roughly describes the size of the skyrmion. In general, the number of poles (counted with multiplicity) equals the topological charge. A general solution can be written as

$$w(z) = \prod_{i=1}^{N} \left( \frac{z - a_i}{z - b_i} \right), \tag{120}$$

where $a_i$ and $b_i$ are complex numbers.

Using Eq. (89), we get from Eq. (98)

$$w = \frac{\sin\theta e^{i\phi}}{1 + \cos\theta} = \tan\left(\frac{\theta}{2}\right) e^{i\phi}. \tag{121}$$

Let us now write $z = re^{i\varphi}$, and consider the simple solution

$$w(z) = \lambda / \bar{z}. \tag{122}$$

We obtain $\varphi = \phi$, and $\tan\left(\frac{\theta}{2}\right) = \frac{\lambda}{r} \rightarrow \Theta = 2\arctan(R/r)$,

the solution found before. We have

$$n_1(r, \Phi) = \sin\left(2\arctan\frac{R}{r}\right)\cos(\phi + \gamma), \quad n_2(r, \Phi) = \sin\left(2\arctan\frac{R}{r}\right)\sin(\phi + \gamma)$$

$$n_3(r) = \cos\left(2\arctan\frac{R}{r}\right), \tag{123}$$

which can be written as

$$n_1 = \frac{2Rr}{r^2 + R^2}\cos(\phi + \gamma) \qquad n_2 = \frac{2Rr}{r^2 + R^2}\sin(\phi + \gamma), \qquad n_3 = \frac{r^2 - R^2}{r^2 + R^2}. \tag{124}$$

An example of a skyrmion with topological charge $m$ is [24]

$$w(z) = \left(\frac{\lambda}{\bar{z}}\right)^m, \qquad \Theta(r) = 2\arctan\left(\frac{\lambda}{r}\right)^m, \qquad \Phi = m\phi + \gamma. \tag{125}$$

An elliptical skyrmion has the same topological nature as the standard skyrmion but is stretched along one axis, giving it an oval shape. This distortion can arise from magnetic anisotropy (different magnetic properties in different directions), material strain, external magnetic fields, crystal-structure effects, and geometric confinement in nanodevices. They may move more easily in one direction than another. Their shape can often be adjusted using external fields or material engineering. The rescaling $x \rightarrow x/a$, $y \rightarrow y/b$, where $a$ and $b$ *are* the semi-major and semi-minor axes, turns circular contours into ellipses. An elliptical skyrmion breaks the conformal symmetry of the Belavin-Poliakov solution and is not a minimum-energy configuration of the $Q = 1$ class. The Cauchy-Riemann equations are no longer satisfied. The spatial stretching changes the shape but not the degree of the mapping, provided $a$, $b > 0$.

As mentioned early, meromorphic functions generate topologically nontrivial spin textures. The Weierstrass elliptic functions $\wp$ are the periodic version of those solutions and naturally describe a skyrmion lattice, rather than an isolated skyrmion. We have

$$\wp(z, w_1, w_2) = \frac{1}{z^2} + \sum_{m^2+n^2\neq 0} \left\{ \frac{1}{(z - mw_1 - mw_2)^2} - \frac{1}{(mw_1 + nw_2)^2} \right\}, \tag{126}$$

where *m* and *n* are integers. Different choices of the periods $(w_1, w_2)$ give square, rectangular, or triangular skyrmion lattices.

The static solution for the antiferromagnet is the same as the one for the ferromagnet. So, we can use the Belavin-Polyakov bound given by Eq. (48): $E \geq 4\pi JS^2 |Q|$. Skyrmion solutions saturate this bound, so their energy is $E = 4\pi JS^2 |Q|$. In the pure nonlinear sigma model (no anisotropy, Zeeman, DMI, etc.), the energy of a skyrmion does not depend on its size; any rescaling of the skyrmion profile keeps the energy the same. Ferromagnets and antiferromagnets break this scale invariance via additional terms (anisotropy, dipolar, DMI, Zeeman), which then fix a finite skyrmion radius and modify the energy away from the pure $E = 4\pi JS^2 |Q|$ value. The skyrmion that solves the exchange-energy Hamiltonian is generally called, in the literature, a Belavin-Polyakov (BP) skyrmion.

Biskyrmions with a topological charge $Q = \pm 2.$ are formed by two partially overlapping skyrmions, with a topological charge $Q = \pm 1$. To be geometrically compatible, both skyrmions often possess a helicity difference of $\pi$, meaning that the in-plane magnetization of both skyrmions is mutually reversed [2]. Because biskyrmions formation requires opposite in-plane magnetization, the DM interaction cannot stabilize such biskyrmions. They can, however, be stabilized by the dipole-dipole interaction in centrosymmetric materials in which the DMI is forbidden.

A biskyrmion is not rotationally symmetric, so there is no universally simple radial form. One common construction uses the stereographic variable $w(z)$ and places two skyrmions at positions $\pm d/2$:

$$w(z) = \frac{\lambda^2}{(z - d/2)(z + d/2)}. \tag{127a}$$

For large separation, this describes two skyrmions; for moderate *d*, it becomes a biskyrmion with $Q = 2$. We also have

$$w(z) = \frac{(z - d/2)(z + d/2)}{\lambda^2}. \tag{127b}$$

Taking the reciprocal of Eq. (127a) does not change the skyrmion topological charge, but it reverses the magnetization orientation at the cores and at infinity.

More generally, a skyrmion characterized by a size $\lambda$, chirality $\gamma$, and topological charge *Q*, centered at $\bar{z} = \bar{z}_0$ can be expressed by

$$w = e^{i\gamma} \left( \frac{\lambda}{\bar{z} - \bar{z}_0} \right)^Q. \tag{128}$$

For two skyrmions of topological charge $Q = 1$, separated by a distance *d*, we have a solution with $Q = 2$, given by

$$w = \frac{\lambda e^{i\gamma_1}}{\bar{z} - d/2} + \frac{\lambda e^{i\gamma_2}}{\bar{z} + d/2}. \tag{129}$$

For $\gamma_1 = \gamma_2 = \gamma$ we get

$$w = \frac{\lambda d e^{i\gamma}}{\bar{z}^2 - (d/2)^2} . \tag{130}$$

When two skyrmions with opposite polarities are combined, we obtain a skyrmionium with vanishing topological charge. If both skyrmions have the same vorticity but opposite polarity, bulk or interfacial DMI can stabilize them [2]. Note that the sum of two skyrmions as given by Eq. (39) is not a solution of the Euler-Lagrange equation because the equation is nonlinear. For skyrmions, this nonlinearity is even more fundamental because the field is constrained to lie on a sphere, so $|\vec{n}_1 + \vec{n}_2| \neq 1$. For the Belavin-Polyakov equation $\partial_{\bar{z}} w = 0$, we can sum holomorphic functions. But the superposition occurs in the complex field *w*, not in the magnetization vector $\vec{n}$. After mapping back to $\vec{n}$, the resulting configuration is nonlinear.

To finish this section, I will show another way to find the solutions for a static configuration. Using the relations $\vec{n}^2 = 1 \rightarrow \vec{n}.\partial_\mu \vec{n} = 0$ and $\vec{a} \times \vec{b} \times \vec{c} = \vec{b}.(\vec{a}.\vec{c}) - \vec{c}.(\vec{a} \times \vec{b})$, we can write Eq. (30) as

$$\frac{\partial \vec{n}}{\partial x} = \pm \vec{n} \times \frac{\partial \vec{n}}{\partial y}, \qquad \frac{\partial \vec{n}}{\partial y} = \mp \vec{n} \times \frac{\partial \vec{n}}{\partial x}. \tag{131}$$

Using now Eq. (131), Eq. (98) gives

$$\frac{\partial w}{\partial x} = \pm i \frac{\partial w}{\partial y}. \tag{132}$$

Which is the Cauchy-Riemann condition that is satisfied for any analytic function.

## 7. Merons.

A **meron** is a *topological spin texture* that represents half of a skyrmion [2]. If we picture a skyrmion as a full "whirl" of spins wrapping the sphere once, a meron wraps it halfway, so its topological charge is ±1/2. Because the boundary spins lie on the equator rather than approaching a single point on $S^2$, the physical plane cannot be compactified to $S^2$. Consequently, the usual integer-value homotopy classification is lost, allowing configurations with half-integer charge. Topological charge is quantized only when boundary conditions force full coverage. An isolated meron with equatorial winding at infinity cannot be represented by a globally single-valued meromorphic function on the Riemann sphere In many real magnetic materials, isolated merons cannot exist freely; they appear as meron–antimeron pairs or as building blocks of larger textures (such as bimerons or skyrmion lattices). A bimeron consists of two merons bound together by a continuous spin texture. In its simplest form, the two merons possess the same topological sign and are connected by a striped-like domain wall. The resulting structure carries an integer topological charge, $Q = \pm 1$, which is identical to that of a skyrmion. For this reason, a bimeron is frequently regarded as the in-plane counterpart of a skyrmion.

Indeed, a global rotation of the spin configuration can transform a skyrmion into a bimeron without changing its topology. While a skyrmion is typically embedded in an out-of-plane ferromagnetic background, a bimeron exists in a predominantly in-plane magnetized environment. The distinction between skyrmions and bimerons matters for transport phenomena. In skyrmion crystals, the finite out-of-plane magnetization produces an anomalous Hall effect that is superimposed on the topological Hall effect generated by the skyrmion texture. In bimeron crystals, the average out-of-plane magnetization can be nearly zero, strongly suppressing the anomalous Hall contribution. Electrons still experience the emergent magnetic field associated with the nontrivial topology, allowing the topological Hall effect to be observed with much less interference from conventional magnetic Hall signals. This property makes bimerons attractive candidates for studying topology-driven transport phenomena

In antiferromagnets and certain frustrated magnets**,** individual merons can be stable, which is why they are getting so much attention in current research. For a meron, the boundary condition is: $\Theta(0) = 0,\ \Theta(\infty) = \pi/2$.

To describe a meron with the correct boundary condition, an ansatz chosen for simplicity is given by

$$\Theta(r) = \arctan(r/R)\,. \tag{133}$$

The Euler Eq.(67) for an easy-plane anisotropy becomes

$$2J\left(\frac{d^2\Theta}{dr^2} + \frac{1}{r}\frac{d\Theta}{dr} - \frac{\sin\Theta\cos\Theta}{r^2}\right) + \frac{2D}{r}\sin^2\Theta + 2A\sin\Theta\cos\Theta = 0\,. \tag{134}$$

For a sufficiently high easy-plane anisotropy, all spins will align in the plane. In Ref. [18], the following ansatz was used

$$\Theta(r) = \pm\frac{\pi}{2} + 2\arctan[\exp(-\zeta r/r_a)]\,, \tag{135}$$

where $r_a = D/A$ and the parameter $\zeta$ was determined by minimizing the energy. The sign + ( - ) corresponds to spin down (up) at the center of the meron. The + sign gives $\theta(0) = \pi$, $\Theta(\infty) = \pi/2$. The – sign gives $\theta(0) = 0,\ \Theta(\infty) = -\pi/2$.

# 8. Vortices.

The classical rotor model is described by the Hamiltonian

$$H = -J\sum_{<i,j>}(S_i^x S_j^x + S_i^y S_j^y)\,, \tag{136}$$

with

$$S_i^x = S\cos\phi_i\,, \qquad S_j^y = S\sin\phi_j\,. \tag{137}$$

Then we can write the Hamiltonian (136) as

$$H = -JS^2\sum_{<i,j>}\cos(\phi_i - \phi_j)\,. \tag{138}$$

For slowly varying configurations, that is, those with adjacent angles nearly equal, we can expand (138) up to quadratic terms and obtain

$$H = \frac{JS^2}{2} \sum_{<i,j>} (\phi_i - \phi_j)^2 . \tag{139}$$

In the continuum limit, Eq. (139) becomes

$$H = \frac{JS^2}{2} \int dxdy \left[ \left( \frac{\partial \phi}{\partial x} \right)^2 + \left( \frac{\partial \phi}{\partial y} \right)^2 \right]. \tag{140}$$

Using $\delta H / \delta \phi = 0$, we get

$$\frac{\partial^2 \phi}{\partial x^2} + \frac{\partial^2 \phi}{\partial y^2} = 0 . \tag{141}$$

A vortex solution is given by

$$\phi(x, y) = \arctan(y / x) . \tag{142}$$

In a vortex configuration, as we go around a closed path containing the center of the vortex, $\phi$ will change by $2\pi$ for each revolution of the spin. The vorticity $q$ is defined as

$$\frac{1}{2\pi} \oint d\phi(x, y) dxdy = q , \qquad \text{where } q = 0, \pm 1, \pm 2, ... \tag{143}$$

We have a map $\pi_1(S^1) = \mathbb{Z}$. The derivatives of $\phi$ are given by

$$\frac{\partial \phi}{\partial x} = -\frac{y}{x^2 + y^2} , \qquad \frac{\partial \phi}{\partial y} = \frac{x}{x^2 + y^2} . \tag{144}$$

The energy of an isolated vortex of vorticity $q$ is then given by

$$E = \frac{JS^2 q^2}{2} \int_a^r \frac{2\pi r dr}{r^2} = \pi J S^2 q^2 \ln(r / a) , \tag{145}$$

where $r^2 = x^2 + y^2$, and $a$ is the lattice spacing. For two vortices at positions $\vec{r}_1$ and $\vec{r}_2$, we write

$$\phi(\vec{r}) = q_1 \phi(\vec{r} - \vec{r}_1) + q_2 \phi(\vec{r} - \vec{r}_2) . \tag{146}$$

The energy interaction is the cross term

$$E_{\text{int}} = JS^2 q_1 q_2 \int d^2 r \vec{\nabla} \phi_1 . \vec{\nabla} \phi_2 . \tag{147}$$

Using the identity $\vec{\nabla} \phi_1 . \vec{\nabla} \phi_2 = \vec{\nabla} . (\phi_1 \vec{\nabla} \phi_2) - \phi_1 \nabla^2 \phi_2$, we get

$$E_{\text{int}} = -2\pi J S^2 q_1 q_2 \ln\left( \frac{|\vec{r}_1 - \vec{r}_2|}{a} \right). \tag{148}$$

The sign here is a convention. Vortices of the same sign repel; opposite signs attract. The in-plane component of a meron is a vortex.

# 9. Frustration.

A key distinction between frustrated and DMI-stabilized skyrmions is the origin of their chirality. In DMI systems, crystal symmetry and spin-orbit coupling fix the skyrmion's handedness. In exchange frustrated magnets, both helicities are generally degenerate because the stabilization mechanism originates from exchange competition rather than chiral interaction (DMI or dipolar terms lift the degeneracy). Magnetic frustration stabilizes skyrmions by creating competing interactions that favor non-collinear spin textures over simple ferromagnetic order**.** In frustrated magnets, these competing exchange interactions make it energetically preferable for spins to twist, rotate, or form multi-Q states, conditions under which skyrmions naturally emerge and remain stable. Frustrated magnets can host attractive skyrmion-skyrmion interactions, in contrast to the predominantly repulsive interactions observed in DMI-stabilized systems. This behavior promotes skyrmion-skyrmion clustering and bound states, opening new possibilities for collective information processing.

Frustration occurs when different exchange interactions cannot all be simultaneously satisfied. It prevents standard long-range ferromagnetic or antiferromagnetic ordering. Instead, it generates exotic phases, such as quantum spin liquids [14]. It allows skyrmions to form even without Dzyaloshinskii–Moriya interactions, which are usually required in chiral magnets. In some frustrated Heisenberg magnets, skyrmions arise solely from competition between exchange interactions (e.g., ferromagnetic vs. antiferromagnetic). Quantum calculations indicate that frustration can bind many magnons into topologically nontrivial quantum states that can be interpreted as quantum skyrmions.

A typical example, used in the skyrmion theory, is the two-dimensional Heisenberg ferromagnet with nearest-neighbor interaction $J_1$ competing with an antiferromagnetic next-nearest-neighbor interaction $J_2$, with the following Hamiltonian:

$$H = -J_1 \sum_{<i,j>} \vec{S}_i.\vec{S}_j + J_2 \sum_{<<i,j>>} \vec{S}_i.\vec{S}_j \,. \tag{149}$$

Where $<i.j>$ denotes nearest-neighbor (NN), and $<<i,j>>$denotes next-nearest-neighbors (NNN) along the diagonals. For $|J_2/J_1| << 1$, the ground state is a uniform ferromagnet [25]. For $|J_2/J_1| > 0.5$, competing spiral states emerge. The compromise is a spin spiral $\vec{S}_i = [\cos(\vec{q}_0.\vec{r}_i), \sin(\vec{q}_0.\vec{r}_i), 0]$. The wavevector $q_0$ is obtained by minimizing the energy, as I will show later. The exchange interaction in the momentum space is given by

$$J(\vec{q}) = J_1(\cos q_x + \cos q_y) + 2J_2 \cos q_x \cos q_y \,. \tag{150}$$

We can obtain the value of $q_0$ using Eq. (150), but I will instead use the continuum model, since the result can be used to study several skyrmion properties. As in Section 3, I will focus on the long-wavelength behavior and study the classical continuum limit. I will assume that the spin textures vary slowly compared with the lattice spacing $a$. As before, I will write $\vec{S}_i = S\vec{n}(\vec{r})$, with $\vec{n}.\vec{n} = 1$. I will expand the neighboring spins, keeping terms up to order $a^4$. For the first term, I will start with the $x$-direction:

$$\vec{n}(\vec{r} + a\hat{x}) = \vec{n} + a\partial_x \vec{n} + \frac{a^2}{2}\partial_x^2 \vec{n} + \frac{a^3}{6}\partial_x^3 \vec{n} + \frac{a^4}{24}\partial_x^4 \vec{n} + ... \tag{151}$$

Because we sum over $\pm$ neighbors, all odd derivative terms cancel, and we get

$$\vec{n}(\vec{r}).\vec{n}(\vec{r} + a\hat{x}) = 1 + \frac{a^2}{2}\vec{n}.\partial_x^2 \vec{n} + \frac{a^4}{24}\vec{n}.\partial_x^4 \vec{n} + ... \tag{152}$$

To proceed, I will use the following relations, which can be easily derived by starting with $\vec{n}.\vec{n}=1$:

$$\vec{n}.\partial_\mu \vec{n}=0\,, \qquad \vec{n}.\partial_\mu^2 \vec{n}=-(\partial_\mu \vec{n})^2\,, \qquad \vec{n}.\partial_\mu^3 \vec{n}=-3\partial_\mu \vec{n}.\partial_\mu^2 \vec{n}\,,$$

$$\vec{n}.\partial_\mu^4 \vec{n}=-4(\partial_\mu \vec{n}.\partial_\mu^3 \vec{n})-3(\partial_\mu^2 \vec{n})^2\,, \qquad \partial_\mu \vec{n}.\partial_\mu^3 \vec{n}=-(\partial_\mu^2 \vec{n})^2\,. \tag{153}$$

We find

$$\vec{n}(\vec{r}).\vec{n}(\vec{r}+a\hat{x})=1-\frac{a^2}{2}(\partial_x \vec{n})^2+\frac{a^4}{24}\left[-4(\partial_x \vec{n}.\partial_x^3 \vec{n})-3(\partial_x^2 \vec{n})^2\right]$$

$$= 1-\frac{a^2}{2}(\partial_x \vec{n})^2+\frac{a^4}{24}(\partial_x^2 \vec{n})^2\,. \tag{154}$$

Adding the $x$ and $y$ bonds, we obtain the term

$$-\frac{a^2}{2}[(\partial_x \vec{n})^2+(\partial_y \vec{n})^2]+\frac{a^4}{24}[(\partial_x^2 \vec{n})^2+(\partial_y^2 \vec{n})^2]\,. \tag{155}$$

Before considering the next term, I will comment briefly on the result (155). The discreteness breaks the lattice-scale invariance of the skyrmion energy. The second term in Eq. (155) can be considered a correction due to lattice discreteness. Using the BP solution given by Eq. (124) and calculating the energy, we obtain [26]

$$E_{dis}=-\frac{2\pi JS^2}{3}\left(\frac{a}{R}\right)^2, \tag{156}$$

that violates the skyrmion's scale invariance. Eq. (156) can be interpreted as a potential energy. It breaks the scalar invariance of the Belavin-Polyakov solution and favors skyrmion shrinkage towards the lattice scale.

Next, I will consider the next-nearest-neighbor term. The four diagonal vectors $\vec{\delta}$ are

$$a(\hat{x}+\hat{y}),\;\; a(\hat{x}-\hat{y}),\;\; a(-\hat{x}+\hat{y}),\;\; a(-\hat{x}-\hat{y})\,.$$

For $\vec{\delta}=a(\hat{x}+\hat{y})$ we have $(\vec{\delta}.\vec{\nabla}\vec{n})^2=a^2(\partial_x \vec{n}+\partial_y \vec{n})^2$. (157)

For $\vec{\delta}=a(\hat{x}-\hat{y})$ we have $(\vec{\delta}.\vec{\nabla}\vec{n})^2=a^2(\partial_x \vec{n}-\partial_y \vec{n})^2$. (158))

Adding both one diagonal direction, we find

$$(\partial_x \vec{n}+\partial_y \vec{n})^2+(\partial_x \vec{n}-\partial_y \vec{n})^2=2(\partial_x^2 \vec{n}+\partial_y^2 \vec{n})\,, \tag{159}$$

which leads to the term

$$2a^2[(\partial_x \vec{n})^2+(\partial_y \vec{n})^2]\,. \tag{160}$$

The biquadratic terms are

$$(\partial_x+\partial_y)^4.\;\; (\partial_x-\partial_y)^4,\;\; (-\partial_x+\partial_y)^4,\;\; (-\partial_x-\partial_y)^4\,. \tag{161}$$

Adding both one-diagonal terms, we obtain

$$2(\partial_x^4+6\partial_x^2\partial_y^2+\partial_y^4) \;\;\rightarrow\;\; 2\vec{n}.(\partial_x^4+6\partial_x^2\partial_y^2+\partial_y^4)\vec{n}\,. \tag{162}$$

Using the relations (153), we get the term

$$\frac{2a^4}{24}[(\partial_x^2 \vec{n})^2+(\partial_y^2 \vec{n})^2+6\vec{n}.\partial_x^2\partial_y^2 \vec{n}]\,. \tag{163}$$

To convert a sum to an integral, the following expression is used

$$\sum_r \rightarrow \frac{1}{a^2}\int d^2r\,, \tag{164}$$

And we use relations like (153) for mixed derivatives to rewrite the last term in Eq. (163). Combining the NN and NNN contributions, the Hamiltonian in the continuum limit becomes

$$H = const + S^2 \int d^2 r \left\{ \rho[(\partial_x \vec{n})^2 + (\partial_y \vec{n})^2] + \kappa_1 [(\partial_x^2 \vec{n})^2 + (\partial_y^2 \vec{n})^2] + \kappa_2 (\partial_x \partial_y \vec{n})^2 \right\}. \qquad (165)$$

Where

$$\rho = \frac{S^2}{2}(J_1 - 2J_2), \qquad \kappa_1 = \frac{S^2 a^2}{24}(-J_1 + 2J_2), \qquad \kappa_2 = \frac{S^2 a^2 J_2}{2}. \qquad (166)$$

If $J_2$ is very small compared to $J_1$, the terms $\kappa_1$ and $\kappa_2$ are negligible, and we use the Hamiltonian derived in section 3. Increasing the value of $J_2$, $\rho$ vanishes at $J_2 = J_1/2$ and changes sign for large values of $J_2$. In this case, all the terms in the Hamiltonian (165) are relevant. We can also include the third-nearest neighbor with an exchange constant $J_3$. The procedure is like the one used for $J_1$, with the replacement $a \to 2a$. We then get a contribution $-4J_3$ to $\rho$, and $16J_3$ to $\kappa_1$.

Let us take a spiral $\vec{n} = (\cos q_x a, \sin q_y a, 0)$. Fourier transforming the derivative terms, we have for small wavevectors

$$(\nabla \vec{n})^2 \sim (q_x^2 + q_y^2), \qquad (\nabla^2 \vec{n})^2 \sim (q_x^4 + q_y^4), \qquad (\partial_x \partial_y \vec{n})^2 \sim q_x^2 q_y^2, \qquad (167)$$

I will write the energy density as

$$E \sim A(q_x^2 + q_y^2) + B(q_x^4 + q_y^4) + C q_x^2 q_y^2. \qquad (168)$$

Where

$$A = \frac{S^2}{2}(J_1 - 2J_2 - 4J_3), \quad B = \frac{S^2 a^2}{24}(-J_1 + 2J_2 + 16J_3), \quad C = \frac{S^2 a^2 J_2}{2}. \qquad (169)$$

Minimizing the energy with respect to $q_x$, we obtain

$$\frac{\partial E}{\partial q_x} = 2Aq_x + 4Bq_x^3 + 2Cq_x q_y^2 = 0, \qquad (170)$$

And with respect to $q_y$

$$\frac{\partial E}{\partial q_y} = 2Aq_y + 4Bq_y^3 + 2Cq_y q_x^2 = 0. \qquad (171)$$

If $A > 0$, the solution $q_x = q_y = 0$ represents a uniform, ordered, ferromagnetic, stable phase. If $A < 0$, we have two kinds of solutions. One with $q_y = 0$, $q_x \neq 0$ ( or $q_x = 0$, $q_y \neq 0$ ) (for spirals along the axes):

$$q_{0,x}^2 = -\frac{A}{2B}, \qquad \text{with energy} \qquad \varepsilon_1 = -\frac{A^2}{4B} \qquad (172)$$

The other with $q_x = q_y = q$, (for spiral along the diagonal) with solution

$$q_0^2 = -\frac{A}{2B+C}, \qquad \text{and energy} \qquad \varepsilon_2 = -\frac{A^2}{2B+C}. \tag{173}$$

Depending on the relation between the parameters *B* and *C*, we find spiral phases with ordering wavevector $\vec{q} = (\pm q_0, 0)$ or $\vec{q} = (0, \pm q_0)$ with $q_0$ given by Eq. (172) or $\vec{q} = (\pm q_0, \pm q_0)/\sqrt{2}$ with $q_0$ given by Eq. (173). In both situations, the ground state can be either a single spiral or a multi-structure, depending on the interaction strengths, magnetic field, and anisotropies [27,28,29]. The ground state energy gain due to the spiral is proportional to $q^4$. The fourth-order gradients of $\vec{n}$ stabilize the spiral state and determine its wavevector, if $B > 0$ and $2B + C > 0$ ($C < 2B$ diagonal, $C > 2B$ axial).

Kharkov et al. [29] numerically studied the Hamiltonian (149) including a $J_3$ interaction and anisotropy (easy axis or easy plane). They found that skyrmions are unstable under nearest-neighbor interactions alone. For $J_2 > 2J_3$, the ferromagnetic phase transforms into the spiral state with wavevector $\vec{q}_0$ parallel to the square lattice axes. For $J_2 < 2J_3$, the wavector $\vec{q}_0$ is along the diagonals. They also found that, for $A < 0$, depending on the distance between skyrmions, they could repel or attract one another.

For a triangular lattice, I will study the exchange interaction $J(\vec{q})$ given by [14]

$$J(\vec{q}) = 2J_1\left[\cos(q_x) + 2\cos\left(\frac{q_x}{2}\right)\cos\left(\frac{\sqrt{3}}{2}q_y\right)\right] + 2J_2\left[\cos\left(\sqrt{3}q_y\right) + 2\cos\left(\frac{3q_x}{2}\right)\cos\left(\frac{\sqrt{3}q_y}{2}\right)\right]$$

(174a)

In the long wavelength limit, we expand $J(\vec{q})$ in terms of *q*, and obtain

$$J(\vec{q}) = 6(J_1+J_2) - \frac{3}{2}(J_1+J_2)q^2 + \frac{3}{32}(J_1+9J_2)q^4 - \frac{J_1+27J_2}{384}q^6 + J_{anis}(q_x,q_y), \tag{174.b}$$

with the anisotropic term given by

$$J_{anis}(q_x,q_y) = -\frac{J_1+18J_2}{128}q^2q_x^2q_y^2 + \frac{-J_1+27J_2}{3840}(q_x^6 - q_y^6). \tag{175}$$

For the triangular lattice, a spatial anisotropy (terms in $q_xq_y$) appears only in the sixth order. This sixfold anisotropy acts as a *locking potential* that stabilizes triple-Q skyrmion crystals, expanding the region where skyrmions are stable in the temperature–field phase diagram. We now have six minima of the energy at $\vec{q} = \pm\vec{q}_1, \pm\vec{q}_2, \pm\vec{q}_3$, where .

$$\vec{q}_1 = q(1,0), \qquad \vec{q}_2 = q\left(-\frac{\sqrt{3}}{2}, -\frac{1}{2}\right), \qquad \vec{q}_3 = q\left(\frac{\sqrt{3}}{2}, -\frac{1}{2}\right). \tag{176}$$

The phase diagrams for this lattice are studied in Refs. [30,31].

Frustrated skyrmions exhibit a characteristic oscillatory behavior that distinguishes them from skyrmions stabilized by the Dzyaloshinskii-Moriya interaction. In frustrated magnetic systems, competing exchange interactions generate damped oscillations in the radial spin profile, leading to multiple reversals of the out-of-plane magnetization around the skyrmion core. Consequently, the skyrmion displays an oscillatory tail that strongly influences skyrmion-skyrmion interactions, stability, and dynamics. These oscillations originate from the finite wave vector associated with magnetic frustration and can be described by an asymptotic profile of the form $m_z(r) \propto e^{-r/\lambda}\cos(kr+\delta)$ , where $\lambda$ is the decay length and $k$ is the characteristic modulation wave vector. Such behavior gives rise to unique phenomena, including helicity-dependent interactions and the coexistence of attractive and repulsive interaction regimes.

## 10. Dipole-dipole interaction.

Dipole-dipole interactions are magnetic interactions that arise between neighboring magnetic moments and play a crucial role in determining the stability and arrangement of magnetic textures [2]. Unlike exchange interactions, which favor parallel or antiparallel alignment at very short distances, dipole-dipole interactions are long-range and depend on both the distance and relative orientation of the magnetic moments. Because they are long-range and anisotropic, dipole-dipole interactions can introduce magnetic frustration, as the energy-minimizing orientation of a magnetic moment depends on the relative positions of all surrounding moments. This frustration competes with exchange and anisotropy energies. In magnetic materials, these interactions contribute significantly to skyrmion formation and stabilization**.** Dipole-dipole interactions can influence skyrmion size, shape, and energy by balancing other magnetic forces, such as exchange and anisotropy energies. They help stabilize skyrmions in thin films and multilayer systems, making these structures promising candidates for future spintronic devices, including high-density data storage and energy-efficient information processing technologies.

The dipolar (magnetostatic) interaction contributes a non-local, anisotropic term to the continuum Hamiltonian of a 2D ferromagnet. In continuum form, it adds an energy functional of the magnetization field $\vec{m}(\vec{r})$ that depends on the long-range dipole–dipole kernel $1/r^3$.

For localized spins $\vec{S}_i$ at lattice sites $\vec{r}_i$, the dipolar interaction is

$$H_{dip} = \frac{\mu_0}{4\pi}\sum_{i\neq j}\frac{\vec{S}_i.\vec{S}_j - 3(\vec{S}_i.\hat{r}_{ij})(\vec{S}_j.\hat{r}_{ij})}{r_{ij}^3}, \tag{177}$$

where: $\vec{r}_{ij} = \vec{r}_i - \vec{r}_j$, $\hat{r}_{ij} = \vec{r}_{ij}/\left|\vec{r}_{ij}\right|$. To obtain the continuum Hamiltonian. I define the magnetization density $\vec{M}(\vec{r}) = M_s\vec{m}(\vec{r})$, with $\vec{m}$ a unit vector. Replacing the sum by an integral, we obtain

$$H_{dip} = \frac{\mu_0 M_s^2}{4\pi}\int d^2r\int d^2r'\frac{\vec{m}(\vec{r}).\vec{m}(\vec{r}') - 3[\vec{m}(\vec{r}).\hat{R}][\vec{m}(\vec{r}').\hat{R}]}{\left|\vec{R}\right|^3}, \tag{178}$$

with $\vec{R} = \vec{r} - \vec{r}'$.

Now, I will consider a thin film with uniform magnetization (no domains) on the *xy*-plane. I also assume the thickness is very small compared to the lateral size. Thus

$$\vec{m}(\vec{r}) = \vec{m} = \text{constant} \tag{179}$$

Then we can factor out the magnetization in Eq. (178) and write

$$E_{dip} = \frac{\mu_0 M_s^2}{4\pi}\left[ m_i m_j \iint d^2r d^2r' \frac{\delta_{ij} - 3\hat{R}_i.\hat{R}_j}{\left|\vec{R}\right|^3} \right]. \tag{180}$$

The demagnetizing tensor is defined as

$$N_{ij} = \frac{1}{4\pi}\iint d^2r d^2r' \frac{\delta_{ij} - 3\hat{R}_i.\hat{R}_j}{\left|\vec{R}\right|^3}. \tag{181}$$

Thus

$$E_{dip} = \mu_0 M_s^2 m_i N_{ij} m_j . \tag{182}$$

For a thin film in the *xy*-plane, the only non-vanishing component is $N_{zz} = 1$. Then

$$E_{dip} = \mu_0 M_s^2 m_z^2 . \tag{183}$$

In the approximation mentioned above, the dipolar interaction reduces to an effective local easy-plane anisotropy.

It can be shown that dipolar interactions generally increase skyrmion diameter. That happens because a larger skyrmion reduces the density of magnetic charges at the domain wall and spreads out the out-of-plane magnetization, lowering the demagnetizing energy. That is why skyrmions in thick films or multilayers can reach 100–1000 nm, while in ultrathin films with weak dipolar effects, they can be less than 10 nm. Unlike the DMI, dipole-dipole interactions are achiral and energetically favor Bloch skyrmions of both helicities [2].

# 11. Skyrmion-skyrmion interaction.

Within the pure Belavin-Poliakov model, the total energy is independent of the separation between the constituent skyrmions implying the absence of an interaction potential. Thus interaction potential arises once additional terms such as DMI, Zeeman coupling, anisotropy, dipolar interactions, or lattice corrections are include. Here, following Ref. [32], I will show that a magnetic field can play this role.

The effective field $\vec{H}_{eff}$ is defined from the energy functional $E$ as

$$\vec{H}_{eff} = -\frac{\delta E}{\delta \vec{n}}, \tag{184}$$

That is like a generalized "force" derived from energy. The torque on $\vec{n}$ is proportional to $\vec{\tau} \propto \vec{n} \times \vec{H}_{eff}$. So, the condition for a static equilibrium configuration is that the torque is zero; that is, $\vec{n} \times \vec{H}_{eff} = 0$. The system sits at an energy extremum (minimum, maximum, or saddle point).

I will first study the behavior of skyrmions for large *r*, and to do that, I will keep only linear terms in the in-plane spin components. At linear order, only the exchange and Zeeman terms contribute, so we can neglect the DMI term. Thus, we have

$$E = \frac{J}{2a}\int dxdy(\nabla\vec{n})^2 - \frac{H}{a^2}\int dxdyn_z \tag{185}$$

We apply the magnetic field *H* in the negative *z*-direction to prevent skyrmion expansion. Using Eq. (184), we get the effective field

$$H_{eff} = Ja^2(\nabla^2\vec{n}) - H\hat{e}_z\,. \tag{186}$$

Now, using $\vec{n}\times\vec{H}_{eff} = 0$ we find

$$n_y(H + Ja^2\nabla^2 n_z) - n_z(Ja^2\nabla^2 n_y) = 0\,, \tag{187}$$

$$n_z(Ja^2\nabla^2 n_x) - n_x(H + Ja^2\nabla^2 n_z) = 0\,. \tag{188 }$$

For large *r* we have $n_z \sim 1$, $\nabla^2 n_z = 0$, so we can write

$$Ja^2\nabla^2 n_x - n_x H = 0\,, \qquad Ja^2\nabla^2 n_y - n_y H = 0\,. \tag{189}$$

Writing the Laplacian in polar coordinates in two dimensions as

$$\nabla^2 = \frac{\partial^2}{\partial r^2} + \frac{1}{r}\frac{\partial}{\partial r} + \frac{1}{r^2}\frac{\partial^2}{\partial\phi^2}\,. \tag{190}$$

and taking

$$n_x = n_p\cos\phi\,, \qquad n_y = n_p\sin\phi\,, \tag{191}$$

we arrive at the equation

$$Ja^2\left(\frac{\partial^2}{\partial r^2} + \frac{1}{r}\frac{\partial}{\partial r} + \frac{1}{r^2}\right)n_p(r) - |H|n_p(r) = 0\,, \tag{192}$$

where we use $|H|$ as $H < 0$. Defining $\delta = a\sqrt{J/|H|}$, we write

$$\frac{\partial^2 n_p}{\partial r^2} + \frac{1}{r}\frac{\partial n_n}{\partial r} + \frac{n_p}{r^2} - \left(\frac{1}{r^2} + \frac{1}{\delta^2}\right)n_p = 0\,. \tag{193}$$

Multiplying the above equation by $r^2$ and defining $x = r/\delta$, we get

$$x^2\frac{d^2 n_p}{dx^2} + x\frac{dn_p}{dx} - (x^2+1)n_p = 0\,. \tag{194}$$

That is the modified Bessel equation of order 1. The general solution of Eq.(194) is

$$n_p = C_1 I_1(x) + C_2 K_1(x)\,, \tag{195}$$

Where $C_1$ and $C_2$ are constants, and $I_1(x)$ is the modified Bessel function of the first kind and order 1, and $K_1(x)$ is the modified Bessel function of the second kind of order 1. We have

$$\text{As } x\to 0 \quad I_1(x) \sim \frac{x}{2}, \qquad K_1(x) \sim \frac{1}{x} \tag{196}$$

$$\text{As } x\to\infty \quad I_1(x) \sim \frac{e^x}{\sqrt{2\pi x}}, \qquad K_1(x) \approx \sqrt{\frac{\pi}{2x}}e^{-x} \tag{197}$$

For large *r* we choose the second term in Eq. (197), and therefore, the magnetic field makes the skyrmion field fall off exponentially as $\exp(-r/\delta)$ at large distances. We can also apply the above linearization near the origin (and then use the first term in Eq. (196)), and we obtain the correct behavior $n_p(r) \sim r$ for the BP skyrmion.

Another way to achieve the same result is as follows. For $\vec{n} \to (0,0,1)$, we write

$$\vec{n} = (\psi_x, \psi_y, \sqrt{1-\psi^2}), \qquad \psi^2 = \psi_x^2 + \psi_y^2. \tag{198}$$

For small $\psi$, $n_z = 1 - \psi^2/2$. We have $\partial_i n_z = -\frac{1}{2}\partial_i(\psi^2)$. Since we keep only up to quadratic order in $\psi$, we can neglect this term. Thus

$$(\nabla\vec{n})^2 = \sum_{i=x,y} [(\partial_i\psi_x)^2 + (\partial_i\psi_y)^2] = |\nabla\vec{\psi}|^2, \tag{199}$$

with $\vec{\psi} = (\psi_x, \psi_y)$.

Expanding to quadratic order, we can write the Hamiltonian (185) as

$$E = \int d^2r[J|\vec{\nabla}\vec{\psi}|^2 + H|\vec{\psi}|^2]. \tag{200}$$

For an asymmetric skyrmion, we write $\vec{\psi}(r,\phi) = f(r)(\cos\phi, \sin\phi)$. Let

$$\kappa = \sqrt{H/J}, \tag{201}$$

the Euler-Lagrange equation then becomes

$$\frac{d^2 f(r)}{dr^2} + \frac{1}{r}\frac{df(r)}{dr} - \left(\frac{1}{r^2} + \kappa^2\right) f(r) = 0, \tag{202}$$

which is Eq. (194).

From a detailed analysis, we find that the interaction energy (leading to a repulsion) between two skyrmions at large separation *d*, behaves like

$$E_{\text{int}}(d) \propto \frac{e^{-d/\delta}}{\sqrt{d}}, \tag{203}$$

up to a prefactor that depends on the skyrmion type, the DMI, and the core structure [32].

## 12.1 Landau-Lifshitz-Gilbert and Thiele equations.

Magnetic skyrmions behave as particle-like topological textures embedded in a continuous magnetization field. Their dynamics are governed by the Landau–Lifshitz–Gilbert (LLG) equation [33-38]. Still, for slowly varying, rigid skyrmions, the full field dynamics can be reduced to an effective equation of motion for the skyrmion's center of mass. This reduction leads to the Thiele equation, the standard tool for describing skyrmion motion under spin torques, magnetic forces, and defects.

The Landau-Lifshitz-Gilbert equation is the fundamental equation describing how the magnetization in a ferromagnetic material evolves under the influence of magnetic fields and damping. It is central to micromagnetic, spintronic, and magnetic memory devices, as well as to the study of magnetic dynamics. In its simplest form, it is given by:

$$\frac{d\vec{M}}{dt} = -\gamma\vec{M} \times \vec{H}_{eff} + \frac{\alpha}{M_S}\vec{M} \times \frac{d\vec{M}}{dt}. \tag{204}$$

Where $\vec{M}$ is the magnetization vector, $M_S$ is the saturation magnetization (the maximum magnetization a magnetic material can achieve), $\vec{H}_{eff}$ is the effective magnetic field, $\gamma$ is the gyromagnetic ratio, and $\alpha$ is the Gilbert damping constant. This term is divided by $M_S$ to match units, making the equation dimensionally correct. The precession term $-\gamma\vec{M}\times\vec{H}_{eff}$ causes magnetization to precess around the effective magnetic field. It represents the conservation of angular magnetic momentum in an effective field. The damping term

$$\frac{\alpha}{M_S}\vec{M}\times\frac{d\vec{M}}{dt}, \tag{205}$$

leads to energy dissipation and gradually aligns magnetization with the effective field. It is a phenomenological term that ensures energy decreases monotonically in the absence of driving fields (see Fig. 11).

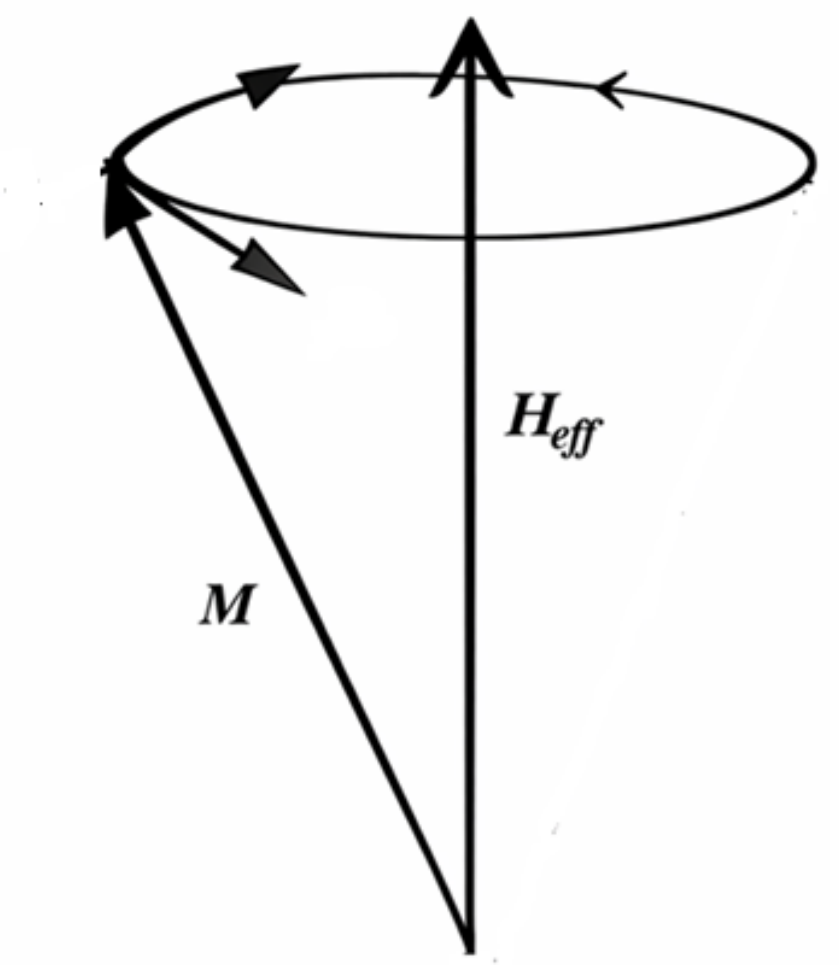

Figure 11. The two terms in Eq. (204).

The effective magnetic field is a combination of several contributions:

$$\vec{H}_{eff}=\vec{H}_{ext}+\vec{H}_{anis}+\vec{H}_{exc}+\vec{H}_{demag}+... \tag{206}$$

These include the external applied magnetic field, magneto-crystalline anisotropy, exchange interaction, and demagnetizing (dipolar) fields. $\vec{H}_{eff}$ is defined as the functional derivative of the magnetic energy with respect to the magnetization:

$$\vec{H}_{eff}(\vec{r})=-\frac{1}{\mu_0}\frac{\delta U[\vec{M}]}{\delta\vec{M}(\vec{r})}. \tag{207}$$

Where $\mu_0$ is the vacuum permeability. The original LLG equation (204) assumed only field-driven motion. It can be generalized to include adiabatic, and nonadiabatic spin transfer torques (STT) from a spin-polarized current [33] (see Appendix E). It becomes:

$$\frac{d\vec{M}}{dt}=-\gamma\vec{M}\times\vec{H}_{eff}+\frac{\alpha}{M_S}\vec{M}\times\frac{d\vec{M}}{dt}+(\vec{u}.\vec{\nabla})\vec{M}-\beta\vec{M}\times(\vec{u}.\vec{\nabla})\vec{M}\,. \tag{208}$$

The STT arises from the transfer of spin current from conduction electrons to the background magnetization and requires a noncollinear magnetic structure. The term $-(\vec{u}.\vec{\nabla})\vec{M}$ has its origin in *s-d* exchange between itinerant electron spins and local magnetization, plus conservation of spin angular momentum in a current-carrying ferromagnet. $\beta$ is the nonadiabatic torque parameter. A spin-polarized current with drift velocity $\vec{u}$ transports spin density; in the adiabatic limit, electron spins follow the local magnetization, leading to a convective derivative:

$$\frac{d\vec{M}}{dt} \rightarrow \frac{d\vec{M}}{dt} + (\vec{u}.\vec{\nabla})\vec{M} . \tag{209}$$

Moving the convective term to the right-hand side gives the adiabatic spin-transfer torque. The term $\beta\vec{M} \times (\vec{u}.\vec{\nabla})\vec{M}$ is the nonadiabatic spin-transfer torque. It originates from the finite spin-relaxation length and the misalignment between conduction-electron spins and local magnetization, microscopically linked to spin-flip scattering and spin dephasing. (A more detailed discussion is presented in Appendix E)..

I will divide Eq. (204) by $M_S$, define $\vec{m} = \vec{M} / M_S$, and write

$$\frac{d\vec{m}}{dt} = -\gamma\vec{m} \times \vec{H}_{eff} + \alpha\vec{m} \times \frac{d\vec{m}}{dt} . \tag{210}$$

The Thiele equation provides a compact, powerful formalism for describing magnetic-domain motion by reducing the LLG equation to a force-balance equation involving gyroscopic and dissipative integrals. This framework remains central to modern spintronics and skyrmion research. The Thiele equation describes the motion of the magnetization texture's center of mass in magnets when driven by an external field. Deriving the Thiele equation from the Landau-Lifshitz-Gilbert equation is an example of reducing a field theory to a collective-coordinate equation for a rigid magnetic texture. Identifying collective coordinates allows work with fewer degrees of freedom, i.e., a minimal set of equations of motion rather than the full field equations. For a skyrmion whose internal profile remains approximately rigid during motion, one introduces the collective coordinate $\vec{X}(t)$ of its center-of-mass, such that

$$\vec{m}(\vec{r},t) = \vec{m}_0(\vec{r} - \vec{X}(t)) , \tag{211}$$

where $\vec{m}_0$ is the static skyrmion configuration. Then

$$\frac{\partial\vec{m}}{\partial t} = -\frac{\partial\vec{m}}{\partial X_j}\frac{\partial X_j}{\partial t} = -(\vec{v}.\nabla)\vec{m} . \tag{212}$$

Where $\vec{v} = \dot{\vec{X}}$ is the drift velocity of the center-of-mass motion. Inserting Eq. (212) into Eq. (210), we get

$$-v_j\frac{\partial\vec{m}}{\partial X_j} = -\gamma\vec{m} \times \vec{H}_{eff} - \alpha v_j\vec{m} \times \frac{\partial\vec{m}}{\partial X_j} . \tag{213}$$

Where repeated indices imply a summation. We project Eq. (213) onto the translational degrees of freedom by multiplying it by $\vec{m}\times\partial\vec{m}/\partial X_i$ and integrating over the sample volume.

$$\int dV\vec{m}\times\frac{\partial\vec{m}}{\partial X_i}.\left[\gamma\vec{m}\times\vec{H}_{eff}-v_j\frac{\partial\vec{m}}{\partial X_j}+\alpha\vec{m}\times\left(v_j\frac{\partial\vec{m}}{\partial X_j}\right)\right]=0 \quad (214)$$

Using $(\vec{a}\times\vec{b}).\vec{c}=\vec{b}.(\vec{c}\times\vec{a})$, and $\vec{m}.(\partial\vec{m}/\partial X_j)=0$, we get for the second term in Eq.(214)

$$\left(\vec{m}\times\frac{\partial\vec{m}}{\partial X_i}\right).\frac{\partial\vec{m}}{\partial X_j}=\vec{m}.\left(\frac{\partial\vec{m}}{\partial X_i}\times\frac{\partial\vec{m}}{\partial X_j}\right). \quad (215)$$

We define the gyroscopic tensor $\vec{G}$ as

$$G_{ij}=\frac{M_S}{\gamma}\int dV\vec{m}.(\partial_i\vec{m}\times\partial_j\vec{m}). \quad (216)$$

The gyrotropic tensor plays the role of an effective Berry curvature in the space of collective coordinates. Using $(\vec{a}\times\vec{b}).(\vec{a}\times\vec{c})=(\vec{a}.\vec{a}).(\vec{b}.\vec{c})-(\vec{a}.\vec{b})(\vec{a}.\vec{c})$ and $\vec{m}.\vec{m}=1$, we obtain for the last term

$$\left(\vec{m}\times\frac{\partial\vec{m}}{\partial X_i}\right).\left(\vec{m}\times\frac{\partial\vec{m}}{\partial X_j}\right)=\frac{\partial\vec{m}}{\partial X_i}.\frac{\partial\vec{m}}{\partial X_j}. \quad (217)$$

The damping tensor $\vec{D}$ is defined as

$$D_{ij}=\frac{M_S}{\gamma}\int\partial_i\vec{m}.\partial_j\vec{m}dV. \quad (218)$$

For the first term in Eq. (214), we get

$$\left(\vec{m}\times\frac{\partial\vec{m}}{\partial X_i}\right).(\vec{m}\times\vec{H}_{eff})=\frac{\partial\vec{m}}{\partial X_i}.\vec{H}_{eff}. \quad (219)$$

Integrating and using Eq. (207), we find

$$\frac{\gamma}{\mu_0 M_S}\int dV\frac{\partial\vec{m}}{\partial X_i}.\frac{\delta U}{\delta\vec{m}}. \quad (220)$$

Using

$$\frac{\partial U}{\partial X_i}=-\int dV\frac{\delta U}{\delta\vec{m}}.\frac{\partial\vec{m}}{\partial X_i}=-F_i. \quad (221)$$

We obtain the term

$$\frac{\gamma}{\mu_0 M_S}F_i. \quad (222)$$

For an isotropic texture, $D_{ij}=D\delta_{ij}$. Collecting all terms, we obtain the Thiele equation

$$F_n+\alpha D_{nm}v_m+G_{nm}v_m=0. \quad (223)$$

In two dimensions, the antisymmetric tensor $G_{ij}$ can be written as

$$G_{ij}=\varepsilon_{ijk}G_k, \quad (224)$$

and we get

$$\vec{G}\times\vec{v}-\alpha\vec{D}\vec{v}+\vec{F}=0. \quad (225a)$$

Depending on the sign convention adopted, we can also write

$$\vec{G}\times\vec{v}+\alpha\vec{\vec{D}}\vec{v}+\vec{F}=0. \quad (225b).$$

Both forms are common in the literature. The Thiele equation for a ferromagnet does not contain the acceleration term with an inertial mass, which is absent if the magnetic texture is not deformed.

For a skyrmion, we have

$$\vec{G}=4\pi Q\frac{M_s t}{\gamma}\hat{z}. \quad (226)$$

where $Q$ is the topological charge and $t$ is the magnetic film thickness. The topological charge generates a gyrocoupling vector that acts as an effective magnetic field in real space. The gyroscopic term (Magnus force) $\vec{G}\times\vec{v}$ is responsible for the transverse deflection known as the skyrmion Hall effect, while the dissipative term $\alpha\vec{\vec{D}}\vec{v}$ governs longitudinal damping. The balance between the gyrotropic force, dissipative forces, and the driving force determines the skyrmion trajectory. The angle between the motion direction and the applied current is called the skyrmion Hall angle. It depends on factors such as damping, skyrmion size, material properties, and driving-current strength. Although the skyrmion Hall effect is a direct consequence of the topological nature of skyrmions, it presents a significant challenge for technological applications. In narrow magnetic nanotracks, the transverse motion can cause skyrmions to drift toward the device edges, where they may become pinned or annihilated. This limits reliable skyrmion transport in proposed memory and logic devices, such as racetrack memories. A skyrmion racetrack is a proposed spintronic memory device in which magnetic skyrmions move along a narrow magnetic strip like tiny, information-carrying particles on a track. A deeper understanding of skyrmion dynamics and better control of their motion are essential to developing next-generation magnetic memory and neuromorphic computing technologies.

Let us consider the case where

$$\vec{G}=G\hat{z}, \quad \vec{v}=(v_x,v_y), \quad \vec{F}=(F_x,F_y), \quad (227)$$

and

$$\vec{\vec{D}}=\begin{pmatrix} D_{xx} & D_{xy} \\ D_{yx} & D_{yy} \end{pmatrix}. \quad (228).$$

The Thiele equation in components becomes

$$-Gv_y+\alpha(D_{xx}v_x+D_{xy}v_y)+F_x=0,$$
$$Gv_x+\alpha(D_{yx}v_x+D_{yy}v_y)+F_y=0. \quad (229)$$

It can be rewritten as

$$\begin{pmatrix} \alpha D_{xx} & -G+\alpha D_{xy} \\ G+\alpha D_{yx} & \alpha D_{yy} \end{pmatrix}\begin{pmatrix} v_x \\ v_y \end{pmatrix}=-\begin{pmatrix} F_x \\ F_y \end{pmatrix}. \quad (230)$$

The determinant is

$$\Delta=\alpha^2 D_{xx}D_{yy}-(-G+\alpha D_{xy})(G+\alpha D_{yx}). \quad (231)$$

We have the velocities

$$v_x = \frac{-F_x \alpha D_{yy} + F_y(-G + \alpha D_{xy})}{\Delta},$$
$$v_y = \frac{-F_y \alpha D_{xx} + F_x(G + \alpha D_{yx})}{\Delta}. \tag{232}$$

For the longitudinal drive $F_y = 0$, and the skyrmion Hall angle is

$$\tan\theta_H = \frac{v_y}{v_x} = -\frac{G + \alpha D_{yx}}{\alpha D_{yy}}. \tag{233}$$

We see that the off-diagonal element $D_{yx}$ renormalizes the effective transverse response.

Now I will treat Eq. (208) following the same procedure as before. We have

$$(\vec{u}.\vec{\nabla})\vec{m} = u_j \frac{\partial \vec{m}}{\partial X_j}. \tag{234}$$

This term can be combined with the term in $v_j$. The term

$$\beta \vec{m} \times (\vec{u}.\vec{\nabla})\vec{m} = \beta u_j \vec{m} \times \left( \frac{\partial \vec{m}}{\partial X_j} \right), \tag{235}$$

leads to an expression similar to the term in $\alpha$. In the literature, it is common to use $\vec{v} = \vec{v}_d, \vec{u} = \vec{v}_s$ (where $\vec{v}_s$ is the spin current velocity of the conduction electrons). Collecting all projected terms, we obtain the modified Thiele equation

$$\vec{G}.(\vec{v}_s - \vec{v}_d) - \vec{D}(\beta \vec{v}_s - \alpha \vec{v}_d) + \vec{F} = 0. \tag{236a}$$

or

$$G_{mn}(v_{sn} - v_{dn}) - D_{mn}(\beta v_{sn} - \alpha_{dn}) + F_m = 0. \tag{236b}$$

Spin-orbit torque is a torque exerted by a spin current generated by the spin-Hall effect due to interface spin-orbit coupling in a magnetic heterojunction system. (Ref. [37] is excellent and contains a lot of information on the subject).

A charge current $\vec{J}_c$ flowing in a heavy metal generates a transverse pure spin current through the spin Hall effect (see Appendix E). The damping-like torque is

$$\vec{\tau} = H_{SH} \vec{m} \times (\vec{\sigma} \times \vec{m}), \tag{237}$$

where

$$H_{SH} = \frac{\gamma \Theta_{SH} j_e}{2eM_S d} = \frac{v_s^{\perp}}{d}. \tag{238}$$

Here $v_s^{\perp}$ is the in-plane vertical spin current, *d* is the ferromagnetic thickness, $\vec{\sigma}$ is a unit vector representing the direction of the spin polarization of the spin current, $\Theta_{SH}$ is the spin Hall angle, and *e* is the electron charge. Projecting the torque onto the translational mode and using the relation $\vec{m} \times [\vec{m} \times (\vec{\sigma} \times \vec{m})] = \vec{m} \times \vec{\sigma}$, we obtain a contribution $J_i^{spin}$ to Eq. (236) where [37]

$$J_i^{spin} = -\frac{v_s^{\perp}}{d} \vec{\sigma}. \int dxdy \left( \frac{\partial \vec{m}}{\partial X_i} \times \vec{m} \right), \tag{239}$$

Sometimes, Eq.(239) is written in the literature as

$$B j \vec{I} \vec{\sigma}, \quad (240)$$

where the torque tensor $\vec{I}$ is given by

$$I_{ij} = \int dxdy[\partial_i \vec{m}(\vec{r}) \times \vec{m}(\vec{r})]_j . \quad (241)$$

The field-like torque is given by $\tau_{FL} \vec{m} \times \vec{\sigma}$. Projecting onto the translational mode and using $\partial_i \vec{m}.[\vec{m} \times \vec{m} \times \vec{\sigma}] = \partial_i \vec{m}.[(\vec{m}.\vec{\sigma})\vec{m} - (\vec{m}.\vec{m})\vec{\sigma}] = -\partial_i \vec{m}.\vec{\sigma}$, as $\partial_i \vec{m}.\vec{m} = 0$, we find

$$T_{FL} \propto \vec{\sigma}.\int dxdy \partial_i \vec{m} . \quad (242).$$

For a localized texture with $\vec{m}$ approaching a constant at infinity, that integral often vanishes so that this term can drop out in the Thiele equation depending on boundary conditions.

As discussed previously, Néel skyrmions have radial in-plane magnetization, whereas Bloch skyrmions have tangential (azimuthal) in-plane magnetization. Under spin-transfer torque (STT), both types of skyrmions exhibit similar dynamics because STT depends primarily on the spatial gradients of the magnetization rather than its absolute in-plane orientation. Consequently, the skyrmion velocity and trajectory are only weakly affected by the helicity, and Bloch and Néel skyrmions respond in a qualitatively similar manner. In contrast, spin-orbit torque (SOT**)** is strongly sensitive to the skyrmion helicity. The torque generated by the spin Hall effect couples directly to the internal magnetization texture, so the direction and magnitude of the resulting force depend on whether the skyrmion is of Bloch or Néel type. Within the Thiele formalism, the SOT-induced force is determined by a helicity-dependent tensor, $F_i \propto I_{ij} \sigma_j$, where $\vec{\sigma}$ denotes the spin polarization generated by the spin Hall effect.

The tensor introduced in Eq. (241) is directly related to the skyrmion helicity $\gamma$ and can be written as

$$\vec{I} = I_0 \begin{pmatrix} \cos\gamma & \sin\gamma \\ -\sin\gamma & \cos\gamma \end{pmatrix}, \quad (243)$$

where $I_0$ is a constant determined by the skyrmion size and radial profile. For a Néel skyrmion ($\gamma = 0$ or $\pi$), the tensor is diagonal, whereas for a Bloch skyrmion ($\gamma = \pm\pi / 2$) it becomes purely off diagonal. Consequently, changing the helicity rotates the direction of the SOT-induced driving force relative to the spin polarization. Therefore, Bloch and Néel skyrmions generally move in different directions under the same applied current. The precise direction of motion results from the competition between the SOT-induced force, the gyrotropic force, and dissipation, leading to different skyrmion Hall angles for different helicities. Rather than being strictly parallel or perpendicular to the current, the trajectory depends on the skyrmion helicity and the specific form of the spin-orbit torque.

## 12.2 Landau-Lifshitz-Bloch equation.

The Landau-Lifshitz-Bloch equation (LLB) [40] is a phenomenological equation that describes magnetization dynamics, including changes in its magnitude, unlike the LLG equation, which assumes the magnetization has constant length. That makes the LLB equation particularly useful for describing ferromagnets at elevated temperatures, especially near and above the Curie temperature, where the magnetization length is no longer conserved. A common form for the LLB equation for a ferromagnet is

$$\frac{\partial \vec{m}}{\partial t} = -\gamma \vec{m} \times \vec{H}_{eff} + \gamma \alpha_{//} \frac{(\vec{m}.\vec{H}_{eff})}{m^2} \vec{m} - \gamma \alpha_{\perp} \frac{\vec{m} \times (\vec{m} \times \vec{H}_{eff})}{m^2}, \tag{244 }$$

Where $\vec{m}$ is the reduced magnetization $\vec{M} / M_S(0)$, $m = |\vec{m}|$ is not fixed, $\gamma$ is the gyromagnetic ratio, $\alpha_{//}$ is the longitudinal damping, and $\alpha_{\perp}$ the transverse damping. The effective field $\vec{H}_{eff}$ iincludes, in addition to the usual terms, a longitudinal restoring term that drives the magnetization towards its thermal equilibrium value. Below the Curie temperature $(T < T_C)$, this is often written as

$$\vec{H}_{eff} = \vec{H}_{mic} + \frac{1}{2\chi_l}\left(1 - \frac{m^2}{m_e^2(T)}\right)\vec{m}, \tag{245}$$

Where $\vec{H}_{mic}$ contains exchange, anisotropy, demagnetizing, and applied fields, $\vec{m}_e(T)$ is the equilibrium magnetization at temperature $T$ and $\chi_l$ is the longitudinal susceptibility. This longitudinal field allows the magnitude of $\vec{m}$ to relax toward the temperature-dependent equilibrium value. In the classical LLB theory, we have for $T < T_C$

$$\alpha_{//} = \frac{2\lambda T}{3T_C}, \qquad \alpha_{\perp} = \lambda\left(1 - \frac{T}{3T_C}\right). \tag{246}$$

Where $\lambda$ is the microscopic spin-bath coupling constant. In Eq. (244), the first term on the right-hand side has the same form as the LLG equation. The second term (longitudinal relaxation) changes the magnetization's magnitude; the third term (transverse relaxation) aligns the magnetization with the effective field. The LLB equation reduces to the LLG equation in the low-temperature limit, where $m \approx 1$. Longitudinal fluctuations are negligible, and only transverse damping remains.

The LLB equation is widely used to model phenomena such as ultrafast laser-induced demagnetization, heat-assisted magnetic recording, and finite-temperature spin dynamics, where changes in magnetization magnitude are essential.

# 13. Antiferromagnet

For the collinear antiferromagnet (as mentioned in section 5), there are two sublattices, A and B. As we saw before, the dynamics of the antiferromagnet is described by the nonlinear sigma model, and the equation of motion (unlike that of the ferromagnet) is second order in time. For a collinear Heisenberg antiferromagnet, we work with the Néel vector $\vec{n}$. The effective Lagrangian density for the Heisenberg antiferromagnet with near-neighbor interaction is given by Eq.(94) (see also Appendix C), written as

$$\mathscr{L} = \frac{\chi}{2}\left[\frac{1}{c^2}\left(\frac{\partial \vec{n}}{\partial t}\right)^2 - (\nabla \vec{n})^2\right]. \quad (247)$$

Where $\chi = JS^2 a^{2-d}$ , $c = 2\sqrt{d}JSa$ , and $a$ is the lattice spacing, and $d$ the lattice dimension. We use a rigid skyrmion ansatz

$$\vec{n}(\vec{r},t) = \vec{n}_0(\vec{r} - \vec{X}(t)), \quad (248)$$

write

$$\dot{\vec{n}} = -\dot{X}_j \partial_j \vec{n}_0, \quad (249)$$

and get the kinetic energy term

$$T = \frac{\chi}{2}\int d^2 r (\dot{\vec{n}})^2 = \frac{1}{2} M_{ij} \dot{X}_i \dot{X}_j, \quad (250)$$

with the mass tensor

$$M_{ij} = \chi \int d^2 r (\partial_i \vec{n}_0).(\partial_j \vec{n}_0). \quad (251)$$

The Gilbert damping in the AFM gives a Rayleigh dissipation functional

$$\Re = \frac{\alpha\chi}{2}\int d^2 r (\dot{\vec{n}})^2. \quad (252)$$

Using Eq.(249) we get

$$\Re = \frac{\alpha}{2} D_{ij}^{AFM} \dot{X}_i \dot{X}_j, \quad (253)$$

where

$$D_{ij}^{AFM} = \chi \int d^2 r (\partial_i \vec{n}_0).(\partial_j \vec{n}_0). \quad (254)$$

Thus $D_{ij}^{AFM} = M_{ij}$, and the damping force is given by $F_i^{damp} = \alpha D_{ij}^{AFM} \dot{X}_j$. Putting it together, the Thiele equation for the antiferromagnet is

$$M_{ij}\ddot{X}_j + \alpha D_{ij}^{AFM} \dot{X}_j + F_i^{drive} = 0. \quad (255)$$

Since my main objective is pedagogical, I will present, in the following, a discussion of the antiferromagnet using the LLG equations from the start, as is sometimes done in some publications [40-43]. I consider an antiferromagnet with two sublattice magnetizations, oriented in opposite directions $\vec{m}_A = -\vec{m}_B$. Each sublattice magnetization can be described by the LLG equations:

$$\frac{d\vec{m}_A}{dt} = -\gamma \vec{m}_A \times \vec{H}_A + \alpha \vec{m}_A \times \frac{d\vec{m}_A}{dt} + \ldots, \qquad \frac{d\vec{m}_B}{dt} = -\gamma \vec{m}_B \times \vec{H}_B + \alpha \vec{m}_B \times \frac{d\vec{m}_B}{dt} + \ldots. \quad (256)$$

Following the procedure in the ferromagnetic case, we derive the Thiele equation as [37]

$$(G_A + G_B) \times v_d + \alpha (D_A + D_B) v_d = 0, \quad (257)$$

with $G_A = -G_B$, $D_A = D_B$. Sublattice A has a gyrovector $+\vec{G}$ , and sublattice B has $-\vec{G}$, and so the Magnus force vanishes.

The two-sublattice model is used when the exchange is not assumed to be very strong or when we are interested in the microscopic modeling of sublattices. In the following, I will consider the Néel vector as the true order parameter. To do that, we define, as before

$$\vec{m} = \frac{\vec{m}_A + \vec{m}_B}{2}, \qquad \vec{n} = \frac{\vec{m}_A - \vec{m}_B}{2}. \tag{258}$$

So
$$\vec{m}_A = \vec{m} + \vec{n}, \qquad \vec{m}_B = \vec{m} - \vec{n}. \tag{259}$$

I use different notations from Appendix D to align with the current literature on skyrmions ( $\vec{m}$ here corresponds to $\vec{l}$ in Appendix D). In an antiferromagnet, a skyrmion is not defined separately on each sublattice. It is a topological texture of the Néel (staggered) order parameter. The topological charge is defined for the Néel order parameter $\vec{n}$, and not for each sublattice separately. Even though the two sublattices twist in opposite directions, the combined staggered field forms a single skyrmion with a single topological charge. In the absence of a magnetic field, antiferromagnetic skyrmions have no net magnetization, making them insensitive to external magnetic fields and eliminating the skyrmion Hall effect. Antiferromagnetic skyrmions couple much more weakly to a uniform magnetic field than ferromagnetic skyrmions.

Adding and subtracting the two Eqs. (256) and using relations such as

$$\vec{m}_A \times \vec{H}_A + \vec{m}_B \times \vec{H}_B = (\vec{m} + \vec{n}) \times \vec{H}_A + (\vec{m} - \vec{n}) \times \vec{H}_B \tag{260}$$

$$= \vec{m} \times (\vec{H}_A + \vec{H}_B) + \vec{n} \times (\vec{H}_A - \vec{H}_B) = \vec{m} \times \vec{H}_m + \vec{n} \times \vec{H}_n, \tag{261}$$

we obtain

$$\dot{\vec{n}} = -\gamma \vec{n} \times \vec{H}_m - \gamma \vec{m} \times \vec{H}_n + \alpha(\vec{n} \times \dot{\vec{n}} + \vec{m} \times \dot{\vec{m}}), \tag{262}$$

$$\dot{\vec{m}} = -\gamma \vec{n} \times H_n - \gamma \vec{m} \times \vec{H}_m + \alpha(\vec{n} \times \dot{\vec{m}} + \vec{m} \times \dot{\vec{n}}). \tag{263}$$

The energy density (considering only the exchange term) is obtained from Eq,(D.12), and is given by

$$E = JS^2 (\nabla \vec{n})^2 + H_E \vec{m}^2. \tag{264}$$

Where $H_E = 4JS^2$. Thus

Thus

$$\vec{H}_m = -\frac{\delta E}{\delta \vec{m}} = -2H_E \vec{m}, \qquad \vec{H}_n = -\frac{\delta E}{\delta \vec{n}} = 2JS^2 \nabla^2 \vec{n}. \tag{265}$$

The second term in Eq. (262) is small compared to the first one and can be neglected. Hence

$$\dot{\vec{n}} \approx -2\gamma H_E \vec{n} \times \vec{m}. \tag{266}$$

Multiplying by $\vec{n}$, and using the cross product, we have

$$\vec{n} \times \dot{\vec{n}} = -2\gamma H_E \vec{n} \times (\vec{n} \times \vec{m}). \tag{267}$$

Using $\vec{n}.\vec{m} = 0$, $|\vec{n}| = 1$, we get $\vec{n} \times (\vec{n} \times \vec{m}) = -\vec{m}$. Thus

$$\vec{m} = \frac{1}{2\gamma H_E} \vec{n} \times \dot{\vec{n}}. \tag{268}$$

Since $\dot{\vec{n}}\times\dot{\vec{n}}=0$, we get

$$\dot{\vec{m}}=\frac{1}{2\gamma H_E}\vec{n}\times\ddot{\vec{n}}. \tag{269}$$

Using the first term of Eq.(262) leads to

$$\frac{1}{2\gamma H_E}\vec{n}\times\ddot{\vec{n}}=-\gamma\vec{n}\times\vec{H}_n. \tag{270}$$

Using $\vec{n}\times(\vec{n}\times\vec{a})=\vec{n}.(\vec{n}.\vec{a})-\vec{a}=-\vec{a}_p$, (this equals minus the component perpendicular to the vector $\vec{n}$, and multiplying once more by $\vec{n}$, we get

$$\frac{1}{2\gamma H_E}\ddot{\vec{n}}=\gamma\vec{H}_n,\quad \chi\ddot{\vec{n}}=-\frac{\delta E}{\delta\vec{n}},\text{ where }\quad \chi=\frac{1}{2\gamma^2 H_E}. \tag{271}$$

Adding the Gilbert damping gives

$$\chi\ddot{\vec{n}}+\alpha\dot{\vec{n}}=-\frac{\delta E}{\delta\vec{n}}. \tag{272}$$

Using Eq. (265), we obtain

$$\chi\ddot{\vec{n}}+\alpha\dot{\vec{n}}-2JS^2\nabla^2\vec{n}=0. \tag{273}$$

If the magnetization texture is rigid and not deformed during the motion, we can, as in the ferromagnetic case, write

$$\vec{n}(\vec{r},t)=\vec{n}_0(\vec{r}-\vec{R}(t)). \tag{274}$$

Then $\dot{\vec{n}}=-\dot{R}_i\partial_i\vec{n}$, and $\ddot{\vec{n}}=\ddot{R}_i\partial_i\vec{n}+\dot{R}_i\dot{R}_j\partial_i\partial_j\vec{n}$.

For slow motion, we neglect the quadratic velocity term. We multiply (272) by $\partial_i\vec{n}$ and integrate over space, obtaining

$$\int\partial_j\vec{n}.\left(\chi\ddot{\vec{n}}+\alpha\dot{\vec{n}}+\frac{\delta E}{\delta\vec{n}}\right)dxdy=0. \tag{275}$$

Using (274), we get

$$\chi\int dxdy(\partial_i\vec{n}.\partial_j\vec{n})\ddot{R}_i+\Gamma_{ij}\dot{R}_i=F_j. \tag{276}$$

Finally

$$M_{ij}\ddot{R}_i+\Gamma_{ij}\dot{R}_i=F_j. \tag{277}$$

With

$$M_{ij}=\chi\int(\partial_i\vec{n}.\partial_j\vec{n})dxdy. \tag{278}$$

That matches the experimental result that antiferromagnetic skyrmions behave like massive quasiparticles with inertia [43], unlike ferromagnetic skyrmions.

For a rotationally symmetric skyrmion $M_{ij}=M\delta_{ij}$, giving

$$M\ddot{\vec{R}}+\Gamma\dot{\vec{R}}=\vec{F}. \tag{279}$$

In this case, we get

$$M=\chi\int d^2r(\partial_i\vec{n})^2. \tag{280}$$

For $\vec{n}$ given by Eq. (31) we have

$$(\partial_i \vec{n})^2 = \left(\frac{d\theta}{dr}\right)^2 + \frac{\sin^2\theta}{r^2},. \tag{281}$$

Thus

$$M = \chi \int 2\pi r dr \left[ \left(\frac{d\theta}{dr}\right)^2 + \frac{\sin^2\theta}{r^2} \right]. \tag{282}$$

For the skyrmion given by Eq. (39) we find

$$\frac{d\theta}{dr} = \frac{2R}{r^2 + R^2}, \qquad \sin\theta = \frac{2rR}{r^2 + R^2}. \tag{283}$$

And the mass term is given by

$$M = 16\pi\chi R^2 \int \frac{rdr}{(r^2 + R^2)^2} = \frac{8\pi\chi}{\gamma^2}. \tag{284}$$

Now we solve the equation of motion for a constant force. For each component, we write

$$M\ddot{R} + \Gamma \dot{R} = F_0. \tag{285}$$

With solution

$$R(t) = R_0 + v_\infty t + \frac{M}{\Gamma}(v_0 - v_\infty)(1 - e^{-\frac{\Gamma}{M}t}), \tag{286}$$

where the terminal velocity $v_\infty$ is given by $v_\infty = F_0 / \Gamma$. We see that the motion is collinear with $\vec{F}_0$, there is no transverse Hall angle. Because they do not suffer transverse deflections, antiferromagnetic skyrmions can be driven at ultra-high velocities (e.g., up to 900 m/s) with a significantly lower current density than in ferromagnets [44].

. To stabilize an antiferromagnetic skyrmion, we need a magnetic field, as with ferromagnets. However, it must be oriented oppositely in the two sublattices, A and B. One way to do that is to grow an antiferromagnetic skyrmion host on top of a collinear antiferromagnet with the same crystal structure at the interface. Thus, the interface exchange interaction simulates a staggered magnetic field [2].

Ferrimagnets consist of at least two magnetic sublattices (A and B) whose spins align antiparallel due to exchange interactions. However, because the moment magnitudes differ, the cancellation is incomplete, leaving a nonzero macroscopic magnetization. This net magnetization behaves as the order parameter. The derivation of the Thiele equation follows a procedure similar to that used for an antiferromagnet. While skyrmions were initially investigated in ferromagnetic systems, recent attention has shifted to ferrimagnetic materials, which combine the advantages of both ferromagnets and antiferromagnets. In particular, ferrimagnetic skyrmions exhibit enhanced dynamical properties, including high propagation velocities and a strongly reduced skyrmion Hall effect, making them attractive for information storage and logic applications [45,46].

## 14. Topological structures in three dimensions.

Magnetic systems host a zoo of three-dimensional topological quasiparticles and defects, such as skyrmion tubes, Bloch points, bobbers, hedgehogs, and hopfions [47]. Skyrmions often stretch vertically as tubes through the depth of thick magnets or multilayer films, behaving as robust string-like tubes. These tubes can twist, intertwine, and have varying helicity along their length. They can interact, merge, begin, end or form closed loops, leading to the occurrence of Bloch points, chiral bobbers, or hopfions. Bloch points are singular, zero-dimensional topological defects where the magnetization direction becomes undefined. A hedgehog describes a particular magnetization arrangement around a Bloch point. In a hedgehog, magnetic moments point radially inward or outward in all directions (Fig.12). (The terms hedgehog and Bloch points are sometimes used interchangeably in the literature). Because the magnetization at the very core of the hedgehog is not well defined, these structures behave as emergent magnetic monopoles (rather than genuine magnetic monopoles )and play a crucial role in how magnetic states shift and evolve. A hedgehog is described by the magnetization field

$$\vec{m}(\vec{r}) = \pm\frac{\vec{r}}{|\vec{r}|}. \tag{287}$$

A hedgehog is characterized by how the magnetization field wraps the spin sphere on a closed surface enclosing the defect point. The total exchange energy of the hedgehog texture enclosed by a sphere of radius $R$ is given by $E = 8\pi JR$.

Chiral bobbers are 3D localized structures that resemble a discontinued skyrmion tube. They generally consist of a skyrmion-like swirl at the surface of the magnetic material that plunges into the bulk but tapers off, capping at a single Bloch point.

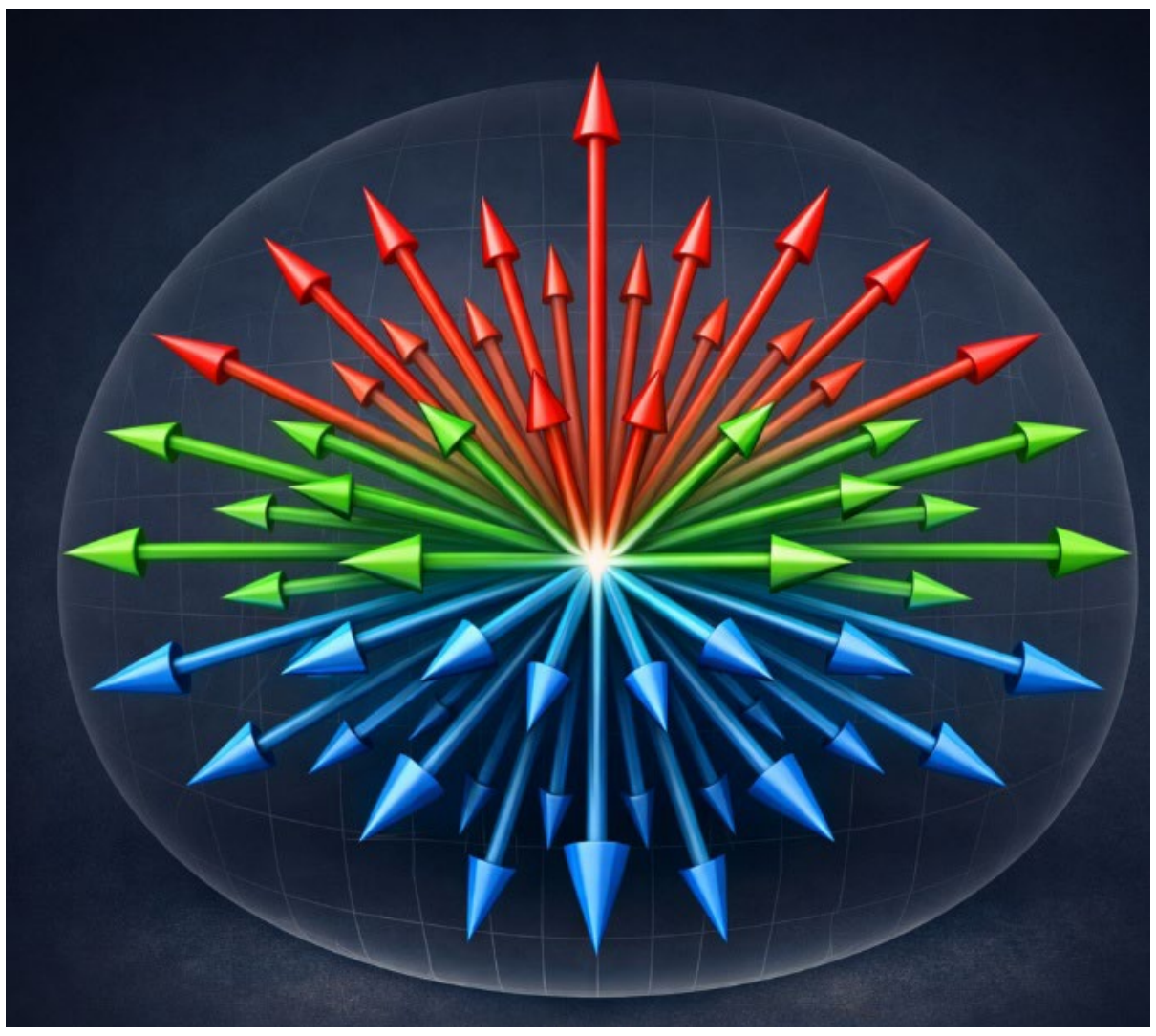

Figure 12. A Hedgehog.

As mentioned in section 2, **hopfions** are three-dimensional localized field configurations characterized by a nontrivial mapping between spatial coordinates and an order-parameter field. It is defined by a continuous vector field $\vec{m}(\vec{r})$ mapping physical space to the unit sphere. Consider a unit vector field $\vec{m}(\vec{r}) = (m_x, m_y, m_z)$ defined in three-dimensional space, satisfying $m_x^2 + m_y^2 + m_z^2 = 1$. Thus, every point in space is assigned to a point on the unit sphere. To obtain finite-energy configurations and define the Hopf map, the field is required to approach a constant direction far from the localized structure, as for instance

$$\lim_{|\vec{r}|\to\infty} \vec{m}(\vec{r}) = (0,0,1).$$

The physical space is non-compact. It extends infinitely in all directions. If we add one single point, denoted $\infty$, and define the topology so that "going infinitely far away in any direction" converges to that point, the resulting space $\mathbb{R}^3 \cup \{\infty\}$ becomes topologically identical to the sphere $S^3$. Consequently, the field becomes a continuous mapping from one compact sphere $S^3$ (in $\mathbb{R}^4$) into the unit sphere $S^2$ (in $\mathbb{R}^3$): $S^3 \to S^2$ Different mappings cannot generally be continuously deformed into one another. They are classified by the Hopf invariant, denoted by $H \in \mathbb{Z}$. Hopf invariant only exists for maps $S^3 \to S^2$. Unlike the winding number of vortices or magnetic monopoles, the Hopf invariant measures field-line linking. Hopfions represent knotted textures whose stability arises because the field cannot be continuously deformed into a trivial state without introducing singularities. In magnetic materials, topological decay can occur through singularities where the magnetization vanishes. Configurations with $H \neq 0$ cannot be unwound without discontinuities, providing topological protection (in the literature $H$ is sometimes referred to as $Q$ or $Q_H$). In Appendix F, I discuss hopfions in more detail from a mathematical point of view (the reader should consult this Appendix for definitions of some terms used here).

A convenient way to define $H$ is through an auxiliary vector field. First, define the antisymmetric tensor

$$F_{ij} = \vec{m}.\left(\frac{\partial \vec{m}}{\partial x_i} \times \frac{\partial \vec{m}}{\partial x_j}\right), \tag{288}$$

with $i$, $j$ = 1, 2, 3. Then we introduce a vector potential $A$ satisfying

$$F_{ij} = \frac{\partial A_j}{\partial x_i} - \frac{\partial A_i}{\partial x_j}. \tag{289}$$

The Hopf invariant is then given by

$$H = \frac{1}{32\pi^2} \int \varepsilon_{ijk} A_i F_{jk} d^3 r. \tag{290}$$

Where $\varepsilon_{ijk}$ is the Levi-Civita symbol.

In ferromagnets, magnetic hopfions arise as stationary points of the energy

$$E = \int [J_{ex}(\nabla \vec{m})^2 + D\vec{m}.(\vec{\nabla} \times \vec{m}) + Af(\vec{m}) + E_{dip}(\vec{m})] d^3 r, \tag{291}$$

where $A$ encodes anisotropy, and $E_{dip}$ is the nonlocal dipolar energy. Hopfions typically require a balance between exchange, DMI, and dipolar interactions. Exchange favors smoothness, DMI favors twisting, and dipolar interactions penalize certain deformations, collectively stabilizing knotted configurations.

The stability of magnetic hopfions depends on both topological constraints and energetic considerations. While the Hopf index prevents topological decay, the detailed field configuration must also minimize the system's free energy. Numerical simulations reveal that hopfions often occupy metastable minima, stabilized by geometric frustration or confinement. Higher-index hopfions $(H > 1)$ exhibit increasingly complex linking structures and may decompose into multiple lower-index solitons under perturbations [48].

For a hopfion, the preimage of any point on $S^2$ is a closed circle in $S^3$. All these circles wrap around a torus, link with each other, and form nested toroidal surfaces. $H$ counts the number of times the circles are linked with each other.

Toroidal coordinates can describe such shapes. Toroidal coordinates $(\eta, \beta, \varphi)$ are built around a ring (the focal circle), and surfaces of constant $\eta$ are tori. Recent works on hopfions use toroidal coordinates because they match the topology and symmetry of hopfions, allowing their fields, magnetization textures, and topological invariants to be expressed in simpler, more symmetric, and often analytically tractable forms. In many physical realizations (e.g., cylindrical nanodots or nanowires), hopfions form ring-like structures. Toroidal coordinates allow natural boundary surfaces, easier enforcement of finite-radius constraints, and more stable numerical simulations. This matters for studies of hopfion stability in ring-like structures, where geometry plays a crucial role in metastability and dynamics. Simulations in Cartesian coordinates often waste resolution in empty space and struggle with curved structures. Toroidal coordinates concentrate grid points where the hopfion structure actually lives, improving computational efficiency, energy-minimization accuracy, and the stability of dynamical simulations. Thus, to study hopfions, it is useful to introduce the toroidal coordinates. The terms toroidal and poloidal refer to directions relative to a reference torus (Fig. 13).

The connection with the cylindrical coordinates $(\rho, \varphi, z)$ is $\varphi = \varphi$,

$$\rho = \frac{a \sinh\eta}{\cosh\eta - \cos\beta}, \qquad z = \frac{a\sin\beta}{\cosh\eta - \cos\beta}. \tag{292}$$

Where the toroidal parameter $\eta$ varies from 0 to $\infty$, the poloidal angle $\beta$ varies from $-\pi$ to $\pi$, the azimuthal angle $\varphi$ varies from 0 to $2\pi$, and $a$ is the focal circle radius of the toroidal coordinates (not exactly the physical hopfion radius) in the plane $z = 0$. The relation to Cartesian coordinates is

$$x = \frac{a \sinh\eta}{\cosh\eta - \cos\beta}\cos\varphi, \quad y = \frac{a \sinh\eta}{\cosh\eta - \cos\beta}\sin\varphi, \quad z = \frac{a\sin\beta}{\cosh\eta - \cos\beta}. \tag{293}$$

The relation between toroidal coordinates and cylindrical coordinates is given by

$$\eta = arc\tanh\left(\frac{2a\rho}{\rho^2+z^2+a^2}\right) \qquad \beta = arc\tan\left(\frac{2az}{\rho^2+z^2-a^2}\right), \tag{294}$$

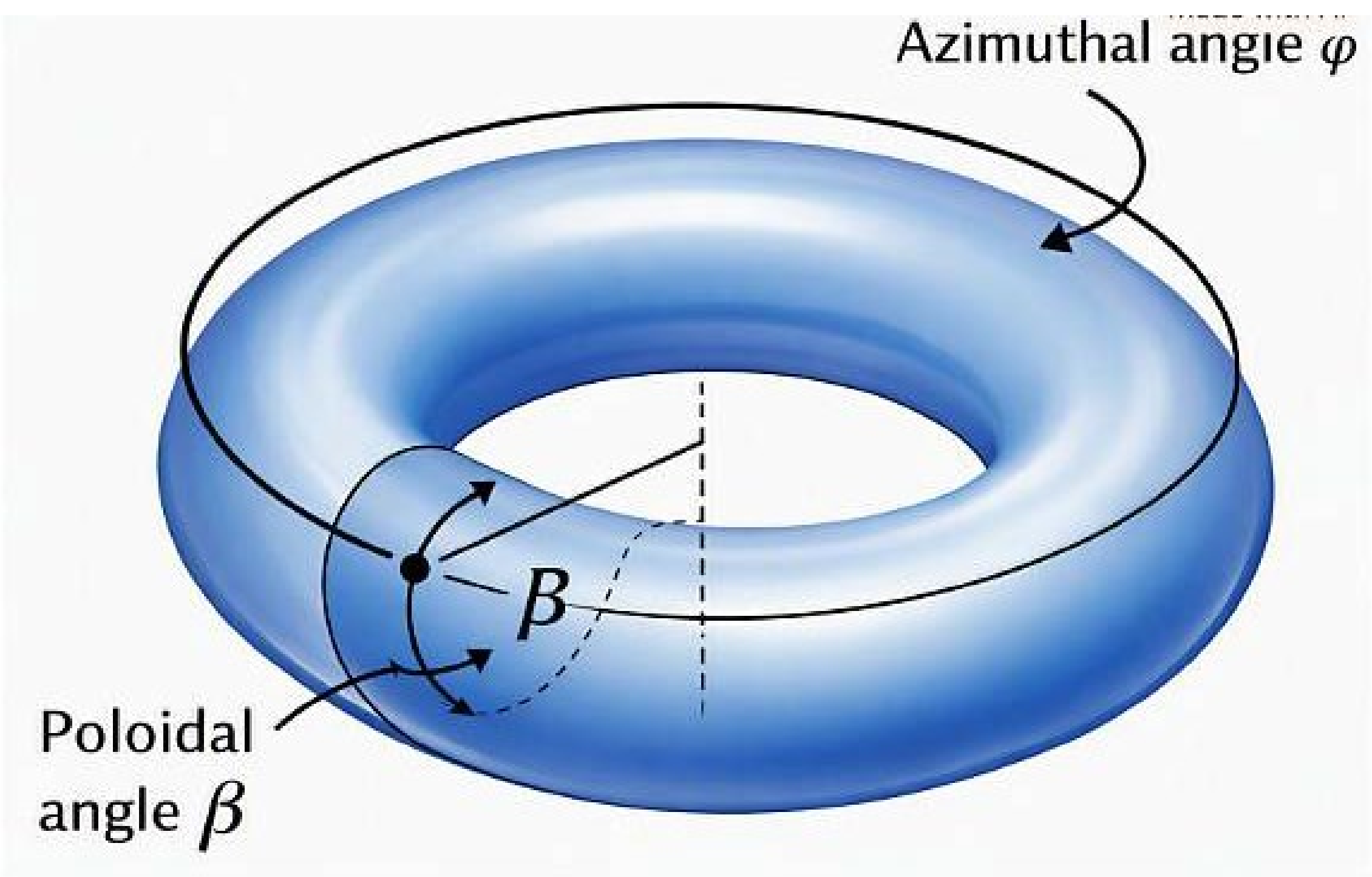

Figure 13. Toroidal coordinates

A 3-sphere consists of the points $(n_1,n_2,n_3,n_4)$ in $\mathbb{R}^4$ with $n_1^2+n_2^2+n_3^2+n_4^2=1$. A map $S^3 \to S^2$ can be defined in several ways. For instance, identify $\mathbb{R}^4$ with $\mathbb{C}^2$ by $(n_1,n_2,n_3,n_4) \leftrightarrow (z_1,z_2)=(n_1+in_2,n_3+in_4)$, and identify $\mathbb{R}^3$ with $\mathbb{C}\times\mathbb{R}$ by writing $(x_1,x_2,x_3) \leftrightarrow (z,x)=(x_1+ix_2,x_3)$. Thus $S^3$ is identified with the subset of all $(z_1,z_2)$ in $\mathbb{C}^2$ such that $|z_1|^2+|z_2|^2=1$, and $S^2$ is identified with the subset of all $(z,x_3)$ in $\mathbb{C}\times\mathbb{R}$ such that $|z|^2+x_3^2=1$.

Because the Hopf map is geometrically unique but *coordinate-dependent*, different authors choose different identifications between $\mathbb{R}^4$ and $\mathbb{C}^2$, different sign conventions, different quaternion bases, and different stereographic projections. All these choices produce different formulas for $z_1(n_1,n_2,n_3,n_4)$ and, $z_2(n_1,n_2,n_3,n_4)$ but they all describe the *same* Hopf fibration. All are valid. They differ by coordinate choices, rotations, and reflections. Even after defining $z_1$and $z_2$, authors may choose different stereographic projections: projection from the north pole, projection from the south pole, projection onto different coordinate axes. Each choice produces a different explicit formula for the Hopf map.

Take a point $(z_1,z_2)\in S^3 \subset \mathbb{C}^2$. That is a normalized quantum spinor

$$\psi = \begin{pmatrix} z_1 \\ z_z \end{pmatrix}, \qquad |z_1|^2 + |z_2|^2 = 1. \tag{295}$$

Spinors transform under $SU(2)$. The space of normalized spinors is the 3-sphere $S^3$.
In quantum mechanics, every spinor defines a point on the Bloch sphere:

$$\vec{r} = \psi^\dagger \vec{\sigma} \psi . \tag{296}$$

where $\sigma_i$ are the Pauli matrices. That is a map $S^3 \to S^2$. Calculating the components of Eq. (296), we obtain:

$$x = z_1^* z_2 + z_2^* z_1 = 2\,\mathrm{Re}(z_1 z_2^*),\ \ y = -iz_1^* z_2 + iz_2^* z_1 = 2\,\mathrm{Im}(z_1 z_2^*),\ \ z = |z_1|^2 - |z_2|^2 . \tag{297}$$

Writing $h(z_1, z_2) = (x + iy, z)$, we see that the Hopf map $h$ (also called Hopf fibration) can be defined by

$$h(z_1, z_2) = (2z_1 z_2^*, |z_1|^2 - |z_2|^2). \tag{298}$$

The Hopf map sends a point $(z_1, z_2) \in S^3$ to a point $(x, y, z) \in S^2$. In Appendix F, I will show how to obtain the Hopf map using quaternions.

If two points on $S^3$ map to the same point on $S^2$, i. e. $h(z_1, z_2) = h(w_1, w_2)$, then $(w_1, w_2)$ must be equal to $(\lambda z_1, \lambda z_2)$ for some complex number $\lambda$ with $|\lambda|^2 = 1$. The converse is also true; any two points on the sphere $S^3$ that differ by a common complex factor $\lambda$ map to the same point on $S^2$. Since the set of complex numbers $\lambda$ with $|\lambda|^2 = 1$ form the unit circle in the complex plane, it follows that for each point $p$ in $S^2$, the inverse image $h^{-1}(p)$ in $S^3$ is a unit circle, i.e. $h^{-1}(p) \simeq S^1$. That is, every single point on $S^2$, in the canonical Hopf map, corresponds to an entire circle on $S^3$. Every pair of distinct circles (fibers) on $S^3$ is linked exactly once. In other words, any two distinct fibers form a Hopf link with linking number 1 (see Fig. 5). All the circles are linked with all the others. The fibers over any circle on $S^2$ form a torus. So, the whole structure is a nested torus, each one threaded through the others.

The forward stereographic projection $S^3 \setminus \{N\} \to \mathbb{R}^3$ from the north pole $N = (0,0,0,1)$ ,is given by

$$\Phi(n_1, n_2, n_3, n_4) = \frac{(n_1, n_2, n_3)}{1 - n_4} \qquad \in \mathbb{R}^3 . \tag{299a}$$

Stereographic projection is conformal (angle-preserving) but not length-preserving**.** Consequently, the images of the fibers in $\mathbb{R}^3$ have different Euclidean radii. Fibers near the projection point become very large. A fiber passing through the projection point becomes a straight line. Fibers farther away become smaller circles.

The inverse stereographic projection $\Phi^{-1}(x, y, z)$ from $\mathbb{R}^3$ to $S^3$ is then given by

$$n_1 = \frac{2x}{r^2 + 1}, \qquad n_2 = \frac{2y}{r^2 + 1}, \quad n_3 = \frac{2z}{r^2 + 1}, \quad n_4 = \frac{r^2 - 1}{r^2 + 1} . \tag{299b}$$

With $r^2 = x^2 + y^2 + z^2$. We can verify that $n_1^2 + n_2^2 + n_3^2 + n_4^2 = 1$. Eq. (299b) sends $(x, y, z) \in \mathbb{R}^3$ to $(n_1, n_2, n_3, n_4) \in S^3$. We can deduce this by intersecting the line from the north pole with the unit 3-sphere and solving for the parameters. Then, we apply the Hopf map to $(n_1, n_2, n_3, n_4)$ to get a point on $S^2$.

The sphere $S^2$ can be stereographically projected onto the complex plane using the standard formula

$$W = \frac{x + iy}{1 - z}. \tag{300}$$

Conversely

$$x + iy = \frac{2W}{1 + |W|^2} \qquad z = \frac{1 - |W|^2}{1 + |W|^2}. \tag{301a}$$

Writing $W = |W| e^{i\theta}$, it is easy to show that

$$x + iy = \sqrt{1 - z^2} e^{i\theta} \tag{301b}$$

Using the Hopf map (297) in Eq. (300), we get

$$W = \frac{2 z_1 z_2}{1 - (|z_1|^2 - |z_2|^2)} \tag{302}$$

We have $x + iy = 2 z_1 z_2^*$, and using $|z_1|^2 + |z_2|^2 = 1$, we obtain $1 - z = 1 - (|z_1|^2 - |z_2|^2) = 1 - |z_1|^2 + |z_2|^2 = 2|z_2|^2$, then we arrive at

$$W = \frac{2 z_1 z_2^*}{2|z_2^2|} = \frac{z_1 z_2^*}{z_2 z_2^*} = \frac{z_1}{z_2}. \tag{303}$$

Eq. (303) is the projection of the Hopf map's output onto the complex plane.

To apply the above results to a magnetic system, I will take $m_x = x,\ m_y = y$ and $m_z = z$, and let the letters *x*, y, and *z* be used for the coordinates of a point in $\mathbb{R}^3$. The physical field is a unit vector $\vec{m}(\vec{r}) \in S^2$.

Now, I will calculate an explicit profile for the magnetization in a hopfion following the diagram.

$$\mathbb{R}^3 \xrightarrow{\Phi^{-1}} S^3 \xrightarrow{h} S^2, \quad \text{or} \qquad (x, y, z) \xrightarrow{\Phi^{-1}} (z_1, z_2) \xrightarrow{h} \vec{m}(\vec{r}).$$

We start with a point (*x*, *y*, *z*) in $\mathbb{R}^3$, and using Eq. (299b) we get

$$z_1 = n_1 + i n_2 = \frac{2x + i2y}{r^2 + 1} . \quad z_2 = n_3 + i n_4 = \frac{2z + i(r^2 - 1)}{r^2 + 1} . \tag{304}$$

Eq.(297) gives

$$m_x = 2 \operatorname{Re} z_1 z_2^* = \frac{4[2xz + y(r^2 - 1)]}{(1 + r^2)^2} \qquad m_y = 2 \operatorname{Im} z_1 z_2^* = \frac{4[2yz - x(r^2 - 1)]}{(1 + r^2)^2} , \tag{305}$$

$$m_z = |z_1|^2 - |z_2|^2 = \frac{4(x^2 + y^2 - z^2) - (r^2 - 1)^2}{(1 + r^2)^2}. \tag{306}$$

We would get the same result using $W = z_1 / z_2$, and Eq.(301a). Some authors use $z_1 = n_3 + in_4$, $z_2 = n_1 - in_2$, and obtain

$$m_x = \frac{4[2xz - y(r^2 - 1)]}{(1+r^2)^2}, \qquad m_y = \frac{4[2yz + x(r^2 - 1)]}{(1+r^2)^2}, \tag{307}$$

$$m_z = 1 - \frac{8(x^2 + y^2)}{(1+r^2)^2}. \tag{308}$$

It is easy to see that $\vec{m} \to (0,0,1)$ as $r^2 \to \infty$, as required. In cylindrical coordinates, Eq. (308) can be rewritten as

$$m_x + im_y = \frac{4\rho e^{i\phi}[2z + i(\rho^2 + z^2 - 1)]}{(1+\rho^2 + z^2)^2}, \qquad m_z = 1 - \frac{8\rho^2}{(1+\rho^2+z^2)^2}. \tag{309a}$$

If we introduce the hopfion radius $a \neq 1$, and a constant angle $\phi_0$,which describes a freedom of choosing the directions of the *Ox* and *Oy* coordinate axes in the *xOy* plane, Eq. (309a) can be rewritten as

$$m_x + im_y = \frac{4a\rho e^{i(\phi+\phi_0)}[2za + i(\rho^2 + z^2 - a^2)]}{(a^2+\rho^2 + z^2)^2}, \quad m_z = 1 - \frac{8\rho^2 a^2}{(a^2+\rho^2+z^2)^2}. \tag{309b}$$

For $\phi_0 = 0, \pi$, we have a Néel hopfion, and for $\phi_0 = \pm\pi/2$, a Bloch one. For topological charge greater than one, i. e., for arbitrary vorticities *m* and *n*, $z_1^n$ and $z_2^m$ no longer lie on the unit 3-sphere. The following normalization puts it back on $S^3$:

We define a map $F_{n,m} : S^3 \to S^3$ by

$$F_{n,m}(z_1, z_2) = \frac{z_1^n}{\sqrt{|z_1|^{2n} + |z_2|^{2m}}} + \frac{z_2^m}{\sqrt{|z_1|^{2n} + |z_2|^{2m}}}. \tag{310a}$$

We can compose $\phi_{n,m} = h(z_1, z_2).F_{n,m} : S^3 \to S^2$. We write

$$\tilde{z}_1 = \frac{z_1^n}{\sqrt{|z_1|^{2n} + |z_2|^{2m}}}, \qquad \tilde{z}_2 = \frac{z_2^m}{\sqrt{|z_1|^{2n} + |z_2|^{2m}}}. \tag{310b}$$

Using Eq. (304), we obtain

$$z_1 = \frac{2(x+iy)}{1+r^2}, \qquad z_2 = \frac{r^2 - 1 + i2z}{1+r^2}. \tag{310c}$$

This can be written as

$$z_1 = \rho_1 e^{i\alpha}, \qquad z_2 = \rho_2 e^{i\beta}, \tag{311}$$

where

$$\rho_1 = \frac{2\sqrt{x^2+y^2}}{1+r^2}, \qquad \rho_2 = \frac{\sqrt{(1-r^2)^2 + 4z^2}}{1+r^2}. \tag{312}$$

$$\alpha = \arg(x+iy), \quad \beta = \arg(1 - r^2 + i2z). \tag{313}$$

We define

$$u_1 = \frac{\rho_1^n e^{in\alpha}}{N}, \quad u_2 = \frac{\rho_2^m e^{im\beta}}{N}, \text{ with } N^2 = \rho_1^{2n} + \rho_2^{2m}. \tag{314}$$

The Hopf map (297) with $z_1 \to u_1$, $z_2 \to u_2$ gives

$$m_x = \frac{2\rho_1^n \rho_2^m}{N^2}\cos(n\alpha + m\beta), \qquad m_y = \frac{2\rho_1^n \rho_2^m}{N^2}\sin(n\alpha + m\beta) \tag{315}$$

$$m_z = \frac{\rho_1^{2n} - \rho_2^{2m}}{N^2}, \qquad m_x + im_y = \sqrt{1-m_z^2}\, e^{i(n\alpha+m\beta)}. \tag{316}$$

The Hopf number is given by $H = mn$. For $H > 1$, the hopfion magnetization is not azimuthally symmetric.

We can also show for arbitrary vorticities $m$ and $n$, using the theory of quaternions (presented in Appendix F), that Eq. (303) becomes [49]:

$$W = \frac{z_1^n}{z_2^m}. \tag{317}$$

As an example, I will consider $z_1$ and $z_2$ written in toroidal coordinates. From Eqs. (293) and (299b), we have [50]

$$r^2 = x^2 + y^2 + z^2 = \frac{\sinh^2\eta + \sin^2\beta}{\Delta}, \text{ where } \Delta = \cosh\eta - \cos\beta, \tag{318}$$

$$1 + r^2 = \frac{2\cosh\eta}{\Delta}, \qquad r^2 - 1 = \frac{2\cos\beta}{\Delta}, \tag{319}$$

$$n_1 = \frac{2x}{1+r^2} = \tanh\eta\cos\varphi, \quad n_2 = \frac{2y}{1+r^2} = \tanh\eta\sin\varphi, \tag{320}$$

$$n_3 = \frac{2z}{1+r^2} = \frac{\sin\beta}{\cosh\eta}, \qquad n_4 = \frac{r^2-1}{1+r^2} = \frac{\cos\beta}{\cosh\eta}. \tag{321}$$

Using now $z_1 = n_1 + in_2$, $z_2 = n_4 + in_3$, we obtain

$$z_1 = \tanh\eta\, e^{i\varphi}, \qquad z_2 = \frac{e^{i\beta}}{\cosh\eta}. \tag{322}$$

From Eq. (310), we get

$$W = \cosh^m(\eta)\tanh^n(\eta)e^{-i(m\beta+n\varphi)}. \tag{323}$$

Employing Eqs.(301a) and (301b), the following expression is obtained

$$m_z = \frac{1-\cosh^{2m}(\eta)\tanh^{2n}(\eta)}{1+\cosh^{2m}(\eta)\tanh^{2n}(\eta)}, \tag{324}$$

and

$$m_x + im_y = \sqrt{1-m_z^2}\, e^{i(m\beta+n\varphi)}. \tag{325}$$

The integers $m$ and $n$ are the hopfion poloidal and azimuthal vorticities, respectively.

To present a physical idea of a hopfion, without using topology, I start with Eq. (308). Suppose we take the magnetization at $\vec{m} = (0,0,-1)$, it is easy to see that this leads to $z$ = 0, $x^2 + y^2 = 1$, which is a circle in the $xy$-plane. Along this circle, the magnetization points downward along the $z$-direction. In the next step, I consider $\vec{m} = (1,0,0)$. With a little algebra, we find $2x^2 + (y-1)^2 = 2$, $x = z$. That is a tilted circle in the $x = z$ plane. Along this circle, the magnetization points in the $x$ -direction. I show these 2 circles on Fig. 14. Solving Eq. (309a) for other magnetization values would yield many more circles. If the set of magnetizations lies along a closed loop on the magnetization sphere, the corresponding $(x, y, z)$ forms a torus. (The point $\vec{m} = (0,0,1)$ is projected to the entire $z$ -axis). Note that I discuss this from a mathematical point of view. In numerical calculations, things are different.

As shown above, if a unitary field in three dimensions has a constant value at infinity, it gives rise to a topological structure called a hopfion. No physical system or Hamiltonian was mentioned. The Hopfion obtained from the Hopf map *is* an exact topological configuration. If we are interested in a magnetic system, it is called an ansatz because it is not an exact solution for the *micromagnetic energy functional* (an analytical expression for the magnetic Hopfion profile is unknown). It is mathematically exact but not physically exact. The configuration is a true solution only if it satisfies the Euler–Lagrange equations derived from the energy functional. The Hopf map does not solve these equations for any realistic material parameters. It is neither an energy minimizer nor a stationary point of the micromagnetic energy. We use the Hopfion from the Hopf map as an ansatz because it has the correct topology (Hopf charge), is smooth and analytic, is easy to parametrize, and provides a good *starting point* for numerical relaxation. But after relaxation, the final micromagnetic hopfion is no longer exactly the Hopf map**.** The material parameters deform it.

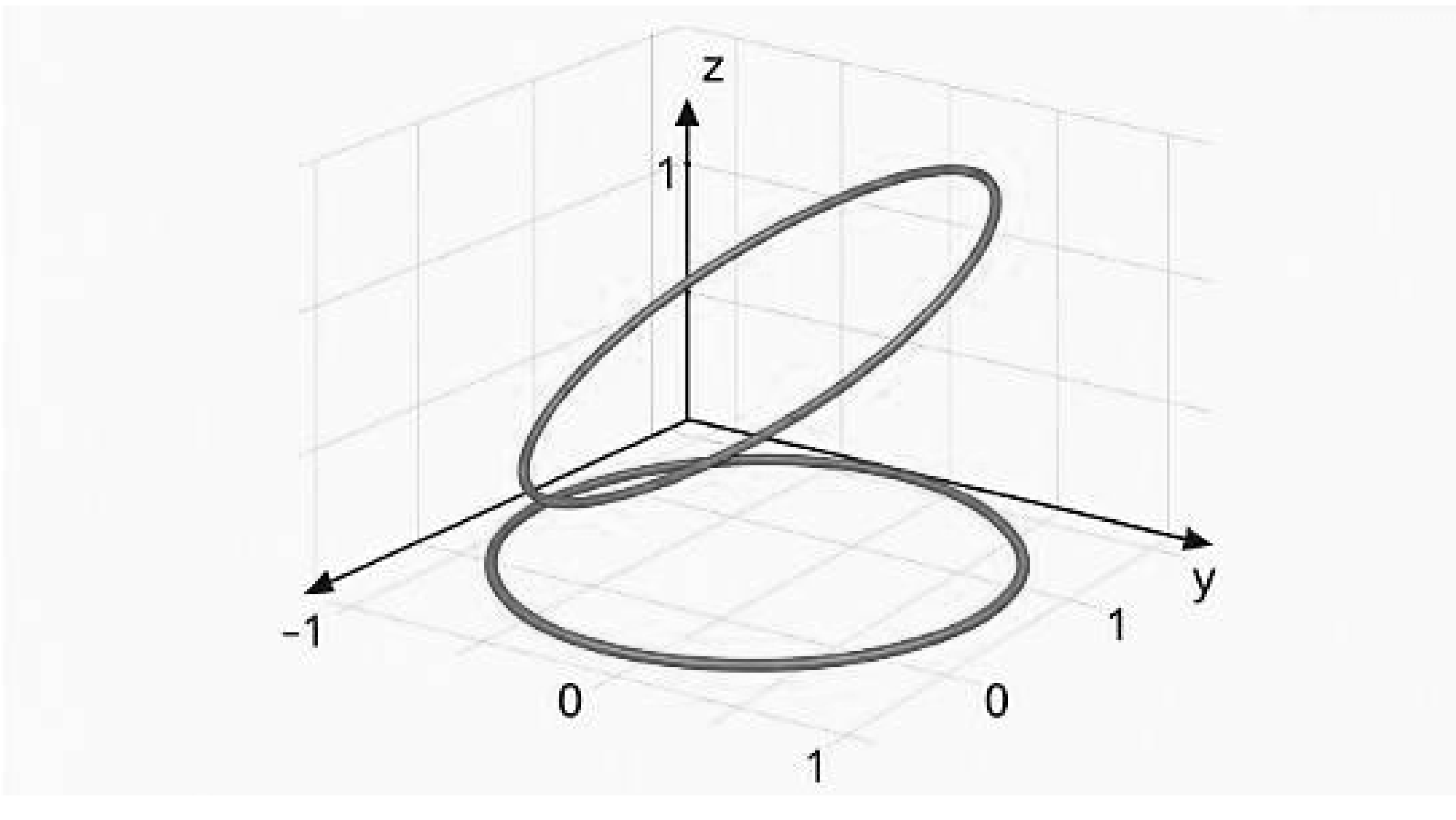

Figure 14. The preimages of (0,0,-1) and (1,0,0) in $\mathbb{R}^3$

Imagine the magnetization field of a magnetic hopfion. At the center, the spins point downward. Far away, they point upward. Intermediate spin directions occur on nested toroidal shells. Each toroidal shell is a Hopf torus corresponding to a constant magnetization value (more precisely, a preimage of a latitude on the target sphere). The individual magnetization fibers on the tori are closed loops, and every pair of fibers is linked according to the Hopf charge. The toroidal isosurfaces commonly plotted in simulations (such $m_z =$ constant) are examples of Hopf tori.

To study the full dynamics of a magnetic hopfion, we use the Landau–Lifshitz–Gilbert equation. To perform a numerical simulation, we start with a hopfion profile (ansatz), then we normalize the magnetization such that $|\vec{m}| = 1$. That gives us $\vec{m}(\vec{r},t) = \vec{m}_{hop}(\vec{r})$. That is, the hopfion profile defines the magnetization at $t$ = 0. Then, we use the LLG equation

$$\frac{d\vec{m}}{dt} = -\gamma \vec{m} \times \vec{H}_{eff} + \alpha \vec{m} \times \frac{d\vec{m}}{dt}. \tag{326}$$

Given $\vec{m}(\vec{r},t) = \vec{m}_{hop}(\vec{r})$, the algorithm (numerically) does:

- Compute $\vec{H}_{eff}(\vec{r},t)$ from the energy terms (exchange, DMI, anisotropy, etc.).
- Evaluate $d\vec{m}/dt$ at each point.
- Update $\vec{m}(t, t+\Delta t)$.

So, the Hopfion is just the starting shape. After that, the LLG equation tells us whether it remains stable (equilibrium or metastable), precesses, deforms, collapses, or responds to fields/currents. We can include spin-orbit torque in the same way as in skyrmions [51,52]. Numerical solutions of the LLG equation show that SOT can split a high-H hopfion into multiple lower-H hopfions, enabling control over their topological properties [51]. Numerical solutions of the LLG equation are well described in the literature. Some researchers skip analytic formulas and instead start with a random field, impose a constraint on the linking number or helicity, and relax the system under the LLG equation until a magnetic hopfion emerges. That is not an analytic ansatz but a numerical one. It is useful when studying spontaneous hopfion formation and allows exploration of non-standard hopfion shapes.

Under the rigid-body approximation, hoppfion dynamics can be reduced to a generalized Thiele equation. However, unlike skyrmions, there is no universally accepted single-particle Thiele equation for hopfions. A generalized Thiele equation for hopfions, can have the same structural form as for other magnetic solitons (skyrmions, vortices, bubbles), but with parameters, gyrovector, dissipation tensor, and force terms computed from the *three-dimensional* hopfion texture [53]. For instance, the gyrovector is given by

$$G_k = \frac{M_S}{\gamma} \int \varepsilon_{kij} \vec{m}.(\partial_i \vec{m} \times \partial_j \vec{m}). \tag{327}$$

Deriving the Thiele equation from the LLG equation using a collective coordinate is more delicate for hopfions because they can deform in 3D. The Thiele equation captures *rigid-body* motion of a soliton's center of mass and breaks down when internal deformation modes become important.

Hopfion dynamics exhibit a richness that far exceeds the behavior of skyrmions, owing to their fully three-dimensional solitonic nature and the intricate internal structure imposed by the Hopf fibration. As three-dimensional topological objects, hopfions evolve under the Landau–Lifshitz–Gilbert (LLG) equation while strictly preserving their Hopf charge. This linking number cannot change under smooth magnetization dynamics. This constraint shapes hopfion motion, deformation, and interaction: although hopfions can translate, twist, expand, or tilt, their internal linked preimage loops remain intact throughout the evolution.

Several external stimuli can drive hopfions through magnetic materials. Magnetic field gradients exert forces that induce translational drift, with velocities determined by the hopfion's toroidal geometry and internal twist. Spin-transfer and spin–orbit torques provide another efficient driving mechanism, particularly in multilayer systems with strong spin–orbit coupling. Under current injection, hopfions follow complex three-dimensional trajectories, including vertical motion, torus expansion or contraction, and tilting of the core relative to the applied torque direction.

Their dynamics also feature characteristic deformation modes. The breathing mode corresponds to periodic oscillations of the torus radius, during which the preimage loops stretch and compress while maintaining their linking. The twisting mode involves rotation of the internal phase of the Hopf map, producing fiber sliding along the torus and dynamic changes in the Villarceau angle; this mode is unique to hopfions because of their double-twist geometry. A third common response is tilting of the toroidal core, often triggered by spin–orbit torques or anisotropy variations.

Hopfion–hopfion interactions arise from the overlap of their magnetization fields and can be repulsive or attractive depending on configuration. Hopfions with identical charges typically repel at short distances due to exchange and dipolar interactions. Attraction becomes possible when their toroidal cores align, their internal phases match, or dipolar flux closure favors coalescence. Under sufficiently strong fields or currents, hopfions can merge or split while conserving total Hopf charge: two $H=1$ hopfions may combine into an $H=2$ structure, and an $H=2$ hopfion may divide into two $H=1$objects. Higher-charge hopfions display even more intricate behavior, including braiding of internal preimage loops, sliding, twisting, and reconfiguration into torus-knot-like structures—phenomena with no analog in two-dimensional skyrmion systems.

External perturbations elicit additional dynamical responses. Thermal fluctuations can destabilize small hopfions, causing shrinkage and collapse, whereas larger or higher-charge hopfions are more robust due to increased linking. Magnetic field pulses can nucleate hopfions, excite breathing modes, or annihilate them if the pulse amplitude exceeds stability thresholds. Geometric confinement in nanowires, nano disks, or toroidal nanostructures plays a crucial role in stabilizing hopfions by suppressing collapse pathways and energetically favoring linked configurations.

Simulations and experiments have identified several nucleation mechanisms. Localized magnetic field pulses can twist the magnetization sufficiently to generate hopfions, which subsequently relax into stable states. Spin–orbit torques can nucleate hopfions at current hotspots in multilayers. In confined geometries, boundary twisting can propagate inward, forming hopfions. Additionally, in three-dimensional systems, collapsing skyrmion tubes may transform into hopfions when the collapse induces the necessary double twist.

Overall, hopfions behave as three-dimensional topological particles characterized by position, internal phase, size (major and minor radii), and Hopf charge. Their dynamics combine translational motion, internal rotation, deformation, and strict preservation of linking, making them compelling candidates for future three-dimensional spintronic technologies. Information could be encoded not only in the hopfion position or charge but also in its internal phase and geometric degrees of freedom, opening pathways toward devices that exploit the full richness of three-dimensional topological magnetism.

Because of their knot-like topology, magnetic hopfions are highly stable against thermal fluctuations and material defects. Researchers are studying them for next-generation computing: Hopfions can be driven through 3D wire networks using electrical currents, acting as information bits in 3D racetrack memories. Their complex, deformable shapes enable them to mimic biological synapses and neurons in response to applied electrical or magnetic pulses. Theoretical studies predict that magnetic Hopfions can emerge in chiral magnetic materials through the interplay of exchange interactions, the Dzyaloshinskii–Moriya interaction, magnetic anisotropy, dipolar interactions, and geometrical confinement. Depending on the material system and external conditions, Hopfions may be stabilized either as equilibrium states or as metastable configurations. Researchers have investigated their existence in bulk chiral magnets, confined nanostructures, multilayer heterostructures, and frustrated magnetic systems, where competition among magnetic interactions favors nontrivial three-dimensional textures.

**Fractional hopfions, heliknotons, and winding hopfions** are all *three-dimensional topological solitons,* stable, knotted field configurations, related to the broader family of hopfions [54,55]. The differences lie in how the field twists, links, and embeds within its background (Fig.15).

The Hopf charge is strictly an integer for ideal maps. But real materials are not perfect mathematical manifolds; the field is not a perfect map $S^3 \to S^2$. Several things break the integer constraint: In chiral magnets, the ground state is a helical phase rather than a constant vector. That means that the field at infinity is not fixed; the domain cannot be compactified to $S^3$. Consequently, the Hopf number is not quantized. Fractional hopfions are hopfion-like textures whose Hopf index is *not an integer* because only part of the field participates in the linking. In standard Hopfions, all preimages of the magnetization map link give an integer Hopf number. Because the field no longer defines a map $S^3 \to S^2$, the usual integer valued Hopf classification breaks down. In fractional hopfions, *not all preimages link*, so the Hopf index becomes non-integer. They arise in systems with mixed topology, where the background magnetization maps only to a subset (e.g., an $S^1$ line) of the full target sphere $S^2$.That partitions the topology into multiple sectors and allows continuously tunable or fractional values of the Hopf index. Fractional Hopfions are often metastable and appear as distortions of skyrmion tubes.

A **heliknoton** is a hopfion embedded in a *helical magnetic background*—a knotted soliton living inside a spiral phase and predicted in chiral magnets, specifically within their spiral (helical) magnetic phase. The helical background modifies the hopfion's structure, producing a variant where the knotting coexists with the underlying helical twist. The term "heliknoton" reflects this hybrid nature: *helix* + *knot*. They are essentially knotted skyrmion tubes embedded in a helix.

**Winding hopfions** refer to hopfions whose topology includes additional integers with toroidal and poloidal windings.. In some physical systems, the Hopf number is not the only topological invariant. Hopfions may carry extra integers describing how the field winds around toroidal or poloidal directions, or how domain walls wrap. These appear in models where the Hopf invariant is expressed as a product or combination of winding numbers (e.g., (P, Q) for toroidal domain walls or other multi-component textures). The key idea is that the hopfion's structure includes multiple independent windings, giving it a richer topology. They arise in bulk chiral magnets under confinement, as well as in nanodisks and nanorings, systems with strong dipolar interactions, and skyrmion strings that close into toroidal loops.

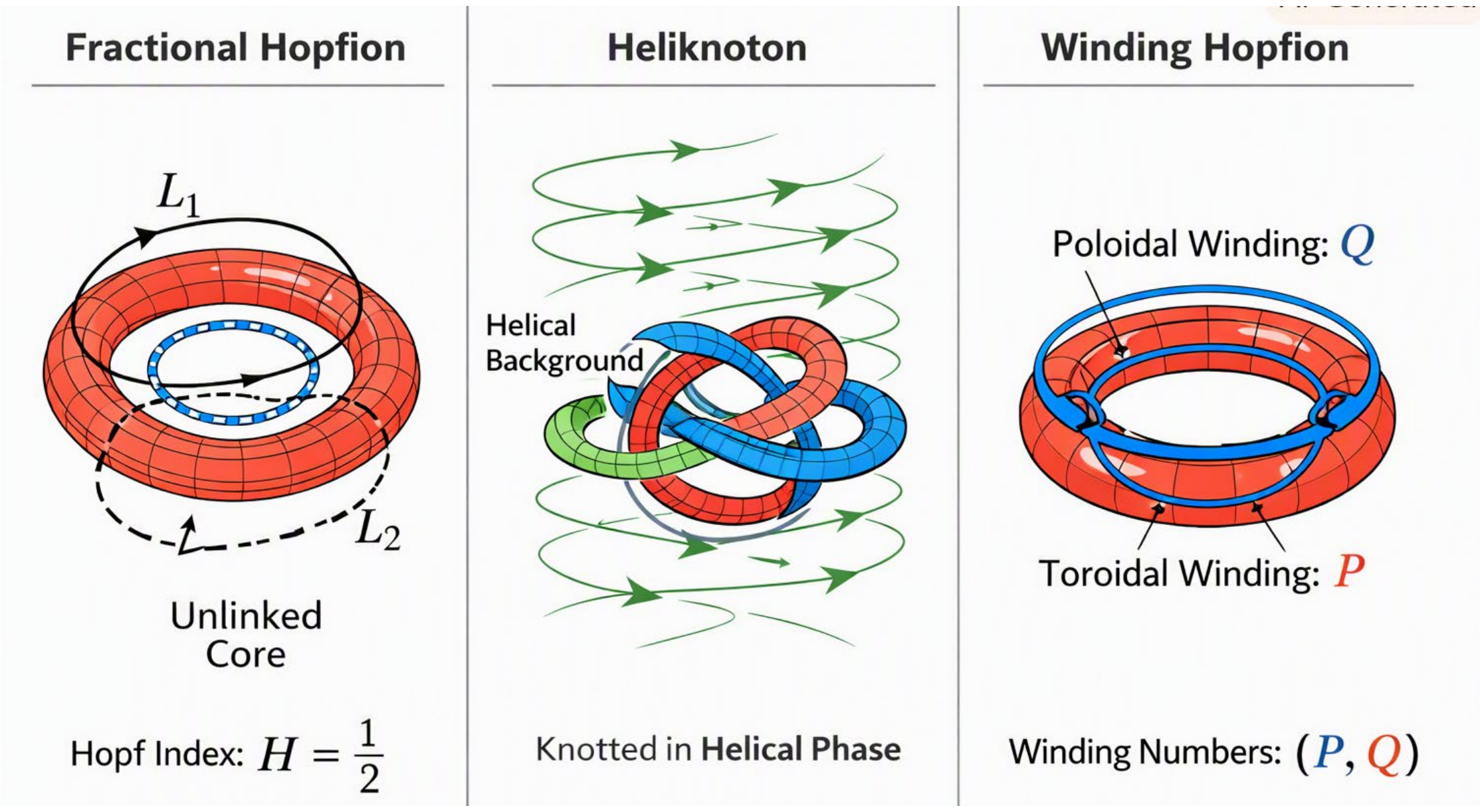

Figure 15. Different kinds of hopfions.

## 15. Quantum Skyrmions.

Quantum skyrmions have emerged as a promising new paradigm at the intersection of topology, magnetism, and quantum information science. While classical skyrmions are understood as nanoscale spin textures stabilized by chiral interactions, driving these objects into the quantum regime reveals qualitatively new physical phenomena When skyrmions shrink to atomic dimensions, when the host magnet exhibits strong quantum fluctuations, or when their collective coordinates acquire discrete spectra, the skyrmion can no longer be regarded as a classical field configuration and instead becomes a quantized topological quasiparticle. In this regime, fundamental degrees of freedom such as the skyrmion's center-of-mass position, radius, and helicity become quantized. Quantum tunneling processes, including transitions between helicity states, macroscopic tunneling of the skyrmion's position, and even tunneling between distinct topological sectors in frustrated magnets, become experimentally relevant and may be observable through low-temperature spectroscopic techniques [56,57].

The quantum nature of skyrmions also enriches their interactions with excitations in the host material. Coupling to magnons leads to bound magnon modes, quantized excitation spectra, and hybrid skyrmion–magnon quasiparticles, providing a natural platform for quantum magnonics. In itinerant magnets, conduction electrons experience emergent electromagnetic fields generated by the skyrmion texture, giving rise to quantized electronic bound states, skyrmion-induced topological Hall responses, and potential pathways toward engineered topological superconductivity. These phenomena can be systematically described trough collective coordinate quantization, Berry-phase effects, and topological quantum field theories, which together capture the interplay between topology and quantum dynamics.

As experimental techniques advance and new material platforms emerge that can stabilize atomic-scale or strongly fluctuating skyrmions, quantum skyrmions are poised to become central building blocks in next-generation quantum technologies. Their topological protection, internal quantum structure, and ability to hybridize with diverse quasiparticles position them as transformative elements in the broader landscape of condensed matter physics and quantum information processing.

In this section, I will develop the semiclassical description based on the spin-coherent states and the path-integral formalism presented in Appendix C. In this approach, the internal and translational degrees of freedom of a skyrmion become quantum variables. Because the underlying spins generate a Berry phase, skyrmion dynamics resembles the motion of a charged particle in an effective magnetic field.

As we saw in section 4, the topological charge of a skyrmion is given by

$$Q = \frac{1}{4\pi}\int \rho d^2x \,. \tag{328}$$

where the topological charge density is

$$\rho = \frac{1}{2}\varepsilon_{ij}\vec{n}.(\partial_i\vec{n}\times\partial_j\vec{n}) = \vec{n}.(\partial_1\vec{n}\times\partial_2\vec{n}) \,. \tag{329}$$

Let us define [56]

$$R_i = \frac{1}{4\pi Q}\int d^2x x_i \rho \,. \tag{330}$$

Under spatial translation $x_i \to x_i + a_i$, and so $R_i \to R_i + a_i$. Therefore, $R_i$ can be interpreted as the skyrmion's position. The Poisson bracket of spin coherent states has the following expression

$$\{n_i(\vec{x}), n_j(\vec{y})\} = -\varepsilon_{ijk}\frac{n_k(\vec{x})}{S}\delta(\vec{x}-\vec{y}) \,. \tag{331}$$

Following Ref. [58], we define the dipole moment of the topological charge distribution as

$$I_i = \int d^2x x_i \rho(\vec{x}) \quad \to \quad R_i = \frac{I_i}{4\pi Q} \,. \tag{332}$$

Using Eq. (329) we obtain

$$\{I_i, n_a\} = \frac{1}{S}\varepsilon_{ij}\partial_j n_a \;\to\; \{I_1, n_a\} = \frac{1}{S}\partial_2 n_a \,, \qquad \{I_2, n_a\} = -\frac{1}{S}\partial_1 n_a \,. \tag{333}$$

We have

$$\{I_1, I_2\} = \int d^2x x_2 \{I_1, \rho(\vec{x})\}. \quad (334)$$

Using $\{I_1, \rho\} = \frac{1}{S}\partial_2\rho$, we get

$$\{I_1, I_2\} = \frac{1}{S}\int d^2x x_2 \partial_2 \rho(\vec{x}). \quad (335)$$

Integrating by parts

$$\int d^2x x_2 \partial_2 \rho = \int d^2x \partial_2 (x_2 \rho) - \int d^2x \rho. \quad (336)$$

For a localized skyrmion, the boundary term vanishes. Thus

$$\{I_1, I_2\} = -\int d^2x \rho = -\frac{1}{S}4\pi Q. \quad (337)$$

$Q$ is a topological invariant, so $\{Q,.\} = 0$. Therefore

$$\{R_1, R_2\} = \frac{1}{(4\pi Q)^2}\{I_1, I_2\} = \frac{1}{4\pi QS}. \quad (338)$$

We can write

$$\{R_i, R_j\} = \frac{\varepsilon_{ij}}{4\pi QS} = -\frac{\varepsilon_{ji}}{4\pi QS}. \quad (339)$$

Comparing Eq. (339) with the canonical Poisson bracket for position and linear momentum $\{R_i, P_j\} = \delta_{ij}$, we obtain

$$P_i = 4\pi SQ\varepsilon_{ji}R_j. \quad (340)$$

As can be verified by the following calculation

$$\{R_i, P_j\} = 4\pi Q\varepsilon_{jk}\{R_i, R_k\} = 4\pi SQ\varepsilon_{jk}\left(-\frac{\varepsilon_{ik}}{4S\pi Q}\right) = \delta_{ij}. \quad (341)$$

Where I have used $\varepsilon_{jk}\varepsilon_{ik} = \delta_{ij}$. We can verify that $\{P_i, n_j\} = \partial_i n_j$. For any function of coordinates only, the following relation holds

$$\{P_i, f(\vec{R})\} = \sum_j \left(\frac{\partial P_i}{\partial R_j}\frac{\partial f}{\partial P_j} - \frac{\partial P_i}{\partial P_j}\frac{\partial f}{\partial R_j}\right). \quad (342)$$

If $V(\vec{R})$ is a potential, the equation of motion $\dot{P}_i = \{P_i, V\}$, using Eq. (342), leads to

$$\dot{P}_i = -\frac{\partial V}{\partial R_i} = F_i. \quad (343)$$

Using $\hat{z} \times \dot{\vec{R}} = -\dot{R}_y\hat{i} + R_x\hat{j}$, defining the gyrocoupling vector as $\vec{G} = -4\pi SQ\hat{z}$, we get combining Eqs. (340) and (343)

$$\dot{\vec{P}} = \vec{G} \times \dot{\vec{R}} = \vec{F}. \quad (344)$$

Eq. (344) has the form of the Thiele equation.

To obtain the effective Lagrangian, I will start with the action for the Berry term, Eq. (C.35), written as

$$S_B = \int\int \vec{n}.(\partial_\tau \vec{n} \times \partial_t \vec{n}) d\tau dt, \quad (345)$$

where we require the motion of $\vec{n}$ to be periodic (as is the case in the calculation of the partition function). Then $S_B$ is gauge invariant. The parameters $t$ and $0 \le \tau \le 1$ parametrize the region $\Gamma$ in the sphere $S^2$ bounded by the curve $\vec{n}(t)$, on which $\tau = 1$ [59]. The area of $\Gamma$ changes due to the variation in its boundary. Thus, we have

$$\delta S_B = \oint \vec{n}.(\delta\vec{n} \times \partial_t \vec{n})dt \, . \tag{346}$$

For a moving skyrmion, we use the ansatz

$$\vec{n}(\vec{r},t) = \vec{n}_0(\vec{r} - \vec{R}(t)) \, , \tag{347}$$

where $\vec{n}_0$ is a static skyrmion profile centered at the origin. We obtain

$$\frac{d\vec{n}}{dt} = \frac{\partial}{\partial R_i}\vec{n}_0(\vec{r} - \vec{R})\frac{\partial R_i}{\partial t} \, , \tag{348}$$

$$\delta\vec{n} = \frac{\partial}{\partial R_i} n_0(\vec{r} - \vec{R})\delta R_i \, . \tag{349}$$

Inserting Eqs. (348) and (349) into Eq. (345) for the continuum model, we find

$$\delta S_B = \int\int n_0(\vec{r} - \vec{R}).[\partial_i(\vec{r} - \vec{R}) \times \partial_j(\vec{r} - \vec{R})]\delta R_i \dot{R}_j dt d^2x \, . \tag{350}$$

Using Eq. (3.28), Eq. (3.50) can be rewritten as

$$\delta S_B = 4\pi Q \varepsilon_{ij} \oint \delta R_i \dot{R}_j dt \, . \tag{351}$$

This is the variation of the effective Lagrangian

$$L_{eff}(\vec{R}, \dot{\vec{R}}) = 2\pi S Q \varepsilon_{ji} \dot{R}_i R_j \, . \tag{352}$$

For the study of the internal dynamics of a skyrmion, such as breathing modes, we need to go beyond the rigid profile and consider a skyrmion with translational motion and rotationally symmetric states [60]. We start with

$$\vec{n}(\vec{r}) = (\sin\theta(\vec{r})\cos\psi(\phi), \sin\theta(\vec{r})\sin\phi(\psi), \cos\theta(\vec{r})) \, . \tag{353}$$

where $\phi(\psi) = \psi + \eta$, for a skyrmion with vorticity $m = 1$, and helicity $\eta$. The radius $R$ of the skyrmion is defined by the circle $n_z = 0$. I will consider the regime $d << R$, where $\theta(\vec{r})$ is nearly constant. The breathing dynamics is described by $R(t)$, and $\eta(t)$. The Berry action is given by

$$S_B = S\int dt \int d^2r(1 - \cos\theta)\dot{\phi} \, , \tag{354}$$

which can be rewritten as

$$S_B = S\int dt\dot{\eta}\int_0^\infty r dr \int_0^{2\pi} d\psi[1 - \cos\theta(\vec{r}, R)] \, . \tag{355}$$

Avoiding choosing a particular skyrmion profile, for $d << R$, we can use the approximation

$$\theta \simeq \pi \ \ (r < R), \quad \theta \simeq 0 \ \ (r > R) \, .$$

We have $1 - \cos\theta \approx 2$ inside the skyrmion. Then

$$\int_0^R r(1 - \cos\theta)dr = CR^2 \, . \tag{356}$$

Where $C$ is a numerical constant. Thus

$$S_B = aS\int dt R^2(t)\dot{\eta} \quad \text{and} \quad S_{eff} = \int dt(aSR^2\dot{\eta} - H) \, . \tag{357}$$

With $a = 2\pi C$, and I suppose that the Hamiltonian is time independent. For a realistic, smooth skyrmion profile, the numerical coefficient C will be adjusted.

$$p_\eta = \frac{\partial L_B}{\partial \dot{\eta}} \quad \text{gives} \quad \{\eta, p_\eta\} = 1. \tag{358}$$

Hence

$$\{\eta, R^2\} = \partial_R R^2 \{\eta, R\} = 2R\{\eta, R\}, \tag{359}$$

which gives

$$\{\eta, R\} = \frac{1}{2aSR}. \tag{360}$$

A change in radius generates helicity dynamics, and a change in helicity generates radius dynamics.

$$\text{Let } \xi_k = (q_1, ..., q_N, p_{1,...,} p_N), \tag{361}$$

then $\{\xi_k, \xi_l\} = w_{kl}$, where

$$w_{kl} = \begin{pmatrix} 0 & I \\ -I & 0 \end{pmatrix}. \tag{362}$$

For any two functions $f(\xi)$ and $g(\xi)$, we have

$$\{f, g\} = \sum_{kl} w_{kl} \frac{\partial f}{\partial \xi_k} \frac{\partial g}{\partial \xi_l}. \tag{363}$$

For $f = \xi_i$, $g = H$, we get

$$\{\xi_i, H\} = \sum_{kl} w_{kl} \frac{\partial \xi_i}{\partial \xi_k} \frac{\partial H}{\partial \xi_l} = \sum_l w_{il} \frac{\partial H}{\partial \xi_l}, \tag{364}$$

Where I have used $\partial \xi_i / \partial \xi_k = \delta_{ik}$. Using $w_{il} = \{\xi_i, \xi_l\}$, Eq. (364) can be rewritten as

$$\{\xi_i, H\} = \sum_l \{\xi_i, \xi_l\} \frac{\partial H}{\partial \xi_l}. \tag{365}$$

From Eq. (365) we find

$$\dot{\eta} = \frac{1}{2aSR} \frac{\partial H}{\partial R}, \qquad \dot{R} = -\frac{1}{2aSR} \frac{\partial H}{\partial \eta}. \tag{366}$$

These equations describe the internal dynamics of a rotationally symmetric skyrmion.

Canonical quantization of the skyrmion can be performed by transforming the position and momentum variables to quantum operators $\hat{R}$ and $\hat{P}$ obeying the commutation rules [57]

$$[\hat{R}_i, \hat{P}_j] = i\hbar \delta_{ij}, \qquad [\hat{R}_i, \hat{R}_j] = \frac{i\hbar \varepsilon_{ji}}{4\pi SQ}. \tag{367}$$

The dynamics of the skyrmion can be described by the collective coordinates $\vec{R}(t)$ and the magnetization fluctuations that arise when the skyrmion is in motion. Using the path-integral formalism and writing

$$\Pi(\vec{r}, \tau) = \Pi_0(\vec{r} - \vec{R}) + \eta(\vec{r} - \vec{R}, \tau), \tag{368}$$

$$\Phi(\vec{r}, \tau) = \Phi_0(\vec{r} - \vec{R}) + \xi(\vec{r} - \vec{R}, \tau), \tag{369}$$

where $\Pi(\vec{r}, t) = \cos\theta$, Psaroudakis et al. [61], integrating out magnon fluctuations, have obtained the effective Lagrangian

$$\mathscr{L}_{eff} = 2\pi SQ\varepsilon_{ji} R_i^{(0)} \dot{R}_j^{(0)} + \frac{M}{2} \dot{R}_i^{(0)} \dot{R}_i^{(0)} + \ldots \tag{370}$$

The mass $M$ accounts for deformations of the moving skyrmion compared with the static solution.

However, Sorn et al. [62] have demonstrated that the quantum dynamics of magnetic skyrmions is governed by a dipole conservation law associated with the topological charge. Consequently, spin fluctuations that deform a rigid magnetic texture do not generate inertia. The skyrmion's vanishing mass results from the cancellation between the motion of the quasiclassical spin configuration and a cloud of quantum fluctuations.

A more radical form of quantum skyrmions occurs when the spins themselves cannot be regarded as classical arrows. The system must then be described by a many-body wave function [63]. Local expectation values may still show a skyrmion-like pattern, but quantum fluctuations and entanglement can reduce their magnitudes. In extreme cases, local magnetization may fail to reveal the state's topology entirely. For this reason, quantum skyrmions are increasingly characterized through spin correlation functions. In this context, an important quantity is the generalization of the classical topological charge to quantum spins on a lattice [56,64]

$$Q_{qua} = \frac{1}{2\pi} \sum_{ijk} \arctan \frac{(1/S^3) < \vec{S}_i.(\vec{S}_j \times \vec{S}_k) >}{1 + (1/S^2)\left( < \vec{S}_i.\vec{S}_j > + < \vec{S}_i.\vec{S}_k > + < \vec{S}_j.\vec{S}_k > \right)} . \tag{371}$$

Where the sum extends over a triangulation of the lattice. Entanglement measures and higher-order correlations can then distinguish a genuine quantum skyrmion from a simple product state arranged in a skyrmionics pattern. Numerical research has identified quantum skyrmion states in frustrated ferromagnets even when their local magnetization differs substantially from the classical texture [56]. Haller et al. [63] have constructed a quantum skyrmion creation operator that represents the object as a classical skyrmion core surrounded by a quantum cloud of correlated spin-flip excitations.

Spin waves are fluctuations around an ordered state, usually taken along the $z$-direction. However, a classical skyrmion is encoded by a local spin rotation at each lattice site. Let the classical texture be $\vec{n}_i = \vec{n}_{sk}(\vec{r}_i)$. Then, we need an SU(2) rotation that maps the ferromagnetic axis $\hat{z}$ to $\vec{n}_i$. That is given by the unitary operator acting on the spin Hilbert space

$$U_i = \exp(-i\phi_i S_i^z)\exp(-i\theta_i S_i^y) , \tag{372}$$

where $(\theta_i, \phi_i)$ are the polar angles of $\vec{n}_i$. The operator that creates the classical skyrmion texture is then given by [57]

$$\hat{R}(\theta,\phi) = \prod_{i \neq 0} U_i \ . \tag{373}$$

Acting on the ferromagnetic vacuum $\left|0\right\rangle = \otimes_i \left|\uparrow\right\rangle_i$, this produces the classical skyrmion spin configuration.

As we saw before, the topological charge of a skyrmion produces an emergent magnetic field

$$\vec{B}_{em} = \frac{\hbar}{2e}\vec{n}.(\partial_x \vec{n} \times \partial_y \vec{n}), \tag{374}$$

that acts on magnons analogously to a real Lorentz force.

As magnons propagate through the skyrmion texture, they acquire Berry phases and experience effective gauge potentials that modify their trajectories, dispersion relations, and scattering amplitudes. The resulting magnon dynamics can be described by a Schrödinger-like equation

$$i\hbar\frac{\partial \psi}{\partial t} = \left[\frac{(\vec{p} - \vec{A}_{em})^2}{2m^2} + V_{eff}\right]\psi . \tag{375}$$

with an emergent vector potential

$$\vec{A}_{em} = \frac{\hbar}{2e}(1 - \cos\theta)\vec{\nabla}\phi , \tag{376}$$

related to the magnetic field (374). This emergent electrodynamics underlies a variety of phenomena including skew scattering, nonreciprocal propagation, and the magnon Hall effect [65].

Researchers have studied magnon scattering by skyrmions extensively, both theoretically and experimentally [66]. Elastic scattering leads to characteristic deflection patterns, with magnons acquiring transverse momentum components due to the skyrmion's gyrovector. When the magnon wavelength approaches the skyrmion diameter, resonant scattering occurs, producing strong interference effects and enhanced mode mixing. Skyrmions also host localized internal modes—such as breathing, gyration, and helicity oscillations—that can hybridize with propagating magnons. This hybridization leads to avoided crossings in the magnon spectrum and enables coherent energy exchange between magnons and the internal degrees of freedom of the skyrmion. In frustrated magnets, where skyrmions can be extremely small and DMI does not rigidly fix helicity, this coupling becomes particularly strong, opening pathways toward helicity control via magnon injection.

The reciprocal effect—magnon-induced skyrmion motion—arises from momentum transfer during scattering. The resulting force can be incorporated into a Thiele-type equation, where the skyrmion's gyrovector produces a transverse velocity component analogous to the skyrmion Hall effect observed under electric-current driving. Magnon-driven motion has been demonstrated experimentally in insulating magnets such as $Cu_2OSeO_3$, where spin waves generated by microwave fields or thermal gradients can propel skyrmions without the Joule heating associated with charge-current manipulation [57]. This mechanism is particularly attractive for low-power spintronic devices, as magnons can be generated and controlled with minimal energy dissipation.

Beyond individual skyrmions, arrays and lattices of skyrmions act as reconfigurable magnonic crystals. The periodic emergent magnetic field modulates the magnon band structure, producing band gaps, flat bands, and topological magnon bands with nonzero Chern numbers. These features enable robust chiral edge modes that propagate along the boundaries of the skyrmion lattice, analogous to electronic edge states in quantum Hall systems. Because skyrmion lattices can be dynamically reconfigured using magnetic fields, electric currents, or optical pulses, they provide a versatile platform for tunable magnonic devices including filters, waveguides, and directional couplers. Recent experiments using Brillouin light scattering and microwave spectroscopy have confirmed magnon-band reconstruction in skyrmion crystals, highlighting the potential of skyrmion-based magnonics for on-chip information processing [56].

# 16. Quantum skyrmions beyond the semiclassical approach

The transition to a pure quantum description immediately raises a conceptual difficulty. For a quantum spin system, a natural local magnetization field can be constructed from expectation values: $\vec{m}_i = \langle \psi | \hat{S}_i | \psi \rangle / S$, but in general $|\vec{m}_i| \le 1$, with equality holding for a local spin-coherent state, or a pure spin state. Entanglement and quantum fluctuations reduce the length of the local expectation value. Thus, the topology of a classical spin texture cannot simply be promoted to an operator by replacing classical spins with spin operators. For example, for the singlet

$$|\psi\rangle = \frac{|\uparrow\downarrow\rangle - |\downarrow\uparrow\rangle}{\sqrt{2}}.$$

We have $< S_i >= 0$, and so $|\vec{m}_i| = 0$. Yet, each spin has quantum fluctuation $< S_i^2 >= \hbar^2 S(S+1)$, which is no zero.

The semiclassical approach is indispensable when the magnetic texture remains well defined, and quantum fluctuations are relatively weak. They do not, however, exhaust the meaning of a quantum skyrmion. The semiclassical construction assumes from the outset that the relevant many-body state remains continuously connected to a classical spin texture. In the deep quantum regime, this assumption becomes questionable. A quantum spin does not have a simultaneously well-defined orientation in all spatial directions, while a strongly correlated many-spin state generally cannot be represented as a product of local spin coherent states. Consequently, one cannot assume a smooth local magnetization field or a classical definition of topological charge a priori. This distinction becomes especially important for nanoscale skyrmions whose dimensions approach only a few lattice constants and for magnets composed of small local moments such as S=1/2. In such systems, the appropriate starting point is not a continuum magnetization configuration but a microscopic quantum Hamiltonian. The central question is then whether particular eigenstates or many-body ground states of the Hamiltonian possess the quantum analog of skyrmionic organization.

Lohani et al. [67] took an important step toward this microscopic perspective by investigating a frustrated spin-(1/2) Heisenberg ferromagnet using exact diagonalization. Rather than starting from a classical skyrmion and quantizing its collective coordinates, they identified the quantum skyrmion directly as a stable many-magnon bound state. A skyrmion in a ferromagnetic background is a bound-state compound of a fixed number of flipped spins. Consequently, the skyrmion possesses genuine many-body quantum numbers, including momentum and angular-momentum-related degrees of freedom. Its translational motion is also unusual: the calculated bandwidth can become exponentially small, reflecting tunneling processes between skyrmionic configurations. They also found that in weak fields the skyrmion accelerates parallel to the force but moves perpendicular to it in stronger fields. This work therefore established an important conceptual alternative to semiclassical quantization: a skyrmion can be regarded as an emergent quantum quasiparticle formed from a correlated collection of spin excitations.

Once the skyrmion is regarded as a many-body state, however, the question of topology becomes less straightforward. For a classical spin configuration, topology follows directly from the map between real space and the order-parameter sphere. For a quantum state $|\psi\rangle$, the local expectation value

$$\vec{m}_i = \langle\psi|\vec{S}_i|\psi\rangle, \tag{377}$$

does not completely characterize the state and can have a magnitude substantially smaller than (S). It may even vanish despite strong nonlocal spin correlations. Constructing a classical winding number from $\vec{m}_i$ can therefore discard precisely the quantum correlations that distinguish the state from its classical counterpart.

Sotnikov et al. [68], addressing this problem, do not find a rigorous integer-valued quantum topological charge analogous to the classical skyrmion number. Instead, they show that a particular three-spin quantum correlation, the scalar spin chirality, behaves as an effective quantum counterpart of the classical skyrmion number. The difficulty is that for a genuine quantum state $|\psi\rangle$ there is generally no definite vector $\vec{m}_i$ attached to each site. A spin-1/2 is an operator $\hat{S}_i$, not a little classical arrow. That immediately destroys the usual construction of $Q[\vec{m}(\vec{r})]$. The authors emphasize that this is not merely a technical inconvenience. A classical topological texture can in principle tunnel quantum mechanically into configurations belonging to another classical topological sector. Thus, the strict classical notion of topological protection does not simply survive quantization. The exact quantum ground state preserves translational symmetry. The skyrmion therefore exists primarily in the correlations between quantum spins, not in the one-spin expectation values. Even more importantly, on a lattice one can approximate the classical skyrmion number as

$$Q_{latt} = \frac{1}{8\pi}\sum_{<ijk>} \vec{m}_i.(\vec{m}_j \times \vec{m}_k). \tag{378}$$

So, the very quantity that measures local noncoplanarity is closely related to the topological density of the classical skyrmion. Now comes the essential step. Unlike a classical direction field, scalar chirality has an immediate quantum version:

$$\chi_{ijk} = \hat{S}_i.(\hat{S}_j \times \hat{S}_k), \tag{379}$$

and therefore, they can calculate $\langle\psi|\hat{\chi}_{ijk}|\psi\rangle$. For their translationally invariant system they define

$$Q_\psi = \frac{N}{\pi} < \hat{S}_1.(\hat{S}_2 \times \hat{S}_3) >, \tag{380}$$

where $N$ counts elementary triangular plaquettes appropriately. This $Q_\psi$ is what they call the quantum scalar chirality/skyrmion indicator. The expectation value of this three-spin operator provides a microscopic quantum observable associated with the local handedness of the spin state. In an appropriate classical limit, sums of such lattice chirality recover the familiar topological characterization of a smooth texture. In a genuine quantum state, however, chirality remains meaningful even when assigning a definite orientation to each spin is impossible.

As the magnetic field is increased, the quantum scalar chirality behaves in three qualitatively different ways. At low fields, it changes continuously with the field. In this regime, the state is interpreted as the quantum counterpart of the helical state becoming progressively noncoplanar. At intermediate fields, something remarkable occurs: $Q_\psi(\vec{B}) \approx \text{constant}$. Meanwhile, the magnetization continues changing. So, the state responds to the field, but this chirality-related quantity is effectively locked. Finally, near saturation, $Q_\psi \to 0$, as the system becomes fully polarized. The plateau is what leads the authors to identify the intermediate regime as a quantum skyrmion phase. Note that $Q_\psi$ is an expectation value. It is not fundamentally forced to be integer-valued, and there is no exact decomposition of Hilbert space into sectors labeled by $Q = 0, \pm 1, \pm 2, \ldots$ that cannot mix. This suggests a broader principle: quantum skyrmion topology should ultimately be formulated in terms of operators and correlation functions rather than exclusively through expectation-value spin textures.

A further departure from the semiclassical picture appears when one explicitly considers quantum entanglement. Haller *et al.* [69] studied spin-(1/2) Heisenberg models with Dzyaloshinskii-Moriya interaction using matrix-product-state and density-matrix-renormalization-group methods. They found zero-temperature quantum skyrmion and skyrmion-lattice ground states whose spatially resolved spin expectation values resemble classical Néel skyrmions. Crucially, however, the many body wavefunction contains substantial entanglement concentrated particularly around the skyrmion boundary. The local polarization magnitude is reduced from the value expected for a classical coherent spin configuration, and connected spin correlations remain finite.

This result shows why replacing the classical magnetization with its expectation value alone does not provide a complete quantum description. Two states may exhibit similar maps of $<\vec{S}_i>$ while possessing fundamentally different correlation and entanglement structures. In this sense, the quantum skyrmion consists of both a recognizable magnetic texture and a correlated “quantum cloud” surrounding or dressing that texture.

Haller et al. [63] recently formalized this interpretation by introducing a quantum magnetic-skyrmion creation operator. Their variational representation combines a component that reproduces the classical magnetic order with local spin-flip excitations that describe quantum corrections. At sufficiently large separation from other skyrmions, the resulting skyrmion can be treated as a bosonic quasiparticle while retaining its microscopic quantum dressing. This construction is particularly important because it suggests a bridge between two previously rather separated levels of description: microscopic quantum-spin models and effective many-skyrmion quantum field theories.

The strongest justification for going beyond semiclassical theory arises when quantum mechanics produces phases for which no corresponding classical state exists. An important recent example is the quantum skyrmion liquid reported by Bhowmick *et al.* [70]. As we have seen before, a classical skyrmion has a well-defined center and corresponds to a product state with essentially no position uncertainty. The authors of Ref. [70] construct a quantum skyrmion by taking a weighted sum of many classical skyrmions centered at slightly different positions $X_0 + n\Delta X_0$:

$$\psi^{\dagger}(X_0) = \frac{1}{N}\sum_{n} \exp\left(-\frac{n^2 \Delta X_0^2}{2\sigma_p}\right) \psi_R(X_0 + n\Delta X_0)\,. \tag{381}$$

Where $N$ is a normalization constant. The Gaussian factor gives the largest weight near $X_0$ and a progressively smaller weight farther away. The parameter $\sigma_p$ therefore measures the spread of the skyrmion position, characterizing the spatial extent of the quantum wave packet and, physically, measures the uncertainty or quantum fluctuations in the skyrmion position. For $\sigma_p = 0$ the state reduces to a single classical skyrmion with a sharply defined center. For $\sigma_p > 0$ the skyrmion center becomes uncertain, and the state is a coherent superposition of translated skyrmions. Larger $\sigma_p$ means stronger quantum fluctuations of the skyrmion position. The authors argue that quantum uncertainty in the skyrmion position generates entanglement concentrated near the skyrmion center. Eq. (381) provides a trial wavefunction that captures this effect by explicitly superposing skyrmions at nearby locations. It gives a simple variational picture of how a classical skyrmion can acquire genuinely quantum character: rather than changing the internal spin texture itself, quantum fluctuations are introduced through the superposition of translated skyrmion configurations. This construction is particularly relevant to the skyrmion-liquid (SkL) phase discussed in the paper. The authors associate this phase with enhanced quantum fluctuations of the skyrmion position and with entanglement concentrated around the skyrmion center.

It is important to distinguish this analytical construction from the matrix product state (MPS) representation used in numerical calculations. Equation (381) is not itself an MPS; rather, it is a trial or illustrative wavefunction designed to provide an intuitive description of a quantum skyrmion with positional fluctuations. In contrast, the MPS is the numerical representation employed within DMRG to obtain the many-body ground state and, where applicable, its excitations. The MPS results can then be compared with the physical picture suggested by Eq. (381). In this way, the Gaussian superposition provides an intuitive interpretation of the quantum skyrmion, while the MPS/DMRG framework supplies a numerically controlled description of the underlying many-body quantum state. Together, these approaches illustrate how a classical, localized skyrmion evolves into a quantum object whose position is delocalized and whose properties are governed by quantum fluctuations and entanglement.

Numerical calculations revealed an intermediate phase separating a quantum skyrmion crystal from the field-polarized state. The analogous classical system instead undergoes a direct transition. This particularly highlights the role of the skyrmion position. In the classical description, the center coordinates ($X$) and ($Y$) label the texture position. In quantum theory, they may behave as noncommuting collective observables,

$$[\hat{X},\hat{Y}]\neq 0, \tag{382}$$

reflecting the Berry-phase structure inherited from the underlying spins. The skyrmion cannot therefore be regarded as a point particle with simultaneously sharply defined coordinates. Quantum delocalization of its center produces a finite kinetic energy and, in appropriate regimes, favors states in which crystalline positional order is destroyed while local skyrmionic correlations survive. The emergence of such a liquid illustrates a qualitative rather than merely quantitative consequence of quantization.

Similar considerations also modify dynamics. Exact real-time evolution of two-dimensional interacting quantum-spin systems is computationally demanding, severely limiting conventional exact-diagonalization approaches. Neural-network quantum states offer one promising route beyond this restriction. Joshi, Peters, and Posske [71] used a restricted Boltzmann machine representation together with time-dependent variational evolution to study quantum skyrmions driven by magnetic-field gradients. Their calculations demonstrate translational motion as well as collision-induced decay and show how machine-learning representations can extend investigations of quantum-skyrmion dynamics to Hilbert spaces inaccessible to straightforward diagonalization.

The microscopic quantum formulation also forces a reconsideration of "topological protection." In a classical continuum theory, configurations with distinct integer winding numbers cannot be continuously deformed into each other without passing through a singular configuration. On an atomic lattice, this distinction is already weakened because the continuum mapping is approximate. In a finite quantum Hilbert space, it becomes still more subtle, since quantum superpositions do not generally belong to sharply separated classical topological sectors.

Consequently, stability of a quantum skyrmion should not automatically be identified with the classical notion of an infinite topological energy barrier. It may instead arise from energetic gaps, symmetry constraints, suppressed matrix elements between competing many-body states, approximate conservation laws, or environmental and measurement effects. Salvati *et al.* [72], for example, investigated how projective measurements can stabilize a quantum skyrmion through the quantum Zeno effect. Their work illustrates that the lifetime of a quantum texture is itself a quantum-dynamical question rather than a straightforward consequence of the classical winding number.

This conceptual shift has practical implications. The appropriate diagnostics of a quantum skyrmion may involve a hierarchy of observables: local magnetization establishes the presence of a skyrmion-like spatial pattern; scalar chirality and higher-order spin correlations probe its topology; entanglement measures test whether its internal structure is genuinely nonclassical; and spectroscopy determines whether collective degrees of freedom are quantized. No single quantity should be assumed to provide a universally sufficient definition.

Experimentally distinguishing a quantum skyrmion from a sufficiently small classical skyrmion remains a central challenge. Imaging a winding magnetization profile alone cannot establish quantum behavior because a classical coherent state can reproduce essentially the same spatial pattern. Evidence for genuine quantum behavior requires an observable that cannot be explained solely through a static classical texture.

Several possibilities have been proposed. Spin correlation functions and scalar chirality provide microscopic probes of many-body organization, while neutron scattering can in principle access correlation signatures associated with quantum skyrmion lattices [68,69]. Spectroscopy of discrete collective levels offers another route, particularly for quantized helicity and breathing modes. More recently, Sharma and Psaroudakis [73] proposed Brillouin light scattering as a direct probe of quantized skyrmion dynamics. In their analysis, vacuum fluctuations of the rotational degree of freedom generate an asymmetry between optical sidebands that is absent in the corresponding classical response, providing a possible quantum witness rather than merely an image of the underlying spin texture.

The quantum character of a skyrmion may become even more evident when superposition itself becomes experimentally accessible. Recent DMRG calculations for monoaxial chiral magnets predict degeneracy between skyrmion and antiskyrmion configurations and the consequent formation of Schrödinger-cat-like superpositions of macroscopically different topological textures [74]. Its main claim is quite striking: The quantum state can be a coherent superposition of the two textures,

$$|\psi\rangle = c|\mathrm{antisk}\rangle + \sqrt{1-c^2}e^{i\phi}|\mathrm{sk}\rangle, \tag{383}$$

which the authors call a skyrmionic Schrödinger-cat state. The important point is that this is much more radical than simply saying that a quantum skyrmion contains entangled spins. Such states represent an extreme departure from semiclassical reasoning because the relevant quantum state can no longer be associated with a single definite classical texture.

The emerging picture therefore suggests that no single quantum correction converts a classical skyrmion into a quantum one. Instead, several regimes should be distinguished. At large spin and large skyrmion radius, semiclassical field theory and collective-coordinate quantization provide a controlled description. At smaller sizes, microscopic quantum fluctuations dress an otherwise identifiable texture. Deeper in the quantum regime, the skyrmion becomes a correlated many-body quasiparticle characterized by reduced polarization, entanglement, chirality, and nontrivial quantum dynamics. Finally, strongly correlated systems may support skyrmion liquids, coherent superpositions of different textures, and other phases with no direct classical counterpart.

A central objective of future theory should therefore be to construct a systematic connection between microscopic spin Hamiltonians and an effective second-quantized theory of quantum skyrmions. Such a theory should identify suitable skyrmion creation and annihilation operators, their statistics and interactions, their coupling to magnons and other excitations, and the conditions under which an approximately conserved skyrmion number emerges.

Equally important is developing an operational definition of a quantum skyrmion. The classical criterion based solely on the winding of the expectation-value magnetization is insufficient once strong correlations are present. A more complete characterization will likely require combining topology, correlation functions, entanglement structure, spectral quantization, and dynamical response. Establishing such criteria would transform the concept of the quantum skyrmion from the quantization of a classical soliton into a genuine problem of quantum many-body topology.

In this view, the transition from classical to quantum skyrmions is not simply replacing classical coordinates with operators. It represents a bigger ontological shift: the skyrmion ceases to be only a texture traced by individual spin directions. It becomes an emergent state encoded in the correlations of the many body wavefunction.

# 17. Quantum hopfions

The extension of magnetic hopfions into the quantum regime is considerably less developed than the corresponding theory of quantum skyrmions. This difference is not merely quantitative. A classical magnetic hopfion, as we saw before, is defined by a smooth unit-vector field $\vec{m}(\vec{r})$ in three spatial dimensions and a map $\vec{m}: S^3 \to S^2$. As mentioned in section 16, the expectation value $\vec{m}_i = \langle \psi | \hat{S}_i | \psi \rangle / S$ of a magnetic spin is generally less than 1, so the topology of the classical spin texture cannot simply be promoted to an operator by replacing classical spins with spin operators. For hopfions, the difficulty is even more pronounced because the Hopf invariant is intrinsically non-local and involves the gauge potential associated with the emergent field. It is therefore useful to distinguish two meanings of a *quantum hopfion*. A *quantum-corrected hopfion* is a many-body spin state whose local expectation values retain the recognizable geometry and Hopf topology of a classical texture, while quantum fluctuations, correlations, and entanglement modify its local properties. A more ambitious *intrinsically quantum hopfion* would instead require defining Hopf topology directly at the level of the many-body quantum state, rather than through the topology of its semiclassical expectation-value texture. Most present microscopic studies of magnetic hopfions fall into the first category.

Kuchkin and Schmidt recently took an important step toward a microscopic theory [75], studying $S = 1/2$ chiral magnets using a cluster mean-field approximation combined with density-matrix renormalization group calculations. Their approach addresses a fundamental computational obstacle: a direct three-dimensional many-body treatment of a hopfion is presently prohibited by the exponential growth of the Hilbert space. Exploiting cylindrical symmetry, the system is decomposed into effectively one-dimensional radial quantum clusters, which are treated with DMRG, while interactions between clusters in the azimuthal and axial directions are treated at mean-field level. Thus, the calculation incorporates intra-cluster entanglement but not the full three-dimensional entanglement structure. The approximation should consequently be understood as an intermediate description between classical micromagnetics and a fully quantum three-dimensional spin model.

The states considered in Ref. [75] are particularly instructive because they consist of a skyrmion string decorated by zero, one, or two hopfion rings. These three configurations share the same transverse skyrmion charge, $Q = -1$, while their Hopf charges differ with the number of hopfion rings. In this geometry, the Hopf degree of freedom can therefore be regarded as an additional three-dimensional winding localized along an otherwise topologically identical skyrmion string. The most striking quantum signature is obtained from the local reduction of the magnetization length,

$$\delta\vec{m}_i = 1 - \left|<\hat{S}_i>\right| / S . \tag{384}$$

For a classical product state, $\delta\vec{m}_i = 0$, whereas $\delta\vec{m}_i > 0$, signal a departure from a local spin-coherent description. For the undecorated skyrmion string, the largest reduction is concentrated near the radius at which the magnetization rotates most strongly. After a hopfion ring is introduced, however, the strongest fluctuations become localized around the ring itself. Within the cluster mean-field calculation, the quantum fluctuations of the hopfion ring are approximately three times larger than those of the corresponding pure skyrmion string [75]. Since the approximation neglects entanglement in two spatial directions, the authors argue that a fully three-dimensional quantum treatment may yield still stronger corrections. This result suggests that the three-dimensional topology is accompanied by a characteristic quantum-fluctuation profile. A quantum hopfion should therefore not be pictured simply as the three-dimensional extension of a quantum skyrmion. Its additional winding generates localized regions where the many-body state departs much more strongly from the classical coherent-state manifold. The reduction of $|\vec{m}|$| may consequently provide one of the most accessible microscopic signatures distinguishing quantum and classical hopfions.

The variable magnetization length has deeper topological implications. Conventional micromagnetic assumes $|\vec{m}(\vec{r})| = 1$, which restricts the order parameter to S². Under this restriction, continuously changing the Hopf number is impossible without either introducing a singularity or modifying the boundary conditions. Creation and annihilation processes therefore commonly involve Bloch points.

Microscopic calculations of quantum Bloch points provide a useful analogy. For an $S = 1/2$ Heisenberg model, Kuchkin *et al.* [76] found that the local magnetization length decreases continuously and can reach $|\vec{m}(\vec{r})| = 0$ at the Bloch point, removing the singularity inherent in the fixed-length classical description [76,77,78]. This also connects to recent work on quantum Bloch points. These results motivated a regularized micromagnetic description where the order parameter is extended to four dimensions by introducing a vector $\vec{v} \in S^3$, with $|\vec{v}| = 1$, and

$$v_x = m_x,\ v_2 = m_y,\ v_3 = m_z \ \text{ and } \ v_4^2 = 1 - |\vec{m}|^2 . \tag{385}$$

This approach describes magnetization length dynamics while preserving the constraint. $|\vec{m}| \le 1$. To obtain a unique solution to the two possible fields which differ by the sign of $v^4$, we require the generalized energy functional to satisfy $E[\vec{m}, v^4] = E[\vec{m}, -v^4]$. The Heisenberg exchange term is generalized as follows [78]

$$E_{ex}(\vec{v}) = A\sum_{i=1}^{4}(\nabla v_i)^2 + kv_4^2 , \tag{386}$$

where $k$ is a temperature-dependent phenomenological parameter. Note that there is no direct physical observable associated with $v^4$. The vector $\vec{v}$ can be written as

$$\vec{v} = (\sin\theta\cos\phi\cos\psi, \sin\theta\sin\phi\cos\psi, \cos\theta\cos\psi, \sin\psi), \quad (387)$$

where $\theta(\vec{r})$ and $\phi(\vec{r})$ are spherical angles and the function $\psi(\vec{r})$ defining the deviation from unit magnetization length satisfies the boundary conditions $\psi(0) = \pi/2$, $\psi(\vec{r}) = 0$ for $r \to \infty$.

The additional component parametrizes departure from the classical unit-magnetization manifold [76,77]. In this enlarged space, a texture can locally leave the manifold when quantum fluctuations reduce the magnetization length. Consequently, the standard protection based exclusively on the homotopy group $\pi_3(S^2) = \mathbb{Z}$ can no longer be applied unchanged whenever the dynamics explore states with $|\vec{m}| < 1$.

This result may become particularly important for the quantum nucleation and annihilation of hopfions. Classical topology produces large barriers between different Hopf sectors because the magnetization must pass through singular configurations. Quantum fluctuations provide an additional route through Hilbert space in which the local expectation value can shrink without the microscopic quantum state itself becoming singular. Whether this mechanism produces appreciable quantum tunneling between different Hopf sectors remains an open question; the existing calculations establish the local reduction of magnetization, rather than a tunneling rate between states of different $H$.

The same issue appears dynamically. A hopfion ring attached to a translationally invariant skyrmion string has a zero-mode corresponding to displacement of the ring along the string. In a classical Landau–Lifshitz description,

$$\frac{\partial \vec{m}}{\partial t} = -\gamma \vec{m} \times \vec{H}_{eff}, \quad (388)$$

the length of magnetization is conserved and $\partial|\vec{m}|^2 / \partial t = 0$.

This is insufficient for a quantum-corrected hopfion because translation must transport not only the orientation texture but also the spatial profile of reduced |m|. Kuchkin and Schmidt therefore show that a regularized Landau–Lifshitz dynamics formulated in terms of the enlarged $S^3$ order parameter can accommodate the motion of the fluctuating region, whereas conventional fixed-length dynamics cannot reproduce this aspect of the quantum texture [75,77]. It was proposed [75] that, without damping, the dynamics of the $\vec{v}$ should be given by

$$\partial_t v_1 = \delta h v_2 + \varepsilon\delta h^2 v_1 v_4, \quad \partial_t v_2 = -\delta h v_1 + \varepsilon\delta h^2 v_2 v_4$$
$$\partial_t v_3 = 0, \quad \partial_t v_4 = \varepsilon\delta h^2 (v_1^2 + v_2^2). \quad (389)$$

Where the Zeeman field is written as $h \to h + \delta h$, with $\delta h << 1$, and $\varepsilon$ is a phenomenological parameter. For $\varepsilon = 0$, Eqs. (389) transform into the standard LLG equation. In the case $\varepsilon \neq 0$, Eqs. (389) allow for time-dependent magnetization dynamics.

This distinction is important for understanding classical versus quantum soliton dynamics. Quantum corrections do not merely renormalize parameters such as the effective mass or resonance frequency; they can require a larger dynamical manifold because a degree of freedom that is frozen in classical micromagnetics—the magnitude of the local magnetization—becomes dynamical.

Quantum hopfions also have an older, complementary meaning in nonlinear field theory. Hopf solitons of the Faddeev–Skyrme model can be quantized semiclassically by promoting symmetry-generated collective coordinates to quantum variables. Su [79], for example, studied the leading contribution of quantum excitations around the classical ground state and treated the model's rotational symmetry modes quantum mechanically using collective coordinates. He found that the quantum Hamiltonian coincides with that of a symmetric top rotating in SU(2).

Related collective-coordinate approaches have explored the quantization and spin-statistics properties of Skyrme–Faddeev-type solitons [80]. These approaches demonstrate that a classical Hopf soliton can possess discrete quantum rotational states, but they should be distinguished from the microscopic many-body quantum magnetism discussed above.

A promising bridge between the two approaches is quantizing the collective excitations of magnetic hopfions. Classical calculations already identify translational, rotational, breathing, and deformation modes. Recent micromagnetic simulations, for example, reveal a coherent sub-GHz breathing excitation in which the hopfion diameter oscillates while the Hopf charge remains unchanged [78]. Spin waves interacting with a hopfion also experience its emergent gauge field, leading to magnonic deflection and focusing [81].

Quantized breathing, twisting, or rotational excitations would constitute quantum states *of* a hopfion without requiring the entire spin texture to enter a strongly quantum regime. At present, however, experiments have not resolved such quantized internal modes in magnetic hopfions. The observed and calculated resonance modes should therefore be regarded as candidate degrees of freedom for future quantization rather than evidence for quantum hopfion spectroscopy.

The most fundamental unresolved problem is defining topology for a genuinely quantum hopfion. A quantum diagnostic may require non-local spin correlations, many-body Berry phases, correlation functions that reconstruct an emergent gauge field, or topology defined directly in an appropriate manifold of many-body states. No generally accepted magnetic many-body Hopf operator has emerged from the microscopic quantum-hopfion literature.

# 18. Conclusions

Topological spin textures have evolved from mathematical concepts within nonlinear field theories into central objects of contemporary condensed matter physics. Among them, skyrmions and hopfions represent particularly rich realizations of topology in magnetic systems, combining fundamental interest with potential technological relevance. Studying these textures has established deep connections between geometry, topology, statistical physics, quantum many-body theory, and non-equilibrium dynamics.

In this review, we have presented a systematic overview of the theoretical foundations underlying skyrmions and hopfions in magnetic materials. Beginning with the continuum description of ferromagnets and the nonlinear sigma model, we discussed how nontrivial topology emerges from the mapping between physical and order-parameter spaces. We reviewed the mathematical concepts needed to characterize topological textures and examined the mechanisms that stabilize them, including Dzyaloshinskii-Moriya interactions, frustration, and dipole-dipole interactions. These ingredients give rise to a remarkably diverse landscape of topological structures, ranging from vortices and merons to isolated skyrmions and skyrmion crystals.

The dynamical properties of magnetic solitons constitute another important aspect of the field. The Landau-Lifshitz-Gilbert, Thiele, and Landau-Lifshitz-Bloch descriptions provide complementary frameworks for understanding their motion, deformation, and response to external perturbations. The resulting collective-coordinate picture reveals the profound role of topology in determining the dynamics of magnetic textures, giving rise to phenomena such as the skyrmion Hall effect and emergent electrodynamics.

We further discussed topological textures in antiferromagnets, where the absence of net magnetization modifies both static and dynamical properties. Antiferromagnetic skyrmions exhibit distinct transport characteristics and enhanced robustness against external magnetic fields, making them promising candidates for future spintronic applications. More generally, antiferromagnetic systems illustrate how topological phenomena transcend the specific microscopic details of a given magnetic order.

A major theme of this review has been the extension of topology from two to three spatial dimensions. Hopfions represent the natural three-dimensional counterparts of skyrmions, characterized by the Hopf invariant rather than the skyrmion number. Their intricate structure, involving linked preimages and nontrivial topology of maps from $S^3$ to $S^2$, places them among the most fascinating magnetic solitons currently under investigation. Although significant theoretical progress has been achieved, stabilization, manipulation, and experimental observation of hopfions remain active areas of research.

The final part of this review addressed the quantum regime of topological textures. While classical skyrmions are by now well established both theoretically and experimentally, the quantum description of skyrmions and hopfions remains comparatively less developed. Quantum fluctuations, tunneling processes, finite-spin effects, and many-body correlations challenge the conventional semiclassical picture and require new theoretical methods. Recent studies suggest that quantum skyrmions and quantum hopfions should be regarded not merely as quantized versions of their classical counterparts, but as genuinely collective quantum objects whose properties emerge from strongly correlated spin systems.

Despite the considerable progress achieved over the last decades, numerous open questions remain. The microscopic origin and stability of quantum topological textures, their interaction with magnons, electrons, and phonons, the role of entanglement and quantum coherence, and the realization of stable three-dimensional topological structures remain major theoretical and experimental challenges. At the same time, advances in material synthesis, numerical techniques, and quantum simulation platforms are creating new opportunities to explore these questions.

The study of skyrmions and hopfions is gradually expanding beyond the traditional framework of classical magnetism toward a broader understanding of topological collective phenomena in interacting quantum systems. In this sense, skyrmions and hopfions are not only fascinating magnetic excitations but also paradigmatic examples of how topology organizes complex behavior across different areas of physics.

Continuing exploration of these structures promises to deepen our understanding of topological matter while opening new avenues for future technologies based on robust, controllable topological degrees of freedom.

Finally, the field's future development will increasingly require a deeper synthesis of topology, geometry, and quantum many-body physics. Beyond characterizing topological charges and classical soliton dynamics, emerging concepts such as quantum geometry, information-theoretic measures, entanglement structure, and quantum simulation may offer new perspectives on topological magnetic textures. In particular, understanding the geometric and quantum properties of skyrmion and hopfion states, their collective-coordinate manifolds, and their interactions with complex quantum environments remain largely unexplored. Exploring these directions may ultimately lead not only to a more complete theoretical description of topological matter but also to novel paradigms for information processing and quantum technologies based on topologically protected collective excitations.

# Appendix A. Solution for a skyrmion.

We want to solve

$$\frac{d^2\Theta(r)}{dr^2}+\frac{1}{r}\frac{d\Theta(r)}{dr}-\frac{1}{2r^2}\sin[2\Theta(r)]=0\,. \tag{A.1}$$

Set $x=\ln r$, and write derivatives with respect to $x$ as dots. Then

$$\frac{d\Theta}{dr}=\frac{1}{r}\dot{\Theta}\,,\quad \frac{d^2\Theta}{dr^2}=\frac{d}{dr}\left(\frac{1}{r}\dot{\Theta}\right)=-\frac{1}{r^2}\dot{\Theta}+\frac{1}{r^2}\ddot{\Theta}=\frac{\ddot{\Theta}-\dot{\Theta}}{r^2}\,. \tag{A.2}$$

Also,

$$\frac{1}{r}\frac{d\Theta}{dr}=\frac{\dot{\Theta}}{r^2}\,. \tag{A.3}$$

Insert Eq. (A.3) into (A.2) and multiply by $r^2$

$$\frac{\ddot{\Theta}-\dot{\Theta}}{r^2}+\frac{\dot{\Theta}}{r^2}-\frac{1}{2r^2}\sin 2\Theta=0 \quad\rightarrow\quad \ddot{\Theta}-\frac{1}{2}\sin 2\Theta=0\,. \tag{A.4}$$

So, in the variable $x = \ln r$, this equation becomes

$$\ddot{\Theta} = \frac{1}{2}\sin 2\Theta . \tag{A.5}$$

To obtain a first integral, we multiply Eq.(A.5) by $\dot{\Theta}$, and get

$$\dot{\Theta}\ddot{\Theta} = \frac{1}{2}\dot{\Theta}\sin 2\Theta . \tag{A.6}$$

The left side is $\dot{\Theta}\ddot{\Theta} = \frac{1}{2}\frac{d}{dx}(\dot{\Theta}^2)$, so

$$\frac{1}{2}\frac{d}{dx}(\dot{\Theta}^2) = \frac{\dot{\Theta}}{2}\sin 2\Theta . \tag{A.7}$$

Now, we multiply by 2 and integrate. We obtain

$$\dot{\Theta}^2 = \int \sin 2\Theta d\Theta = -\frac{1}{2}\cos 2\Theta + C , \tag{A.8}$$

Where $C$ is a constant. Using $\dot{\Theta} = r d\Theta / dr$, we can write

$$\left( r\frac{d\Theta}{dr} \right)^2 + \frac{1}{2}\cos 2\Theta = C . \tag{A.9}$$

Which gives

$$\frac{d\Theta}{dr} = \pm \frac{1}{r}\sqrt{C - \frac{1}{2}\cos 2\Theta} . \tag{A.10}$$

Or, in terms of $x$

$$\frac{d\Theta}{dx} = \pm \sqrt{C - \frac{1}{2}\cos 2\Theta} . \tag{A.11}$$

Thus

$$\int \frac{d\Theta}{\sqrt{C - (1/2)\cos 2\Theta}} = \pm x + C_1 = \pm \ln r + C_1 . \tag{A.12}$$

For general $C$, this integral is an elliptic integral, so the solution is given implicitly in terms of elliptic functions. For the special case $C = 1/2$, we have

$$\frac{d\Theta}{dx} = \pm \sin\Theta . \tag{A.13}$$

With solution

$$\ln\left|\tan(\Theta/2)\right| = \pm x + C_2 , \tag{A.14}$$

which leads to

$$\tan(\Theta/2) = A r^{\pm 1}$$

Where $A$ is a non-zero constant. Thus, an explicit family of solutions is

$$\Theta(r) = 2\arctan(Ar) \quad \text{or} \quad \Theta(r) = 2\arctan(A/r). \tag{A.15}$$

For a skyrmion, we want a solution that satisfies the topological boundary conditions. If, for instance, we take the condition $\Theta(0) = \pi$, $\Theta(\infty) = 0$, a particular choice is

$$\Theta(r) = 2\arctan(R/r), \tag{A.16}$$

Where $R > 0$ is a length scale that we can take to be the skyrmion radius. Another solution is

$$\Theta(r) = 2\arctan(r/R). \tag{A.17}$$

depending on which boundary we want at $r = 0$ and $r \to \infty$. Both solve the same Euler equation; they correspond to different choices of core vs far-field orientation.

# Appendix B. Ansatz for a Skyrmion

I start with Eq. (68)

$$J\frac{d^2\Theta}{dx^2} = A\sin(\Theta)\cos(\Theta). \tag{B1}$$

We multiply both sides of Eq.(B1) by $d\Theta(x)/dx$

$$\frac{d^2\Theta}{dx^2}\frac{d\Theta}{dx} = \frac{\sin\Theta\cos\Theta}{w^2}\frac{d\Theta}{dx}, \tag{B2}$$

Where $w = \sqrt{J/A}$. Integrating Eq. (B2) we get

$$\frac{1}{2}\left(\frac{d\Theta}{dx}\right)^2 = \frac{1}{2w^2}\sin^2\Theta + C. \tag{B3}$$

For a domain wall connecting $\Theta = 0$ and $\Theta = \pi$, the derivative vanishes at the boundary, so $C = 0$. Therefore

$$\frac{d\Theta}{dx} = \frac{\sin\Theta}{w}. \tag{B4}$$

This equation has a solution

$$\ln\tan\left(\frac{\Theta}{2}\right) = \frac{x - x_0}{w}. \tag{B5}$$

Or

$$\tan\left(\frac{\Theta}{2}\right) = e^{(x-x_0)/w}. \tag{B6}$$

A 360° wall is two 180° walls superposed, so its profile is twice that of the 180° solution. To obtain a $0 \to 2\pi$ solution, we start with

$$\tan\left(\frac{\Theta}{2}\right) = f(x). \tag{B7}$$

And choose $f(x)$ so that $\Theta$ solves the sine-Gordon equation and has the right asymptotic. A convenient choice is

$$f(x) = \frac{1}{\sinh[(x - x_0)/w]}.$$

(B8)

Then

$$\Theta(x) = 2\arctan\frac{1}{\sinh[(x - x_0)/w]}.$$

(B9)

To see why this expression works, let

$$u = \frac{x - x_0}{w}, \qquad y = \tan\left(\frac{\Theta}{2}\right) = \operatorname{csch} u = \frac{1}{\sinh u}. \tag{B10}$$

Then

$$\Theta = 2\arctan y. \tag{B11}$$

So

$$\frac{d\Theta}{dx} = \frac{2}{1 + y^2}\frac{dy}{dx}. \tag{B.12}$$

We have

$$\frac{dy}{dx} = \frac{dy}{du}\frac{du}{dx} = -\operatorname{csch} u \coth u\left(\frac{1}{w}\right). \tag{B.13}$$

Now

$$1 + y^2 = 1 + \operatorname{csch}^2 u = \coth^2 u. \tag{B.14}$$

Then

$$\frac{d\Theta}{dx} = \frac{2}{\coth^2 u}\left(-\frac{\operatorname{csch} u \coth u}{w}\right) = -\frac{2(1/\sinh u)}{(w\cosh u/\sinh u)}$$

$$= -\frac{2}{w\cosh u \sinh u} = -\frac{1}{w\sinh(2u)}. \tag{B.15}$$

Differentiating once more and using

$$\sin\Theta = 2\tan(\Theta/2)/(1 + \tan^2(\Theta/2)). \tag{B.16}$$

With $y = \operatorname{csch} u$, we get

$$\frac{d^2\Theta}{dx^2} = \frac{1}{w^2}\sin\Theta. \tag{B.17}$$

So the profile satisfies the wall equation. A skyrmion is a 360° wall wrapped around a circle of radius *R*. So we replace $x - x_0 \to r - R.$ . Naively we can write

$$\Theta(r) = 2\arctan\left[\frac{1}{\sinh((r - R)/w)}\right]. \tag{B.18}$$

But this solution does not satisfy the skyrmion boundary conditions. The simple substitution $x - x_0 \to r - R,$ does not give the correct asymptotic. Wang et al. [82] replaced the linear coordinate by a radial hyperbolic coordinate, and modeled the skyrmion profile by the ansatz

$$\Theta(r) = 2\arctan\left[\frac{\sinh(R/w)}{\sinh(r/w)}\right]. \tag{B.19}$$

We can see that now: $\Theta(0)=\pi$, $\Theta(\infty)=0$, $\Theta(R)=\pi/2$. This ansatz has been used to simulate various stable magnetic states, as discussed in Ref. [82].

# Appendix C: Coherent states and path integral.

### C1. Spin coherent states.

Coherent spin-state methods have become standard tools in quantum magnetism because they offer a direct bridge between classical spin textures and their quantum counterparts. In the context of skyrmions, they play a central role in several aspects of quantum theory. First, coherent states enable the quantization of skyrmion textures by promoting the classical spin field to quantum operators. This approach allows one to derive effective quantum actions and path-integral formulations for the skyrmion's collective coordinates, such as its position and helicity, thereby providing a controlled route from classical topological solitons to quantum quasiparticles. Second, coherent states form the natural language for describing magnon–skyrmion interactions. Magnons, treated as coherent spin-wave excitations, can couple to the skyrmion background in a way that preserves the geometric and topological structure of the underlying spin configuration, enabling analysis of scattering processes, bound states, and the dressing of skyrmions by quantum fluctuations. Third, coherent-state techniques are essential for studying macroscopic quantum phenomena associated with skyrmions, including quantum tunneling of their collective coordinates and helicity quantization.

The usefulness of coherent states in these contexts stems from several key features. They provide a semiclassical representation of the skyrmion texture that retains the geometric intuition of classical magnetism while systematically incorporating quantum corrections. They also provide a natural framework for incorporating Berry-phase effects, which are crucial for topological spin configurations and strongly influence skyrmion dynamics at both the classical and quantum levels. Moreover, coherent-state path integrals allow derivation of effective low-energy theories, such as Thiele-like equations for quantum skyrmion dynamics, in which the interplay between topology, geometry, and quantum mechanics becomes explicit. Finally, coherent states provide a controlled method for treating quantum fluctuations around the skyrmion, including the spin-wave "cloud" that recent work has shown cancels classical motion and fundamentally modifies the dynamical behavior of quantum skyrmions.

Spin states of maximal projection along some direction in space are called spin coherent states [83-89]. They form an infinite set of states $|\vec{n}\rangle$, labeled by the continuous vector that most closely resembles a classical spin.

Let us start with the linear momentum. We have

$$p_x\psi(x) = \frac{\hbar}{i}\frac{d\psi(x)}{dx} = \lim_{\delta x\to 0}\frac{\hbar}{i}\frac{\psi(x)-\psi(x+\delta x)}{\delta x} = \lim_{\delta x\to 0} i\hbar\frac{P_{\delta x}-1}{\delta x}\psi(x), \tag{C.1}$$

where

$$P_{\delta x} = 1 - \frac{i}{\hbar}\delta x p_x . \tag{C.2}$$

Similarly, we define the $x$ component of the angular momentum as the operator whose conservation follows from the invariance of the Hamiltonian under rotation about the $x$-axis. In analogy with the linear momentum, we define for infinitesimal rotations $\delta\theta_x$

$$J_x \psi = \lim_{\delta\theta_x \to 0} i\hbar \frac{P_{\delta\theta_x} - 1}{\delta\theta_x}\psi , \tag{C.3}$$

or

$$P_{\delta\theta_x} = 1 - \frac{i}{\hbar}\delta\theta_x J_x , \tag{C.4}$$

where $P_{\delta\theta_x}$ is the operator that rotates the contour of the function $\psi$ by $\delta\theta_x$ around the $x$-axis in the counterclockwise direction. For a finite rotation, we write $\theta_x = n\delta\theta_x$. Taking $n \to \infty$ with $\delta\theta_x \to 0$ such that $\theta_x$ is a constant, we have

$$P_{\theta_x}\psi = \lim_{\delta\theta_x \to 0}(P\delta\theta_x)^{\theta_x/\delta\theta_x}\psi = \lim_{\delta\theta_x \to 0}\left[1 - \frac{i}{\hbar}\delta\theta_x J_x\right]^{\theta_x/\delta\theta_x}\psi = e^{-i\theta_x J_x/\hbar}\psi . \tag{C.5}$$

For an arbitrary rotation, we use the three Euler angles and get

$$U(\theta,\phi,\beta) = e^{i\phi S^z} e^{i\theta S^y} e^{i\beta S^z} , \tag{C.6}$$

where we have set $\hbar = 1$. Applying the rotation operator $U$ to the maximally polarized state $|S,S\rangle$ (where $|\ \rangle$ is an eigenvector of $S^z$), we get the spin coherent states

$$|\vec{n}\rangle = e^{i\phi S^z} e^{i\theta S^y} e^{i\beta S^z} |S,S\rangle . \tag{C.7}$$

The operator $e^{i\beta S^z}$ in (C.7) can be replaced by $e^{iS\beta}$. The angle $\beta$ is arbitrary and corresponds to a gauge freedom. $|\vec{n}\rangle$ is obtained from $|S,S\rangle$ by a rotation of an angle $\theta$ about the $y$ axis followed by a rotation of angle $\phi$ about the $z$ axis. We have

$$e^{i\phi S^z} e^{i\theta S^y} |S,S\rangle = \sum_{m=-S}^{S}\left[\frac{(2S)!}{(S+m)!(S-m)!}\right]^{1/2} \sin^{S-m}\frac{\theta}{2}\cos^{S+m}\frac{\theta}{2} e^{-i(S+m)\phi}|S,m\rangle . \tag{C.8}$$

If $|m\rangle = |S,m\rangle$ is an eigenstate of $S^z$ with eigenvalue $m$ we have

$$\sum_{m=-S}^{S}|m\rangle\langle m| = 1 . \tag{C.9}$$

We can write (C.7) as

$$|\vec{n}\rangle = \sum_{m=-S}^{S}\langle m|e^{i\phi S^z} e^{i\theta S^y}|S\rangle|m\rangle = \sum_{m=-S}^{S} C_m(\theta,\phi)|m\rangle , \tag{C.10}$$

where

$$C_m(\theta,\phi)=e^{-im\phi}\left[\frac{(2S)!}{(S+m)!(S-m)!}\right]^{1/2}\left(\sin\frac{\theta}{2}\right)^{S-m}\left(\cos\frac{\theta}{2}\right)^{S+m}. \tag{C.11}$$

The angle $\beta$ is determined such that the coherent state be single value under $\phi\to\phi+2\pi n$, $n=0,\pm1,...$ So $\beta$ can take only the values $\beta=(2n+1)\phi$. For $n$ = -1, we have the north-pole gauge, and for n = 0, the south-pole gauge. The two-gauge choices are equivalent.

The coherent states are not orthogonal. We find that [83]

$$\langle\vec{n}'|\vec{n}\rangle=\sum_{m=-S}^{S}\frac{(2S)!}{(S+m)!(S-m)!}\left[e^{i(\phi-\phi')}\sin\frac{\theta}{2}\sin\frac{\theta'}{2}\right]^{S-m}\left[\cos\frac{\theta}{2}\cos\frac{\theta'}{2}\right]^{S+m}$$
$$=\left[\cos\frac{\theta}{2}\cos\frac{\theta'}{2}+e^{i(\phi-\phi')}\sin\frac{\theta}{2}\sin\frac{\theta'}{2}\right]^{2S}=\left(\frac{1+\vec{n}.\vec{n}'}{2}\right)^{S}e^{i\Phi(\vec{\Omega}_1,\vec{\Omega}_2,\vec{\Omega}_3)}, \tag{C.12}$$

Here $\Phi(\vec{n}_1,\vec{n}_2,\vec{n}_3)$ is the (oriented) area of the spherical triangle with vertices at $\vec{n}_1$, $\vec{n}_2$ and $\vec{n}_3$. Now I will discuss the case of $N$ spins. The many-spin coherent states are given by

$$|\vec{n}\rangle=\prod_{i=1}^{N}|\vec{n}_i\rangle. \tag{C.13}$$

The expression for their overlap is

$$\langle\vec{n}|\vec{n}'\rangle=\prod_i\left(\frac{1+\vec{n}_i.\vec{n}_i'}{2}\right)^{S}\exp\left(-iS\sum_i\Phi[\vec{n}_i,\vec{n}'_i]\right).$$

(C.14)

The resolution of identity in the product space becomes

$$\int\prod_i\left(\frac{2S+1}{4\pi}d\vec{n}_i\right)|\vec{n}_i\rangle\langle\vec{n}_i|=I. \tag{C.15}$$

The integral of $\vec{n}$ is over the unit sphere. If $\vec{n}$ is infinitely separated from $\vec{n}'$, we can take $\theta=\theta'$ and define $\Delta\phi=\phi'-\phi$. Hence Eq.(C.12) leads to

$$\langle\vec{n}'|\vec{n}\rangle=1+iS\Delta\phi(\cos\theta-1)\approx e^{-i\Phi S}. \tag{C.16}$$

where here

$$\Phi=(1-\cos\theta)\Delta\phi. \tag{C.17}$$

The area element in a unit sphere is given by $dA=\sin\alpha d\sigma d\phi=-d\cos\alpha d\phi$. If we consider a spherical triangle with a vertex at the north pole and points with the same $\theta$ but different $\phi$, its area is given by

$$A=-\int_0^{\theta}\int_{\phi'}^{\phi}d\cos\alpha d\phi=(1-\cos\theta)\Delta\phi, \tag{C.18}$$

which is equal to (C.17).

Writing $d\vec{n}=\sin d\theta d\phi$, we have

$$\frac{2S+1}{4\pi}\int d\vec{n}\,|\vec{n}\rangle\langle\vec{n}| =$$

$$\frac{(2S+1)}{2}\int_{-1}^{1} d\cos\theta \sum_{m=-S}^{S}\left(\frac{1+\cos\theta}{2}\right)^{S+m}\left(\frac{1-\cos\theta}{2}\right)^{S-m}\frac{(2S)!}{(S+m)!(S-m)!}|S,m\rangle\langle S,m|$$

$$= \sum_{m=-S}^{S}|S,m\rangle\langle S,m| = I\,. \tag{C.19}$$

Because the coherent states are nonorthogonal, they are called overcomplete.

It can be shown that [86]

$$\langle\vec{n}|\vec{S}|\vec{n}\rangle = S\vec{n}\,. \tag{C.20}$$

The classical limit is reached by taking $S\to\infty$. In this limit, $|\vec{n}\rangle$ reduces to a classical spin in the $\vec{n}$ direction. This property implies that the vector $S\vec{n}$ is a classical approximation to the operator $\vec{S}$. Usually it is possible to arrange the operators such that

$$\langle\vec{n}|H(\vec{S})|\vec{n}\rangle = H(\vec{n})\,, \tag{C.21}$$

i. e. $H(\vec{n})$ has the same functional dependence upon $\vec{n}$ as the original Hamiltonian has on $\vec{S}$ [86].

Spin coherent states can be used to evaluate the trace of any operator [76]. If $|m\rangle$ is any orthonormal basis, then

$$TrA = \sum_{m}\langle m|A|m\rangle = \left(\frac{2S+1}{4\pi}\right)^{2}\int d\vec{n}\int d\vec{n}'\sum_{n}\langle m|\vec{n}\rangle\langle\vec{n}|A|\vec{n}'\rangle\langle\vec{\Omega}'|m\rangle$$

$$= \left(\frac{2S+1}{4\pi}\right)\int d\vec{n}\,\langle\vec{n}|A|\vec{n}\rangle. \tag{C.22}$$

**C.2. Path integral for spins.**

Here, I will calculate the imaginary-time path-integral representation for spin systems. The partition function can be written as

$$Z = Tre^{-\beta H} = (2S+1)\int\frac{d\vec{n}}{4\pi}\langle\vec{n}|e^{-\beta H}|\vec{n}\rangle\,, \tag{C.23}$$

as a functional integral over spin coherent states. We split the imaginary "time" $\beta$ into $N$ infinitesimal steps. Substituting the resolution of identity (C.15) $N-1$ times into the time-sliced partition function, we get

$$Z = \left(\frac{2S+1}{4\pi}\right)^{N}\int\prod_{k=0}^{N-1}d\vec{n}_k\,\langle\vec{n}_{k+1}|e^{-\varepsilon H}|\vec{n}_k\rangle\,, \tag{C.24}$$

with $\vec{n}_0 = \vec{n}_N$. Using

$$\left\langle \vec{n}_{k+1} \right| e^{-\varepsilon H} \left| \vec{n}_k \right\rangle = \left\langle \vec{n}_{k+1} \right| (1-\varepsilon H) \left| \vec{n}_k \right\rangle + \mathrm{O}(\varepsilon^2) = 1 - \left\langle \vec{n}_k \middle| \vec{n}_k \right\rangle + \left\langle \vec{n}_{k+1} \middle| \vec{n}_k \right\rangle - \varepsilon \left\langle \vec{n}_{k+1} \right| H \left| \vec{n}_k \right\rangle =$$
$$\exp\left( \left\langle \vec{n}_{k+1} \middle| \vec{n}_k \right\rangle - \left\langle \vec{n}_k \middle| \vec{n}_k \right\rangle - \varepsilon \left\langle \vec{n}_{k+1} \right| H \left| \vec{n}_k \right\rangle \right), \tag{C.25}$$

where I have used $\left\langle \vec{n}_k \middle| \vec{n}_k \right\rangle = 1$. Setting $\tau = k\varepsilon$, considering $\tau \in [0, \beta]$ as a continuous variable in the limit $\varepsilon = \beta / N \to 0$, and using

$$\frac{\left\langle \vec{n}_{k+1} \middle| \vec{n}_k \right\rangle - \left\langle \vec{n}_k \middle| \vec{n}_k \right\rangle}{\varepsilon} = \left\langle \partial_\tau \vec{n}_k \middle| \vec{n}_k \right\rangle, \tag{C.26}$$

we obtain

$$Z = \lim_{N\to\infty} \left( \frac{2S+1}{4\pi} \right)^N \int_{\vec{n}(\beta)=\vec{n}(0)} \prod_{k=0}^{N-1} d\Omega_k \exp\left[ -\varepsilon \sum_{k=0}^{N-1} \left( -\left\langle \partial_\tau \vec{n}_k \middle| \vec{n}_k \right\rangle + \left\langle \vec{n}_{k+1} \right| H \left| \vec{n}_k \right\rangle \right) \right] \tag{C27}$$

In writing (C.26), I implicitly assume, as in the calculation of path integrals, that Z is dominated by smooth paths. In the limit $N \to \infty$ we can write Eq .(C.27) as

$$Z = \int_{\vec{\Omega}(\beta)=\vec{\Omega}(0)} D[\vec{n}] \exp\left\{ S_B[\vec{n}] - \int_0^\beta d\tau \left\langle \vec{n} \right| H \left| \vec{n} \right\rangle \right\} = \int d[\vec{n}] e^{-S[\vec{n}]}, \tag{C .28}$$

where $S$ is the action, and

$$S_B[\vec{n}] = \int_0^\beta \left\langle \partial_\tau \vec{n} \middle| \vec{n} \right\rangle d\tau . \tag{C.29}$$

The function $\left| \vec{n}(\tau) \right\rangle$ is the continuum limit of $\left| \vec{n}_k \right\rangle$. The phase $S_B[\vec{n}]$ is purely a quantum effect and is called the Berry phase. It is a geometric phase determined by the overlap of the coherent states and is well-defined only for a closed path.

We have

$$\left\langle \partial_\tau \vec{n} \middle| \vec{n} \right\rangle = \frac{\left\langle \vec{n}(\tau+\varepsilon) \middle| \vec{n}(\tau) \right\rangle - \left\langle \vec{n} \middle| \vec{n} \right\rangle}{\varepsilon} = \frac{\left\langle \vec{n}(\tau+\varepsilon) \middle| \vec{n}(\tau) \right\rangle - 1}{\varepsilon} . \tag{ C.30}$$

The assumption of $\left| \vec{n}(\tau+\varepsilon) \right\rangle - \left| \vec{n}(\tau) \right\rangle$ being of $O(\varepsilon)$ keeps the slowly varying motions but ignores the short-range fluctuations. Using Eq. (C.16), we can write

$$\left\langle \vec{n}(\tau+\varepsilon) \middle| \vec{n}(\tau) \right\rangle = 1 + iS(1-\cos\theta)[\phi(\tau+\varepsilon) - \phi(\tau)], \tag{C.31}$$

which leads to

$$\left\langle \partial_\tau \vec{n} \middle| \vec{n} \right\rangle = iS(1-\cos\theta) \frac{d\phi}{d\tau} . \tag{C.32}$$

Hence, Eq. (C.29) can be rewritten as

$$S_B[\vec{n}] = iS \int_0^\beta d\tau (1-\cos\theta) \frac{d\phi}{d\tau} . \tag{C.33}$$

The path integral is restricted to paths that are periodic in the imaginary time, i. e. with boundary condition $\vec{n}(\tau=0)=\vec{n}(\beta)$. We introduce a dummy variable $u$ ($0 < u < 1$) and define a continuous vector function $\vec{n}(\tau,u)$ satisfying the following boundary conditions:

$$\vec{n}(\tau,u=1)=\vec{n}(\tau)\,, \quad \vec{n}(\tau,u=0)=(1,0,0)\,. \tag{C.34}$$

Then it can be shown that [72]

$$S_B[\vec{n}]=iS\int_0^\beta d\tau\int_0^1 du\vec{n}.(\partial_u\vec{n}\times\partial_\tau\vec{n})\,. \tag{C.35}$$

This term is defined only up to $4\pi$ (which is the area of the unit sphere). Thus, the net contribution of $e^{S_B}$ has an uncertainty up to a factor $e^{i4\pi S}$. However, for $2S$ an integer, this factor equals unity. The integral in Eq. (C.35) is similar to the one in Eq. (41) defining the topological charge. This occurs because both integrals describe the area in a sphere. However, in Eq. (C.35) the integral is in the parameter space, while in Eq. (41) it is in the real space.

Eq. (C.29) can be rewritten using a real $t$ parameter as

$$S_B[\vec{n}]=\int_0^T\langle\vec{n}(t)|\partial_t\vec{n}(t)\rangle dt\,. \tag{C.36}$$

We define the Berry connection on the sphere by

$$\vec{A}(\vec{n})\equiv i\langle\vec{n}|\nabla_n|\vec{n}\rangle\,. \tag{C.37}$$

By the chain rule, we have

$$\langle\vec{n}|\partial_t|\vec{n}\rangle=\langle\vec{n}|\nabla_n|\vec{n}\rangle.\frac{\partial\vec{n}}{\partial t}=-iS\vec{A}(\vec{n}).\frac{\partial\vec{n}}{\partial t}\,. \tag{C.38}$$

For a closed path, $S_B[\vec{n}]$ becomes a gauge-invariant physical quantity. Then we can rewrite eq. (C.36) as

$$S_B[\vec{n}]=S\oint dt\vec{A}(\vec{n}).\frac{d\vec{n}}{dt}\,. \tag{C.39}$$

Note that integration is over the space parameter, not over real space.

Using $\vec{n}=(\sin\theta\cos\phi,\sin\theta\sin\phi,\cos\theta)$, we have in spherical coordinates

$$\partial_t|\vec{n}\rangle=(\partial_\theta|\vec{n}\rangle)\dot{\theta}+(\partial_\phi|\vec{n}\rangle)\dot{\phi}\,, \tag{C.40}$$

$$A_\mu(\vec{n})=i\langle\vec{n}|\partial_\mu|\vec{n}\rangle\,, \qquad \mu=\theta,\phi\,, \qquad i\langle\vec{n}|\partial_t|\vec{n}\rangle=A_\theta\dot{\theta}+A_\phi\dot{\phi}\,. \tag{C.41}$$

Choosing $A_\phi=(1-\cos\theta)/\sin\theta$, $A_\theta=0$, we recover Eq. C (32).

In the continuum field theory, we introduce the continuum field $\vec{n}(x,\tau)$, and write for the partition function

$$Z=\int D\vec{n}(x,\tau)\delta(\vec{n}^2-1)\exp\left(-\int_0^\beta d\tau\int d^dx\mathscr{L}\right), \tag*{(C.42).)}$$

where the density Lagrangian is

$$\mathscr{L}=iS\int_0^1 du\vec{n}.\left(\frac{\partial\vec{n}}{\partial u}\times\frac{\partial\vec{n}}{\partial\tau}\right)-\frac{JS^2}{2}(\nabla\vec{n})^2\,. \tag{C.43}$$

Other terms, such as the MD interaction and anisotropy terms, can be added to the right-hand side of Eq. (C.43).

We can write

$$\mathcal{L} = \mathcal{L}_B - \mathcal{H} \,. \tag{C.44}$$

The Euler-Lagrange equations are

$$\frac{d}{dt}\left(\frac{\partial \mathcal{L}}{\partial \dot{\theta}}\right) - \frac{\partial \mathcal{L}}{\partial \theta} = 0, \qquad \frac{d}{dt}\left(\frac{\partial \mathcal{L}}{\partial \dot{\phi}}\right) - \frac{\partial \mathcal{L}}{\partial \phi} = 0. \tag{C.45}$$

Which for $\mathcal{L}_B = S(1-\cos\theta)\dot{\phi}$, gives

$$S\sin\theta\dot{\theta} + \frac{\partial \mathcal{H}}{\partial \phi} = 0, \qquad S\sin\theta\dot{\phi} - \frac{\partial \mathcal{H}}{\partial \theta} = 0. \tag{C.46}$$

This is equivalent to the Bloch equation

$$\frac{\partial \vec{n}}{\partial t} = \vec{B}_{eff} \times \vec{n} \,. \tag{C.47}$$

It is interesting to compare this with the case where the Hamiltonin in the phase space is given by $H(q_i, p_i)$. The action is then given by

$$S = \int dt \left( \sum_i p_i \dot{q}_i - H \right). \tag{C.48}$$

The Poisson bracket for time-dependent functions $A(t)$ and $B(t)$ is

$$\{A, B\} = \sum_i \left( \frac{\partial A}{\partial q_i}\frac{\partial B}{\partial p_i} - \frac{\partial A}{\partial p_i}\frac{\partial B}{\partial q_i} \right). \tag{C.49}$$

The time derivative of a function $f(\vec{q}, \vec{p}, t)$ is given by

$$f(\vec{q}, \vec{p}, t) = \sum_i \left( \frac{\partial f}{\partial q_i}\dot{q}_i + \frac{\partial f}{\partial p_i}\dot{p}_i \right) + \frac{\partial f}{\partial t} = \{f, H\} + \frac{\partial f}{\partial t}. \tag{C.50}$$

If $A(t) = A[f_1(\vec{q}, \vec{p}, t), f_2(\vec{q}, \vec{p}, t), ...]$, and similar for B($t$), we have

$$\{A, B\} = \sum_{i,j} \left( \frac{\partial A}{\partial f_i}\frac{\partial B}{\partial f_j} \right) \{f_i, f_j\}. \tag{C.51}$$

The Poisson bracket between scalar fields $A(x)$ and $B(x')$ corresponding to an action of the form

$$S = \int dt \int d^2 r \left( \sum_i \Pi_i \partial_i \Phi_i - H \right), \tag{C.52}$$

is given by [60]

$$\{A(x), B(x')\} = \sum_i \int dy \left( \frac{\delta A(x)}{\delta \Phi_i(y)}\frac{\delta B(x')}{\delta \Pi_i(y)} - \frac{\delta A(x)}{\delta \Pi_i(y)}\frac{\delta B(x')}{\delta \Phi_i(y)} \right). \tag{C.53}$$

Using Eq. (C.49), written as $\{f, g\} = \partial_x f \partial_y g - \partial_y f \partial_x g$, we obtain

$$\vec{n}.(\partial_x \vec{n} \times \partial_y \vec{n}) = \frac{1}{2}\varepsilon_{abc} n_a \{n_b, n_c\}. \tag{C.54}$$

Thus, the topological charge for a skyrmon can be written as

$$Q = \frac{1}{8\pi}\int d^2 r \varepsilon_{abc} n_a \{n_b, n_c\}\,. \tag{C.55}$$

We have also

$$\vec{n}.(\partial_x \vec{n} \times \partial_y \vec{n}) = \sin\theta(\partial_x\theta\partial_y\phi - \partial_y\theta\partial_x\phi) = \sin\theta\{\theta,\phi\}\,. \tag{C.56}$$

Note that $\frac{\partial(\theta,\phi)}{\partial(x,y)} = \{\theta,\phi\}$ is the Jacobian determinant of the map $\mathbb{R}^2 \to S^2$.

I will now show how to obtain the equation of motion for the collective coordinates, introduced in Sect. 12.1, using the Hamiltonian approach.

Let the position of a rigid magnetic texture (e.g. a skyrmion) be described by collective coordinates $\vec{\xi} = (\xi_1, \xi_2, ..., \xi_N)$. For a skyrmion in two dimensions, these are usually $\xi_1 = X,\ \xi_2 = Y.$ The effective Lagrangian can be written as

$$L = A_i(\vec{\xi})\dot{\xi}_i - H(\vec{\xi}), \tag{C.57}$$

where $A_i$ is the Berry-phase vector potential. The canonical coordinates may be taken as collective coordinates: $q_i = \xi_i,\ p_i = A_i(\vec{\xi})$ [60]; The Euler-Lagrange equation gives

$$\frac{d}{dt}\frac{\partial L}{\partial \dot{\xi}_i} - \frac{\partial L}{\partial \xi_i} = 0, \tag{C.58}$$

Since $\frac{\partial L}{\partial \dot{\xi}_i} = A_i$, we have $\frac{dA_i}{dt} = \frac{\partial A_i}{\partial \xi_j}\dot{\xi}_j$. (C.59)

Also

$$\frac{\partial L}{\partial \xi_i} = \frac{\partial A_j}{\partial \xi_j}\dot{\xi}_j - \frac{\partial H}{\partial \xi_i}\,. \tag{C.60}$$

Therefore

$$\left(\frac{\partial A_i}{\partial \xi_j} - \frac{\partial A_j}{\partial \xi_i}\right)\dot{\xi}_j = \frac{\partial H}{\partial \xi_i}\,. \tag{C.61)}$$

We define the gyrocoupling tensor as

$$G_{ij} = \frac{1}{S}\left(\frac{\partial A_i}{\partial \xi_j} - \frac{\partial A_j}{\partial \xi_i}\right). \tag{C.62}$$

From Eq. (C.61) we get

$$G_{ij}\dot{\xi}_j = \frac{1}{S}\frac{\partial H}{\partial \xi_i}\,. \tag{ C.63}$$

The tensor $G_{ij}$ is antisymmetric: $G_{ij} = -G_{ji}$. If it is invertible, we multiply (C.63) by $(G^{-1})_{ij}$ and obtain

$$\dot{\xi}_i = (G^{-1})_{ij}\frac{1}{S}\frac{\partial H}{\partial \xi_j}\,. \tag{C.64}$$

Now using $\dot{\xi}_i = \{\xi_i, H\}$, and the chain rule

$$\{\xi_i, H\} = \{\xi_i, \xi_j\}\frac{\partial H}{\partial \xi_j}, \tag{C.65}$$

we get

$$\{\xi_i, \xi_j\}\frac{\partial H}{\partial \xi_j} = (G^{-1})_{ij}\frac{\partial H}{\partial \xi_j}.$$

(C.66)

Since this must hold for an arbitrary *H*, we obtain

$$\{\xi_i, \xi_j\} = (G^{-1})_{ij}. \tag{C.67}$$

In addition to the energy conservation part, we can introduce a phenomenological damping term directly into the Euler-Lagrange equation using a Rayleigh dissipation potential [60].

For the commonly used gauge choice

$$A_i(\vec{\xi}) = \int d^2r(1-\cos\theta)\frac{\partial \phi}{\partial \xi_i}, \tag{C.68}$$

the gyrotropic tensor is

$$G_{ij} = \int d^2r \vec{n}.(\partial_{\xi_i}\vec{n} \times \partial_{\xi_j}\vec{n}). \tag{C.69}$$

For a translationally invariant skyrmion with $\vec{\xi} = (X, Y)$, the gyrocoupling tensor has the form

$$G = \begin{pmatrix} 0 & G \\ -G & 0 \end{pmatrix}, \tag{C.70}$$

with $G = G_{XY}$. Thus

$$\{X, Y\} = \frac{1}{G}. \tag{C.71}$$

Or with a minus sign depending on the convention used for the definition of $G_{ij}$.

Using the relation

$$\{n_i(\vec{x}), n_j(\vec{x}\,')\} = -\frac{1}{S}\sum_k \varepsilon_{ijk} n_k(\vec{x})\delta(\vec{x} - \vec{x}\,'), \tag{C.72}$$

the time evolution of the component $n_l(\vec{x})$ is given by

$$\begin{aligned} \dot{n}_l = \{n_l(\vec{x}), H\} &= \sum_{ij}\int d\vec{y}d\vec{z}\frac{\delta n_l(\vec{x})}{\delta n_i(\vec{z})}\frac{\delta H}{\delta n_j(\vec{y})}\{n_i(\vec{z}), n_l(\vec{y})\} \\ &= -\frac{1}{S}\sum_{jk}\varepsilon_{ljk}\frac{\delta H}{\delta n_j(\vec{x})}n_k(\vec{x}) = \frac{1}{S}\left[\vec{n}(\vec{x}) \times \frac{\delta H}{\delta \vec{n}(\vec{x})}\right]_l. \end{aligned} \tag{C.73}$$

This is the Bloch equation.

# Appendix D. Quantum antiferromagnet.

I will start with a one-dimensional antiferromagnetic spin chain with an even number of sites, $N$, under periodic boundary conditions. Its generalization to two dimensions is straightforward, except for the topological term.

$$H = J\sum_{j=1}^{N}\vec{S}_j.\vec{S}_{j+1}. \tag{D.1}$$

I will consider an antiferromagnet with a bipartite lattice and nearest-neighbor interactions. In the classical limit of large $S$, as we have seen, the ground state has spins oriented in opposite directions on the two sublattices. For smaller $S$ this orientation should survive, at least over a few lattice spacings, such that a continuum description of the quantum antiferromagnet may apply. The results presented here can be easily generalized to any dimensions and other types of lattices. Using the result of Appendix C, we have

$$\langle\vec{m}|H|\vec{m}\rangle = J\sum_j\langle\vec{m}|\vec{S}_j.\vec{S}_{j+1}|\vec{m}\rangle = JS^2\sum_j\vec{m}_j.\vec{m}_{j+1} \tag{D.2 }$$

We divide $\vec{m}$ into a slowly varying part $\vec{n}$ and a small but fast fluctuation part $\vec{l}$:

$$\vec{m}_j = (-1)^j\sqrt{1-a^2\vec{l}_j^{\,2}}\,\vec{n}_j + a^d\vec{l}_j. \tag{D.3}$$

Here $a$ is the lattice constant, and $d$ is the dimension of the system, which is $d = 1$, for the case treated here. This construction works only for two-sublattice systems close to a Néel state. Here $\vec{n}_j$ is the order parameter field. Although $\vec{n}_j$ varies slowly on the scale of a lattice spacing, values of $\vec{n}_j$ at distant points can differ considerably, so a quantum paramagnetic phase with no magnetic order can exist. The square root factor in (D.3) is to preserve the normalization of $\vec{m}$ to order $a^2$. We drop it in what follows since it always gives order $a^3$ corrections. The constraint

$$\vec{m}_j^2 = \vec{n}_j^2 = 1, \tag{D.4}$$

demands that $\vec{n}_j$ and $\vec{l}_j$ be orthogonal vectors: $\vec{n}_j.\vec{l}_j = 0$ (so, $\vec{l}$ is the magnetization density field perpendicular no $\vec{n}$). We also assume $\vec{l}_j^{\,2} << a^{-2}$.

For small lattice spacing, we can write

$$\vec{n}_{j+1} \approx \vec{n}_j + a\partial_x\vec{n}_j, \qquad \vec{l}_{j+1} \approx \vec{l}_j + a\partial_x\vec{l}_j, \tag{D.5}$$

and take $\sqrt{1-a^2\vec{l}_j^{\,2}} \approx 1 - a^2\vec{l}_j^{\,2}/2$.

Using Eq. (D.3), we get

$$\vec{m}_j = (-1)^j\vec{n}_j + a\vec{l}_j \tag{D.6}$$

$$\vec{m}_{j+1} = (-1)^{j+1}\vec{n}_{j+1} + a\vec{l}_{j+1} = (-1)^{j+1}[\vec{n}_j + a\partial_x\vec{n}_j] + a(\vec{l}_j + a\partial_x\vec{l}_j) \tag{D.7}$$

We can write

$$\vec{m}_j\cdot\vec{m}_{j+1} = \frac{1}{2}(\vec{m}_j + \vec{m}_{j+1})^2 - 2, \tag{D.8}$$

which leads to

$$\vec{m}_j + \vec{m}_{j+1} = (-1)^j\vec{n}_j + (-1)^{j+1}\vec{n}_j + (-1)^{j+1}a\partial_x\vec{n}_j + 2a\vec{l}_j + a^2\partial_x\vec{l}_j. \tag{D.9}$$

But

$$(-1)^j \vec{n}_j + (-1)^{j+1} = 0, \qquad \text{and} \quad a^2 \partial_x \vec{l}_j \to 0. \tag{D.10}$$

We have

$$[\vec{m}_j - \vec{m}_{j+1}]^2 - 2 = [2a\vec{l}_j - (-1)^j a\partial_x \vec{n}_j]^2 = 4a^2\vec{l}_j^{\,2} + a^2\vec{n}_j\partial_x^2\vec{n}_j - 4(-1)^j a^2\vec{l}_j\partial_x\vec{n}_j. \tag{D.11}$$

The last term vanishes due to the alternating sum. Using now Eq. (D.8), and replacing the lattice sums by integrals $\left(\sum_j \to \int dx/a\right)$, we obtain the continuum expression (up to an additive constant):

$$Z = \int D\vec{n} \exp\left\{S_B - \frac{JS^2}{2}\int dx \int_0^\beta d\tau a\left[4\vec{l}^{\,2} + (\partial_x\vec{n})^2\right]\right\}. \tag{D.12}$$

To proceed, we consider the Berry term. Substituting (D.7) into (C.35) and keeping terms only up to order *a*, we find

$$S_B = S' + iS\int_0^\beta d\tau \int dx \int_0^1 du\left\{\vec{n}.(\partial_u\vec{n}\times\partial_\tau\vec{l}) + \vec{n}.(\partial_u\vec{l}\times\partial_\tau\vec{n}) + \vec{l}.(\partial_u\vec{n}\times\partial_\tau\vec{n})\right\}, \tag{D.13}$$

where

$$S' = iS\sum_j (-1)^j \int_0^\beta d\tau \int_0^1 du\vec{n}_j.(\partial_u\vec{n}_j\times\partial_\tau\vec{n}_j)\cdot \tag{D,14}$$

Which can be approximated by

$$S' = iS\sum_j \int_0^\beta d\tau \int_0^1 du(\vec{n}_{2j} - \vec{n}_{2j-1}).\partial_u\vec{n}_{2j}\times\partial_\tau\vec{n}_{2j}\,\cdot \tag{D.15}$$

Using

$$\sum_j (f_{2j} - f_{2j-1})F_{2j} \approx \int \frac{dx}{a}F(x)a\partial_x f(x), \tag{D.16}$$

we obtain

$$S' = -i\frac{S}{2}\int d\tau dx\vec{n}.(\partial_x\vec{n}\times\partial_\tau\vec{n})\cdot \tag{D.17}$$

The vectors $\vec{l}$, $\partial_u\vec{n}$ and $\partial_\tau\vec{n}$ are all perpendicular to $\vec{n}$, and so have a vanishing triple product. Hence, we can write

$$S_B = S' + iS\int dsdud\tau\{\partial_\tau(\vec{n}.\partial_u\vec{n}\times\vec{l}) - \partial_u(\vec{n}.\partial_\tau\vec{n}\times\vec{l})\}. \tag{D.18}$$

All the fields are periodic in $\tau$, so the total $\tau$ derivative vanishes. The total *u* derivative gives a surface contribution at *u* = 1. Hence, we can write

$$S_B = S' - iS\int dxd\tau\vec{l}.(\vec{n}\times\partial_\tau\vec{n}). \tag{D.19}$$

Inserting (D.19) into (D.12) we find

$$S = S' + \int dx\int_0^\beta d\tau\left[-i\frac{S}{2}\vec{n}.(\partial_x\vec{n}\times\partial_\tau\vec{n}) + \frac{JS^2}{2}(\partial_x\vec{n})^2 - S\vec{l}.(\vec{n}\times\partial_\tau\vec{n}) + 2JS^2\vec{l}^{\,2}\right]. \tag{D.20}$$

Since we are interested in the lo ng-range order, we integrate over the fast fluctuation component $\vec{l}$. To show how to do that, suppose we want to integrate the following expression using the saddle point method

$$I=\int_{-\infty}^{\infty}e^{if(x)}dx\,. \tag{D.21}$$

The saddle point is given by $f'(x)\big|_{x=x_0}=0$. $f(x)$ changes slowly around this point. We write

$$f(x)\approx f(x_0)+\frac{1}{2}\frac{\partial^2 f}{\partial x^2}\Big|_{x=x_0}(x-x_0)^2\,. \tag{D.22}$$

$$I=e^{if(x_0)}\int_{-\infty}^{\infty}dx e^{\frac{i}{2}f''(x_0)(x-x_0)^2}=e^{if(x_0)}\sqrt{\frac{2\pi i}{f''(x_0)}}$$

Now we take

$$f(\vec{l})=-S\vec{l}.(\vec{n}\times\partial_\tau\vec{n})+2JS^2\vec{l}^{\,2}\,. \tag{D.23}$$

Let $c=2JS$ , $\vec{b}=\vec{n}\times\partial_\tau\vec{n}$ .

Then

$$f(\vec{l})=-c\vec{l}^{\,2}+\vec{l}.\vec{b}S\,,\qquad f'(\vec{l})=0\quad\rightarrow\quad\vec{l}_0=S\vec{b}/2c\,,$$

which gives

$$f(\vec{l}_0)=-\frac{S^2}{2c}\vec{n}\times\partial_\tau\vec{n}.(\vec{n}\times\partial_\tau\vec{n})+2JS^2\left(\frac{S}{2c}\vec{n}\times\partial_\tau\vec{n}\right)^2\,. \tag{D.24}$$

Using the identity $(\vec{n}\times\partial_\tau\vec{n})^2=(\partial_\tau\vec{n})^2$, we get

$$Z=\sqrt{\frac{4\pi}{c}}\int D\vec{n}\exp\left\{-\int dx\int_0^\beta \mathscr{L}(x,\tau)\right\}=\sqrt{\frac{4\pi}{c}}\int D\vec{n}e^{-S_{eff}}\,. \tag{D.25}$$

Where $S_{eff}$ is the effective long-wavelength action that retains the full rotation symmetry, and the Lagrangian density of the nonlinear sigma model is given by

$$\mathscr{L}(x,\tau)=\frac{c^2}{2g}\left[\frac{1}{c^2}(\partial_\tau\vec{n})^2+(\partial_x\vec{n})^2\right]-i\frac{S}{4}\varepsilon^{\mu\nu}\vec{n}.(\partial_\mu\vec{n}\times\partial_\tau\vec{n})\,. \tag{D.26}$$

The Lagrangian is quadratic in the field $\vec{n}$, but it does not describe a free field theory because of the constraint $\vec{n}^2=1$ imposed at each point in spacetime.

It can be shown that the last term in Eq. (D.26) can be written as

$$S'=i2\pi SQ\,, \tag{D.27}$$

where $Q$ is an integer. We have $e^{i2\pi SQ}=(-1)^{2SQ}$. Therefore, if $S$ is an integer, the term can be omitted from the path integral, and the antiferromagnetic spin chain is described at low energies by the nonlinear sigma model. For half-integer spins $\exp(-S')=(-1)^Q$ ,and the situation is more complicated [86]. The continuum theory in the presence of alternating Berry phases is hard to solve. The integer- and half-integer-spin antiferromagnetic spin chains belong to different universality classes. In two spatial dimensions, it has been shown that $S'$ is identically zero for all smooth configurations [86]. However, there are important singular configurations of $\vec{n}$ that do yield a nonvanishing contribution to $S'$.

Thus, the density Lagrangian for the two-dimensional antiferromagnet is given by

$$\mathscr{L} = \frac{1}{2g}\left[\frac{1}{v}\left(\frac{\partial \vec{n}}{\partial t}\right)^2 - v(\nabla\vec{n})^2\right], \tag{D.28}$$

with $v = 2JS$, $g = 2/S$. For a two-dimensional antiferromagnet, the Berry phase of the two sublattices cancels at leading order. The model has an intrinsic dynamic that is inertial, not precessional.

Consider the antiferromagnet in the presence of an external field $\vec{B}$. It couples with the field $\vec{l}$. It easy to show that the Lagrangian density is given by

$$\mathscr{L} = \frac{v^2}{2g}\left[\frac{1}{v^2}\left(\partial_t\vec{n} - \gamma\vec{B}\times\vec{n}\right)^2 - (\nabla\vec{n})^2\right]. \tag{D.29}$$

Expanding the square term in Eq. (D.29) gives

$$\dot{\vec{n}}^2 - 2\gamma\vec{B}.(\vec{n}\times\dot{\vec{n}}) + \gamma^2(\vec{B}\times\vec{n})^2. \tag{D.30}$$

The second term is linear in the time derivative and acts like a vector potential. It breaks time-reversal symmetry and can generate forces analogous to Lorentz or Magnus forces on magnetic textures. The last term is a field-induced anisotropy and favors configurations $\vec{n}\,//\,\vec{B}$.

For a ferrimagnet with spin $S_A$ in sublattice $A$ and spin $S_B$ in sublattice $B$, we can follow the same steps as were done for the antiferromagnet [45]. We get a term like Eq. (D.28), with $g' = 4/(S_A+S_B)$ and $v\text{l} = 4JS_AS_B/(S_A+S_B)$. But now the Berry density Lagrangian does not vanish. It is given by

$$\mathscr{L}_B = (S_A - S_B)\vec{A}(\vec{n}).\dot{\vec{n}}\,. \tag{D.31}$$

With the action

$$S_B = (S_A - S_B)\int dt(1-\cos\theta)\dot{\phi}\,. \tag{D.32}$$

The Berry term generates the Magnus force, while the quadratic kinetic term generates inertia. Together, inertia and the Magnus force play important roles in determining the ultrafast dynamics and transport properties of ferrimagnetic materials, making them attractive for next-generation spintronic devices. Near the ferrimagnetic compensation point, where the angular momenta of the sublattices balance, the Magnus force can be significantly reduced, allowing more linear and controllable motion of magnetic textures.

# Appendix E. Torques.

## Spin transfer torque.

Every electron generates a local magnetic field $\vec{B}$ with magnetic dipole momentum $\vec{\mu}$ given by $\vec{\mu} = \gamma\vec{s}$, where $\gamma$ is the gyromagnetic ratio ($\gamma = g_e\mu_B$, with $g_e$ the electron $g$ factor and $\mu_B$ the Bohr magneton). A second electron feels this magnetic field, and this field exerts a torque $\vec{\tau} = -\vec{\mu}\times\vec{B}$ on the electron. When an electric current flows through a ferromagnet, it moves more easily when the electrons' spins align with the material's magnetization. However, the current can also disturb magnetization. When an electron with misaligned spin enters a magnetized layer, the mismatch creates a small twisting interaction between its spin and the bulk magnetic moment. This interaction is known as *spin-transfer torque*. The Zhang-Li spin-transfer torque [33] arises from the interaction between itinerant conduction-electron spins and the slowly varying local magnetization. The central idea is that when a spin-polarized current flows through a nonuniform magnetic texture (such as a domain wall or a skyrmion), the electron spins tend to align with the local magnetization. If they follow perfectly, one obtains the adiabatic torque. If the spin-relaxation time is finite, the electron spins lag behind the magnetization, producing a nonadiabatic torque (Fig. E.1). Spin-transfer torque enables control of magnetization with electric current rather than magnetic fields, enabling faster, smaller, and more energy-efficient memory technologies.

The conduction electrons (in general, *s* electrons) interact with the localized magnetic moment (*d* electrons) through the Hamiltonian

$$H_{sd} = -\frac{J_{sd}}{\hbar}\vec{s}.\vec{m}(\vec{r},t), \tag{E.1}$$

where $\vec{s}$ is the conduction electron spin, $\vec{m} = \vec{M}/M_S$ is the unit magnetization (localized electrons), and $J_{sd}$ is the exchange coupling strength. The spin continuity equation for the conduction electrons is

$$\frac{\partial\vec{s}}{\partial t} + \nabla\vec{J}_s = -\frac{J_{sd}}{\hbar}\vec{s}\times\vec{m} + \vec{\Gamma}. \tag{E.2}$$

where $\vec{J}_S$ is the spin current operator, and $\vec{\Gamma}$ is the spin relaxation. We define the electron spin density $\vec{n}(\vec{r},t) = <\vec{s}>$ and the spin current density $\vec{\Im}(\vec{r},t) = <\vec{J}>$, where <.> means the average over all occupied electronic states. We have

$$\frac{\partial\vec{n}}{\partial t} + \vec{\nabla}\vec{\Im} = -\frac{\vec{n}\times\vec{m}}{\tau_{ex}} - <\vec{\Gamma}>. \tag{E.3}$$

Where $\tau_{ex} = \hbar / J_{sd}$. Note that, while the charge current is a vector, the spin current $\vec{\Im}$ is a tensor. i.e.

$$\frac{\partial\vec{n}}{\partial t} + \vec{\nabla}\vec{\Im} = \frac{\partial n_i}{\partial t} + \partial_j\Im_{ij}. \tag{E.4}$$

The first index tells us which component is carried, and the second tells us in which direction it moves. For instance, $\Im_{xz}$ means *x*-polarized spin flowing in the *z*-direction.

We decompose the induced spin density as

$$\vec{n}(\vec{r},t) = \vec{n}_0(\vec{r},t) + \delta\vec{n}(\vec{r},t). \tag{E.5}$$

Where $\vec{n}_0$ is the local equilibrium spin density whose direction is parallel to the magnetization). Using a simple relaxation time approximation, we write $<\vec{\Gamma}> = \delta\vec{n}/\tau_{sf}$, where $\tau_{sf}$ is the spin-flop relaxation time. The spin current caused by the electrons is given by

$$\vec{\vec{\Im}} = \vec{u} \otimes \vec{m}, \qquad \text{where} \qquad \vec{u} = \frac{\mu_B P}{eM_S}\vec{j}_C . \tag{E.6}$$

Here, $e$ is the electron charge, $\vec{j}_C$ the current density, and $P$ the spin polarization (see end of this section for a calculation of $\vec{u}$). We have

$$\vec{\nabla}\vec{\vec{\Im}}_s - \vec{\nabla}.(\vec{u} \otimes \vec{m}) = (\vec{\nabla}.\vec{u})\vec{m} - (\vec{u}.\vec{\nabla})\vec{m} . \tag{E.7}$$

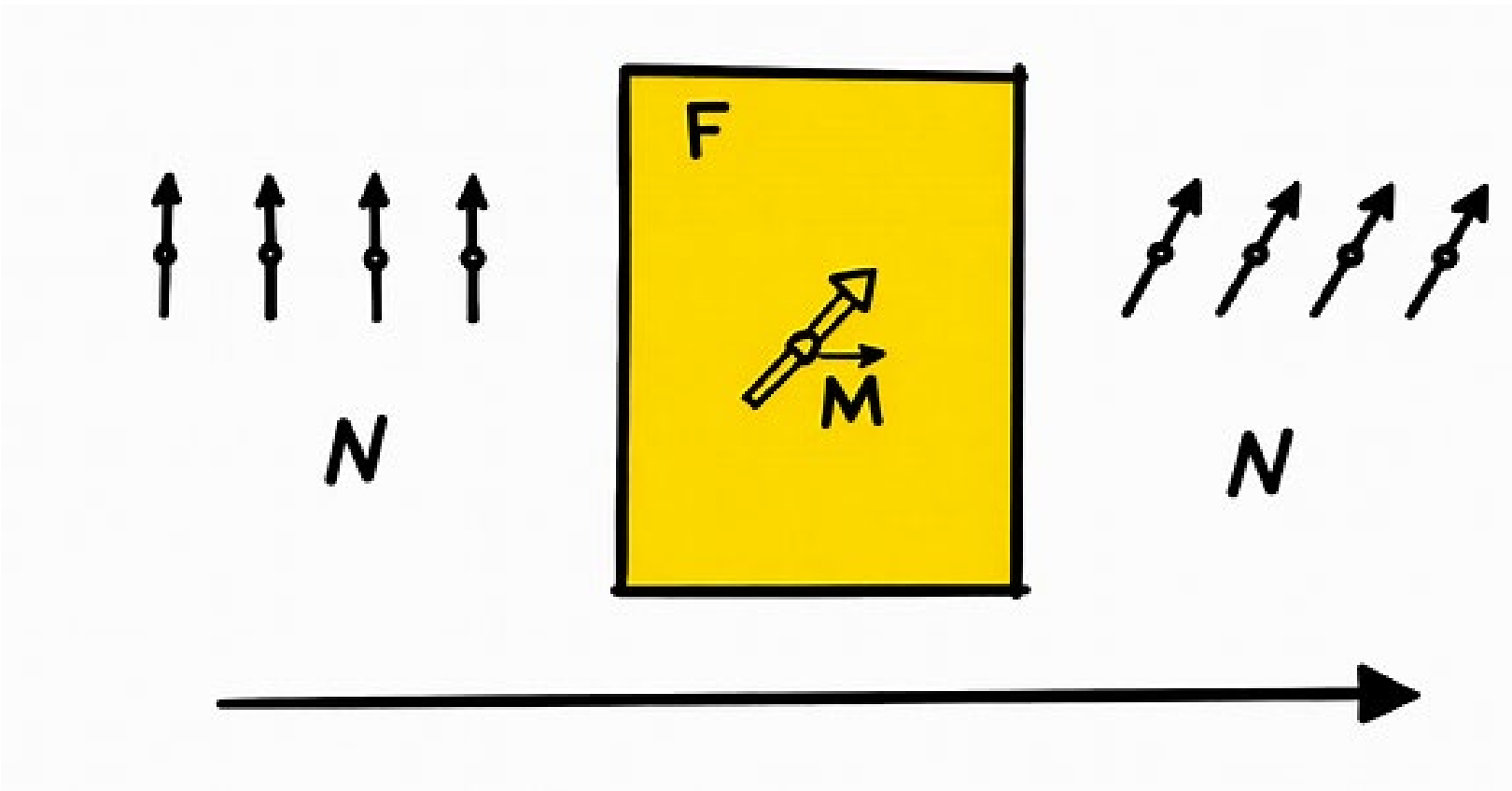

Figure E1. Spin-transfer torque occurs when a current of spin-polarized carriers from the left enters a ferromagnet with a different magnetization direction. The electrons eventually become polarized along the magnetization, meaning the ferromagnet has exerted a torque on them, rotating their spin angular momentum. Due to conservation of angular momentum, the electrons exert an equal but opposite torque on the right magnetic domain, the spin transfer torque. In an experimental setup, the charge current first passes through a ferromagnet on the left-hand side, where it becomes polarized.

If the current is uniform, we have $\vec{\nabla}.\vec{u} = 0$. For slow dynamics and steady state

$$\frac{\partial \delta \vec{n}}{\partial t} \approx 0 . \tag{E.8}$$

Inserting Eq.(E.5) into Eq. (E.3), we get

$$(\vec{u}.\vec{\nabla})\vec{m} = -\omega_e \delta\vec{n} \times \vec{m} - \frac{\delta\vec{n}}{\tau_{sf}} . \tag{E.9}$$

Where $\omega_e = J_{sd} / \hbar$. Let

$$\vec{A} = (\vec{u}.\vec{\nabla})\vec{m} . \tag{E.10}$$

Thus

$$\vec{A} = -\omega_e \delta\vec{n} \times \vec{m} - c\delta\vec{n} . \tag{E.11}$$

Where $c = 1/\tau_{sf}$. Since $\delta\vec{n}$ is perpendicular to $\vec{m}$, we have $\delta\vec{n}.\vec{m} = 0$. Also $\vec{m}$ is a unit vector, so $(\delta\vec{n} \times \vec{m}) \times \vec{m} = -\delta\vec{n}$. Multiplying Eq. (E.11) by $\vec{m}$ (cross product), we get

$$\vec{A} \times \vec{m} = -\omega_e (\delta\vec{n} \times \vec{m}) \times \vec{m} - c\delta\vec{n} \times \vec{m} = \omega_e \delta\vec{n} - c\delta\vec{n} \times \vec{m} . \tag{E.12}$$

We use Eqs. (E.11) and (E.12) to eliminate $\delta\vec{n}$. Multiplying Eq. (E.11) by *c* and Eq. (E.12) by $\omega_e$ and subtracting, we find

$$\omega_e(\vec{A}\times\vec{m}) - c\vec{A} = (\omega_e^2 + c^2)\delta\vec{n}\ . \tag{E.13}$$

Hence

$$\delta\vec{n} = \frac{\omega_e\vec{A}\times\vec{m} - c\vec{A}}{\omega_e^2 + c^2}\ . \tag{E.14}$$

Using now Eq. (E.10)

$$\delta\vec{n} = \frac{1}{\omega_e^2 + c^2}[\omega_e((\vec{u}.\vec{\nabla})\vec{m}\times\vec{m} - c(\vec{u}.\vec{\nabla})\vec{m}]\ . \tag{E.15}$$

The magnetization feels the opposite exchange torque

$$\vec{T} = \frac{J_{sd}}{\hbar}\vec{m}\times\delta\vec{n}\ . \tag{E.16}$$

We have $\vec{m}\times(\vec{m}\times(\vec{u}.\vec{\nabla})\vec{m} = -(\vec{u}.\vec{\nabla})\vec{m}$, since $\vec{m}.(\vec{u}.\vec{\nabla})\vec{m} = 0$ . Let $\beta = \hbar/(J_{sd}\tau_{sf})$. Since experimentally $\beta << 1$, we can write

$$\vec{T} = -(\vec{u}.\vec{\nabla})\vec{m} + \beta\vec{m}\times(\vec{u}.\vec{\nabla})\vec{m}\ . \tag{E.17}$$

The term $-(\vec{u}.\vec{\nabla})\vec{m}$, the adiabatic torque, arises from perfect spin tracking and conserves spin angular momentum. The nonadiabatic torque $\beta\vec{m}\times(\vec{u}.\vec{\nabla})\vec{m}$ arises from finite spin relaxation and reflects the lag of electron spins behind the local magnetizations. If $\tau_{sf}\to\infty$ (no spin relaxation), $\beta\to 0$.

The spin velocity $\vec{u}$ is not a physical velocity of the electrons. Rather, it is an effective drift velocity of the spin angular momentum carried by the conduction electrons. The spin angular momentum of an electron is $\hbar/2$. The spin current is the difference between the spin-up and spin-down charge current converted into angular momentum flow, i. e.

$$\vec{J}_S = \frac{\hbar}{2e}(\vec{J}_\uparrow - \vec{J}_\downarrow)\ . \tag{E.18}$$

We can also write

$$\vec{J}_S = \frac{P\hbar}{2e}\vec{J}_C\ , \tag{E.19}$$

Where *P* is a dimensionless parameter between 0 and 1 that tells us how spin-polarized the conduction electrons are:

$$P = \frac{J_\uparrow - J_\downarrow}{J_\uparrow + J_\downarrow}\ . \tag{E.20}$$

It measures the imbalance between spin-up and spin-down currents. If *P* = 0, there is no spin polarization. If *P* = 1, the current is fully spin-polarized.

The magnetization density stores angular momentum $M_S/\gamma$. To convert the spin current into an equivalent velocity of the magnetic texture, we divide the spin current by the angular momentum density:

$$\vec{u} = \frac{\vec{J}_S}{M_S / \gamma} = \frac{\gamma P\hbar}{2eM_S}\vec{J}_c = \frac{Pg\mu_B}{2eM_S}\vec{J}_c . \tag{E.21}$$

Where $g \approx 2$.

## Spin-orbit torque

Spin–orbit torque (SOT) arises from the interaction between an electron's spin and its orbital motion, and it can be decomposed into two distinct components: *damping-like* SOT and *field-like* SOT [90-92]. These two contributions arise from distinct microscopic mechanisms and influence magnetization dynamics differently. The damping-like SOT is primarily associated with the spin Hall effect (SHE) in heavy metals. Therefore, I will briefly discuss the spin Hall effect.

The charge current density is related to the applied electric field $\vec{E}$ through Ohm's law

$$J_{C,i} = \sigma_{ij} E_j , \tag{E.22}$$

where $\vec{J}_C$ is the charge current. When an electric field is applied, electrons acquire an average momentum $\vec{p} = -e\tau\vec{E}$ , where $\tau$ is the momentum relaxation time. Without spin-orbit coupling, electrons with opposite spin orientations experience the same motion under the applied electric field, and no transverse spin current is generated. Spin-orbit coupling provides the interaction that links an electron's spin to its orbital motion. In its rest frame, an electron moving through the crystal-lattice electric field experiences an effective magnetic field that interacts with its magnetic moment (spin). The spin-orbit coupling Hamiltonian is

$$H_{SO} = \frac{\hbar}{4m^2c^2}(\vec{\nabla}V \times \vec{p}).\vec{\sigma} , \tag{E.23}$$

where *m* is the electron mass, *c* is the speed of light, $V(\vec{r})$ is the crystal potential (arising from different sources, such as impurities, host atoms, and structural confinement [92]), and $\vec{\sigma}$ is the vector of Pauli spin matrices. The term

$$\vec{B}_{eff} \propto \vec{\nabla}V \times \vec{p} , \tag{E.24}$$

acts as an effective magnetic field whose direction depends on the electron momentum. Consequently, electrons with opposite spins experience opposite transverse forces. Because of the SOC, the effective magnetic field becomes momentum-dependent. Electrons with spin-up and spin-down are deflected in opposite transverse directions, producing a pure spin current. The current charge vanishes. Since a pure spin current has no net charge flow, we cannot define its "sign" as we do for an electric current. Note that a spin-down going to the left is equivalent to a spin-up going to the right. The direction of the spin flux sets the sign of the spin polarization in a pure spin current: If spin-up carriers flow in the + $x$ direction while spin-down carriers flow in the - $x$ direction, the spin current $J_{c,x}$ is defined as positive. If the flows reverse, the spin current is negative. Different authors define the sign in slightly different but equivalent ways: In spin Hall effect experiments, the sign is determined by which edge accumulates spin-up vs. spin-down carriers. Pure spin currents produce opposite spin accumulation at opposite edges. We have

$$\vec{J}_S = \theta_{SH} \frac{\hbar}{2e} \vec{J}_c \times \vec{\sigma} . \tag{E.25}$$

Where $\theta_{SH}$ is the spin Hall angle and $\vec{\sigma}$ denotes the spin polarization direction. The sign depends on the sign of the spin Hall angle $\theta_{SH}$. The spin Hall angle (it is basically the efficiency of charge-to-spin current conversion) is defined as

$$\theta_{SH} = \frac{2eJ_S}{\hbar J_c} . \tag{E.26}$$

Although spins separate laterally, the number of electrons moving left and right remains equal, so the transverse charge current is zero. Instead, spins accumulate at the sample edges.

Field-like spin–orbit torque (SOT) originates from the Rashba–Edelstein effect, which appears at interfaces where inversion symmetry is broken. When an electric current passes through such an interface, the Rashba spin–orbit coupling causes charge carriers to experience a momentum-dependent spin splitting. As a result, the current drives a non-equilibrium spin accumulation at the interface. This interfacial spin density acts as an effective magnetic field—often called the *Rashba field*. Because the accumulated spins align with the Rashba interaction, they exert a torque on the adjacent ferromagnet's magnetization. The outcome is a field-like SOT whose magnitude and direction depend on the strength of the Rashba coupling, the current density, and the details of the interfacial electronic structure.

The study of the SOT requires a lengthy discussion [92], so I will only present a brief overview of the physics involved. A Hamiltonian that includes spin-orbit coupling is

$$H = \frac{p^2}{2m} + V(\vec{r}) + \frac{J_{ex}}{\hbar} \frac{\vec{M}.\vec{s}}{M_S} + \frac{\alpha}{\hbar^2} (\vec{L}.\vec{s}) , \tag{E.27}$$

where $J_{ex}$ is the exchange interaction with the ferromagnet, $\vec{M}$ the magnetization with strength $M_S$, and $\vec{L}$ and $\vec{s}$ the electron angular momentum and spin operators. As shown above, in the absence of spin-orbit coupling, the torque relates to the divergence of a spin current. However, spin-orbit coupling changes this simple result [92]. Solving the electronic transport problem (using the Boltzmann equation or the Kubo linear response formalism) yields the nonequilibrium spin density $\delta\vec{s}$, and the torque follows as

$$T_{SOT} = \frac{J_{ex}}{\hbar} \vec{m} \times \delta\vec{s} \ . \tag{E.28}$$

The nonequilibrium spin density decomposes into components that generate the field-like torque $\vec{m} \times \vec{\sigma}$ and the damping-like torque $\vec{m} \times (\vec{m} \times \vec{\sigma})$, where $\vec{\sigma}$ is the spin polarization. Thus

$$T_{SOT} = \tau_{FL} \vec{m} \times \vec{\sigma} + \tau_{DL} \vec{m} \times (\vec{m} \times \vec{\sigma}) . \tag{E.29}$$

Where

$$\tau_{DL} = \frac{\gamma\hbar\theta_{SH}J_C}{2edM_S}, \tag{E.30}$$

With $\theta_{SH}$ the spin Hall angle, and $d$ the thickness of the ferromagnet.

Note that the spin-transfer torque depends on the spatial gradient of the magnetization. If the magnetization is uniform, there is no bulk spin-transfer torque. This torque is strongest for domain walls, skyrmions, and vortices. On the other hand, the spin-orbit torque depends on the spin polarization $\vec{\sigma}$. It requires no magnetic texture and can act even on a uniformly magnetized ferromagnet.

In an experiment, a charge current $\vec{J}_C$ flows through a heavy metal (like platinum). Inside the heavy metal, strong spin-orbit coupling produces a transverse spin current $\vec{J}_S$ (that flows vertically and carries no electric charge) through the spin Hall effect (or in some cases the Rashba-Edelstein effect). That creates a spin accumulation at the heavy-metal/ferromagnet interface. The transverse spin current is fully polarized, so we have

$$v_S^{\perp} = \frac{\gamma\hbar}{2eM_S}J_S = \frac{\gamma\hbar|\theta_{SH}|}{2eM_S}J_c. \tag{E.31}$$

The normalized vertical spin current is always positive by definition [37].

A spin current enters the ferromagnet (for instance, CoFeB) and interacts with the localized magnetic moments in the ferromagnet through the exchange interaction. The charge current remains in the heavy metal, flows horizontally, and does not enter the ferromagnetic layer. Although spin-orbit coupling originates in the heavy metal, the resulting torque acts on the ferromagnet's magnetization. The magnetization absorbs part of the spin current, transferring angular momentum to the ferromagnet (the ferromagnet experiences the spin-orbit torque, not the heavy metal, although the heavy metal generates the spin polarization). SOT devices require efficient spin-current generation in the heavy-metal layer and efficient spin transmission across the heavy-metal/ferromagnet interface.

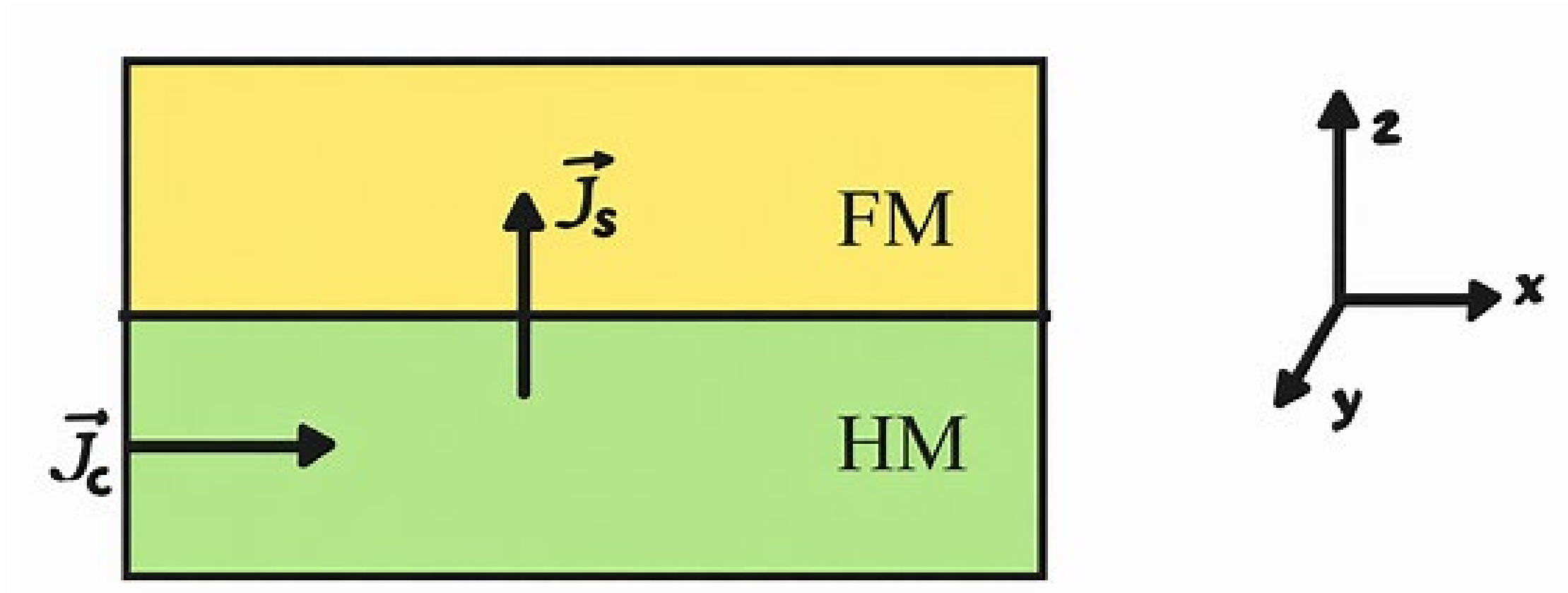

Figure E.2. When an electric current $J_C$ is injected into a heavy-metal layer, it generates a pure spin current $J_S$ flowing perpendicular to the interface, going into the ferromagnet. The current $J_S$ is caused by the spin Hall effect due to the spin-orbit coupling. The polarization $\vec{\sigma}$ is along the $y$–direction.

A third kind of torque is the *interfacial spin-transfer torque* (ISTT). It is the torque produced when spin angular momentum carried by a spin current or spin accumulation is transferred to a ferromagnet at a material interface, causing its magnetization to rotate or switch**.** The three torques are closely related, and many papers present similar equations because they all describe angular momentum transfer to magnetization**.** The main difference is where the spin polarization comes from and how the torque coefficient is written.

## Appendix F. Fiber bundles quaternions and Hopf map.

**Fiber bundles** provide the natural mathematical language for describing the topology, geometry, and field structure of hopfions. Although the formalism is not strictly required for working with magnetic hopfions, it offers deep insight into their structure. For readers interested in the underlying mathematics, this appendix gives a concise introduction and explains the Hopf map from a geometric perspective. See Ref. [13] for more details.

A fiber bundle consists of a total space *E*, a base space *B*, and a projection map: $\pi : E \to B$. If $x \in B$, the set $F_x = \pi^{-1}(x)$ is called the fiber of *E* at point *x*. Locally, every small neighborhood of *B* looks like a simple product $U \times F$, where $U \subseteq B$ is an open neighborhood of *x*. Globally, however, the total space may be twisted. Fig. F.1 illustrates two examples. On the left, a trivial bundle: the total space *E* is a cylinder, and the base space is a circle. The straight-line segment *F* is a fiber. $\pi$ projects the fiber *F* to point *x*. Globally, this space is simply the product $E = B \times F$. On the right we have a nontrivial fiber bundle: the Möbius strip, which cannot be written globally as the product of two spaces.

The Hopf map is a function: $h : S^3 \to S^2$. For every point *p* on the sphere $S^2$, we can ask: which points in $S^3$ get sent to *p* ? This set of points is a *fiber*. $S^3$ is the total space and $S^2$ the base space. The Hopf map sends $(z_1, z_2)$ in $S^3$ to a point on $S^2$ using the formula

$$h(z_1, z_2) = \left(2\,\mathrm{Re}(z_1 z_z^*), 2\,\mathrm{Im}(z_1 z_z^*), |z_1|^2 - |z_2|^2\right), \qquad \text{(F.1)}$$

where $|z_1|^2 + |z_2|^2 = 1$. Let us pick the point $p = (0,0,1)$ (the north pole). That gives $z_1 z_2^* = 0$, $|z_1|^2 - |z_2|^2 = 1$, with solution $z_2 = 0$, $|z_1| = 1$. So, the fiber at the north pole is

$$h^{-1}(0,0,1) = \{(e^{i\theta}, 0); \theta \in [0, 2\pi)\}. \qquad \text{(F.2)}$$

Which is a circle inside $S^3$. For any point in $S^2$, solving the same equation yields a circle for the fiber. So, the Hopf map organizes the entire 3-sphere into a collection of circles. When we stereographically project $S^3$ into $\mathbb{R}^3$, these circles become circles in $\mathbb{R}^3$, arranged in nested tori, with every pair linked exactly once.

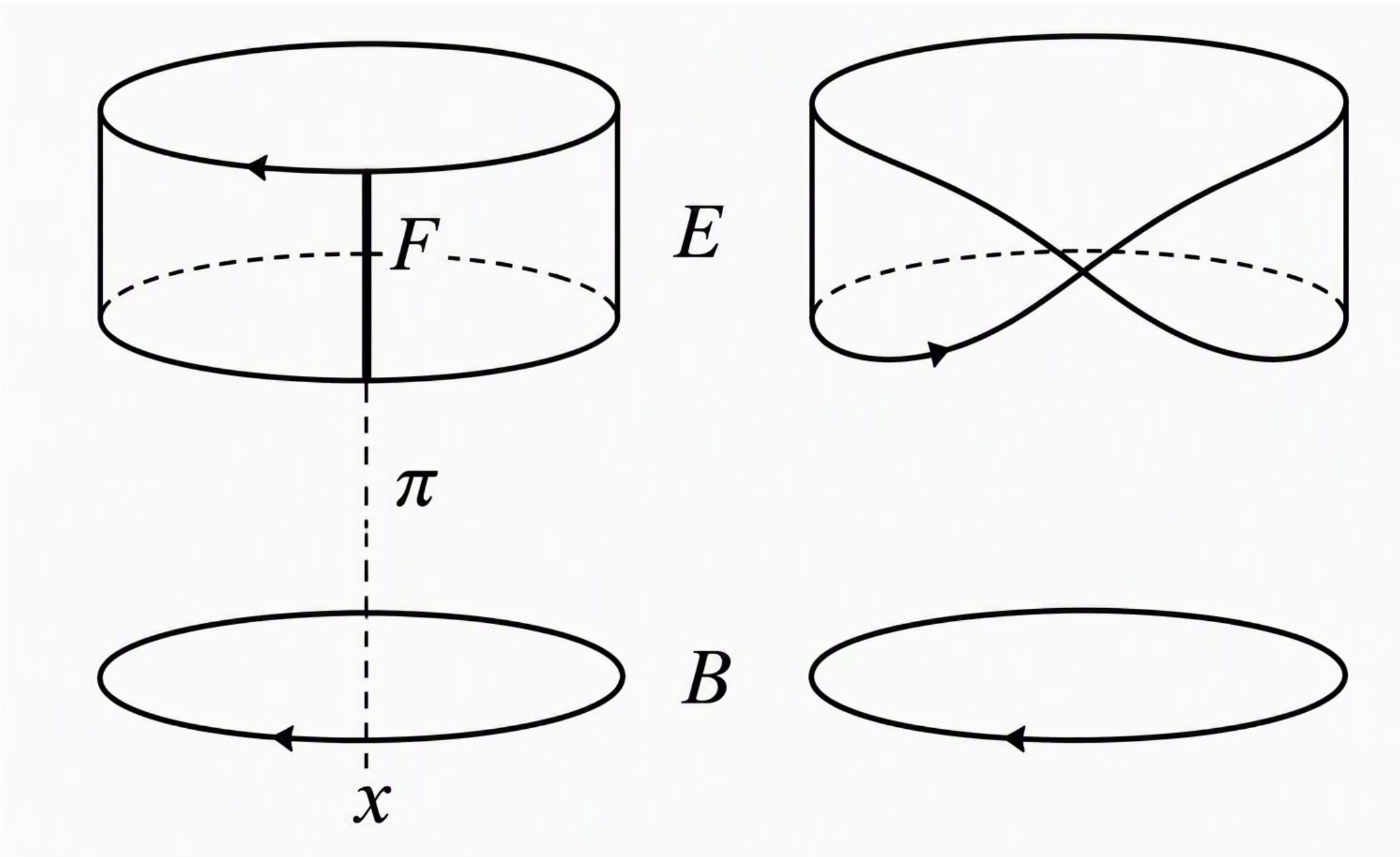

Figure F.1. Examples of fiber bundles.

The Hopf fibration is most elegantly written using a unit quaternion. **Quaternions** are elements $q$ of the vector space $H$. They are a generalization of complex numbers to a 4-dimensional space, with the imaginary basis vectors extended from one ($i$ ) to three ($i, j, k$), with the following property [13]

$$i^2 = j^2 = k^2 = ijk = -1 \quad \text{(F.3)}$$

$$ij = k,\ jk = i\ ,\ ki = j,\quad ji = -k,\ kj = -i\ ,\quad ik = -j\ . \quad \text{(F.4)}$$

They satisfy the commutative and associative laws of addition and the distributive laws of multiplication over addition, but their multiplication is noncommutative: $q_1q_2 \neq q_2q_1$, in general (some pairs do commute). A quaternion has the form

$$q = a + bi + cj + dk\ . \quad \text{(F.5)}$$

With complex conjugate

$$q^* = a - bi - cj - dk\ . \quad \text{(F.6)}$$

The norm is

$$|q|^2 = qq^* = a^2 + b^2 + c^2 + d^2\ . \quad \text{(F.7)}$$

The inverse of any quaternion is

$$q^{-1} = \frac{q^*}{|q|^2}\ . \quad \text{(F.8)}$$

Indeed $q\left(\frac{q^*}{|q|^2}\right) = \frac{qq^*}{|q|^2} = 1$. The unit quaternion satisfy $|q| = 1$, this is $a^2 + b^2 + c^2 + d^2 = 1$, which is the 3-sphere $S^3 = \{q \in H : |q| = 1\}$. So, every point of $S^3$ can be regarded as a unit quaternion. The purely imaginary quaternions are $v = xi + yj + zk$ . Then $v^* = -v$ . Those of norm 1 satisfy $x^2 + y^2 + z^2 = 1$, which is a unity 2-sphre.

The most important application of quaternions is in the algebra of rotations in $\mathbb{R}^3$ [93]. This application originally led William Rowan Hamilton to discover them. Hamilton sought a convenient mathematical framework for describing rotations in a way that lets them be combined, manipulated, and performed sequentially. Traditional approaches based on matrices and linear algebra can often be cumbersome and lack intuitive transparency. A rotation in $\mathbb{R}^3$ is completely determined by an axis of rotation and an angle of rotation. The axis can be represented by a point on the sphere $S^2$, while the rotation angle corresponds to a point on the circle. Consequently, rotation is essentially characterized by three parameters. Quaternions provide an elegant, efficient representation of these rotations, making both computation and composition significantly simpler than many standard methods. The pure quaternions form a three-dimensional space, and it is this fact that allows quaternions to be used to describe rotations in $\mathbb{R}^3$. Biquaternions are quaternions with complex coefficients.

If f $v$ is a pure quaternion corresponding to a point $(x, y, z)$ in $\mathbb{R}^3$, and $q$ is any nonzero quaternion, then the quaternionic conjugation by $q$: $qvq^{-1} = x'i + y'j + z'k$ acts as a rotation in the 3D space of purely imaginary quaternions. After some calculation, we obtain

$$\begin{pmatrix} x' \\ y' \\ z' \end{pmatrix} = \begin{pmatrix} a^2+b^2-c^2-d^2 & 2(bc-ad) & 2(bd+ac) \\ 2(bc+ad) & a^2-b^2+c^2-d^2 & 2(cd-ab) \\ 2(bd-ac) & 2(cd+ab) & a^2-b^2-c^2+d^2 \end{pmatrix} \begin{pmatrix} x \\ y \\ z \end{pmatrix} . \quad \text{(F.9)}$$

It can be shown that the matrix transforming $v$ to $v'$ satisfies $R^T R = I$ and $\det R = 1$, and therefore is a rotation matrix. Hence $R \in SO(3)$.

Let us take a unit quaternion $q$ and a fixed imaginary unit quaternion, usually $i$ , and define

$$h(q) = qiq^{-1} . \quad \text{(F.10)}$$

Because $q$ is unit, $q^{-1} = q^*$ , and so $h(q) = qiq^*$. Quaternion conjugation reverses the order of multiplication: $(q_1q_1q_3)^* = q_1^*q_2^*q_3^*$. Therefore $(qiq^*)^* = (q^*)^*(-i)q^* = -qiq^*$.Hence $qiq^*$ is a purely imaginary quaternion. Thus $q \in S^3 \to qiq^{-1} \in S^2$. This shows that $h(q)$ is a Hopf map. We have $h(qe^{i\theta}) = qe^{i\theta}i(e^{i\theta})^{-1}q^{-1} = qiq^{-1} = h(q)$, because $e^{i\theta}$ commutes with $i$. Therefore all the points $qe^{i\theta}$ $(0 \le \theta \le 2\pi)$ map to the same point of $S^2$ . So, every fiber of the Hopf map is a circle. Using Eq. (F.9) we can write

$$qiq^{-1} = (a^2 + b^2 - c^2 - d^2)i + 2(bc + ad)j + 2(bd - ac)k . \quad \text{(F.11)}$$

Writing $q = z_1 + z_2 j$ , we can rewrite Eq. (F11) as

$$qiq^{-1} = (|z_1|^2 - |z_2|^2)i + 2\operatorname{Re}(z_1 z_2^*)j + 2\operatorname{Im}(z_1 z_2^*)k . \quad \text{(F.12)}$$

Which is equivalent to the standard Hopf map given by Eq. (F.1), with a rotation of the axis. See Fig. F.2.

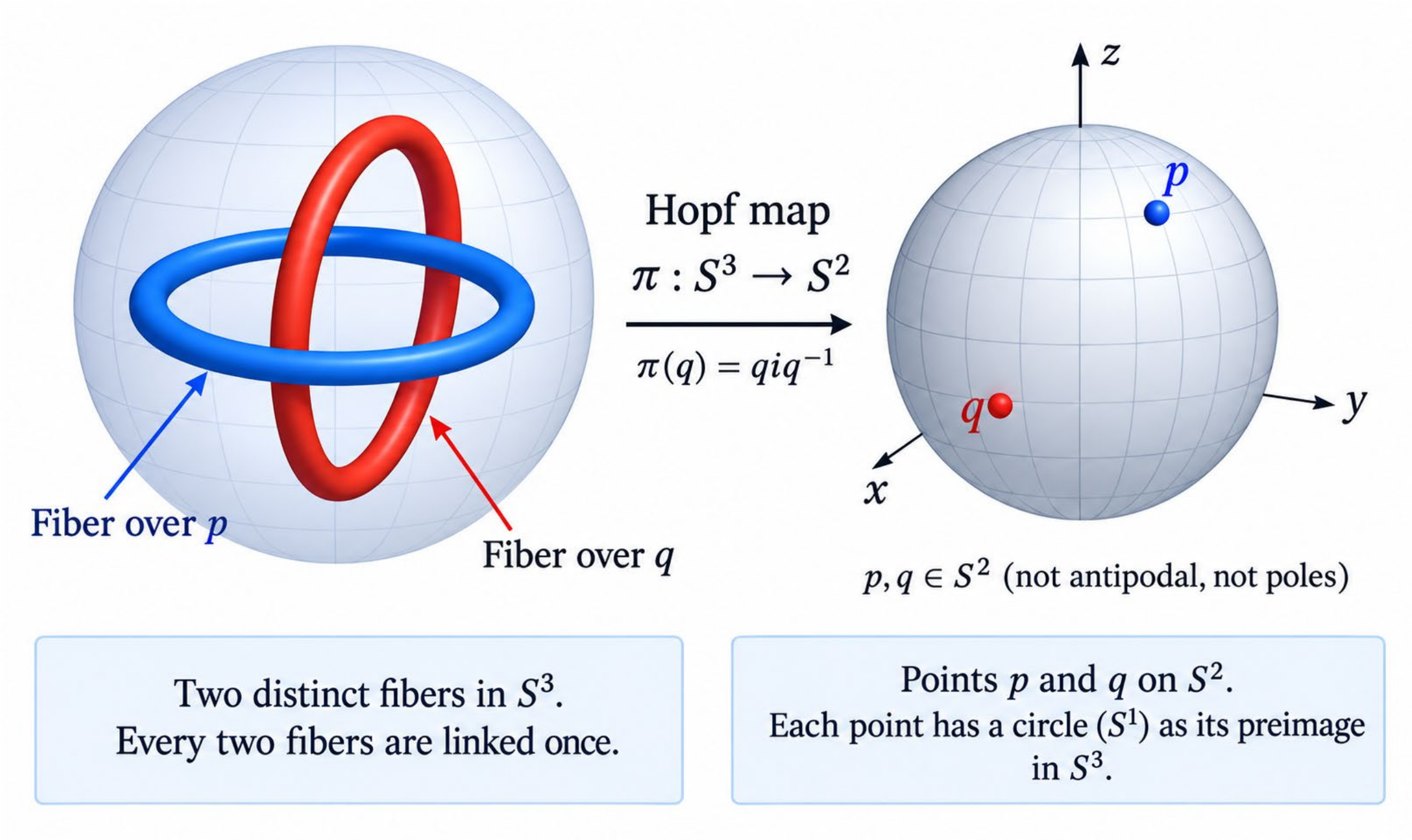

Figure F.2 .The sphere on the left represents $S^3$ (note that it is only a schematic representation, since we cannot visualize a four-dimensional space). The red and blue circles are fibers $h^{-1}(p)$ and $h^{-1}(q)$ over the points $p, q \in S^2$ . The red and blue circles form a Hopf link. This image is only for illustrative purposes.

Next, I present some examples of the Hopf map. Let us start with the equation (297) for the magnetization:

$$m_x = 2\mathrm{Re}(z_1 z_2^*) \quad m_y = 2\mathrm{Im}(z_1 z_2^*) \quad m_z = |z_1|^2 - |z_2|^2, \quad |z_1|^2 + |z_2|^2 = 1. \tag{F.13}$$

Let $m_z = a$, with $-1 < a < 1$, then we have

$$z_1 = \sqrt{\frac{1+a}{2}} e^{i\alpha}, \qquad z_2 = \sqrt{\frac{1-a}{2}} e^{i\beta}, \qquad \alpha, \beta \in [0, 2\pi) \tag{F.14}$$

$$m_x = \sqrt{1-a^2}\cos(\alpha - \beta), \qquad m_y = \sqrt{1-a^2}\sin(\alpha - \beta), \tag{F.15}$$

If we write

$$\vec{m} = (\sqrt{1-m_z^2}\cos\phi, \sqrt{1-m_z^2}\sin\phi, m_z). \tag{F.16}$$

Then $\phi = \alpha - \beta \mod 2\pi$. (F.17)

For a fixed $\phi$, the angles $\alpha$ and $\beta$ can vary satisfying Eq. (F.17). Holding $\alpha - \beta$ constant, we obtain a circle in $S^3$. These circles are the fibers of the Hopf fibration restricted to $m_z = a$. If we take any closed curve $\gamma \subset S^2$, then $T = h^{-1}(\gamma)$ is a torus embedded in $S^3$.

Consider now the equator of the base sphere $S^2$ of the Hopf fibration. Points on the equator correspond to

$$z_1 = \frac{e^{i\alpha}}{\sqrt{2}}, \qquad z_2 = \frac{e^{i\beta}}{\sqrt{2}}. \tag{F.18}$$

So, the fiber over any point in the equator $(\cos\phi, \sin\phi, 0)$ is

$$F_\phi = \left\{ \left( \frac{e^{i\alpha}}{\sqrt{2}}, \frac{e^{i(\alpha-\phi)}}{\sqrt{2}} \right) | \, \alpha \in [0,2\pi) \right\}, \tag{F.19}$$

which is a circle. Let the base point vary along the equator. The union of all these fibers is the Clifford torus in $S^3$, defined by

$$n_1^2 + n_2^2 + n_3^2 + n_4^2 = 1, \qquad n_1^2 + n_2^2 = n_3^2 + n_4^2 = \frac{1}{2}. \tag{F.20}$$

Using the equation

$$n_1 = \frac{2x}{r^2+1}, \quad n_2 = \frac{2y}{r^2+1}, \quad n_3 = \frac{2z}{r^2+1}, \quad n_4 = \frac{r^2-1}{r^2+1}, \tag{F.21}$$

yields a surface whose equation is

$$(x^2 + y^2 + z^2 + 1)^2 = 8(x^2 + y^2). \tag{F.22}$$

Comparing with the equation of a torus of major radius *R* and minor radius *r*, centered around the *z*-axis

$$(x^2 + y^2 + z^2 + R^2 - r^2)^2 = 4R^2(x^2 + y^2), \tag{F.23}$$

we get $R = \sqrt{2}$, $r = 1$. The stereographically projected torus is the familiar donut-shaped object seen in textbooks. However, a perfectly symmetric torus in $S^3$ becomes a torus whose geometry is no longer uniform.

I will repeat the above calculations differently to illustrate the formalism. Let us consider the point (1,0,0) in $S^3$. We have

$$2\,\mathrm{Re}(z_1 z_2^*) = 1, \qquad 2\,\mathrm{Im}(z_1 z_2^*) = 0, \qquad |z_1| - |z_2| = 0, \tag{F.24}$$

which gives

$$z_1 = \frac{e^{i\alpha}}{\sqrt{2}}, \qquad z_2 = \frac{e^{i\beta}}{\sqrt{2}} \qquad \rightarrow \quad e^{i(\alpha-\beta)} \text{ real}, \quad \alpha - \beta = 0 \text{ or } \pi. \tag{F.25}$$

We have

$$h^{-1}(1,0,0) = \left\{ \left( \frac{e^{i\theta}}{\sqrt{2}}, \frac{e^{i\theta}}{\sqrt{2}} \right), \theta \in [0,2\pi) \right\}. \tag{F.26}$$

The stereographic projection into $\mathbb{R}^3$ given by Eq. (299a) gives

$$[x(\theta), y(\theta), z(\theta)] = \left( \frac{\cos\theta/\sqrt{2}}{A}, \frac{\sin\theta/\sqrt{2}}{A}, \frac{\cos\theta/\sqrt{2}}{A} \right), \tag{F.27}$$

where $A = 1 - \sin\theta/\sqrt{2}$. Note that $z(\theta) = x(\theta)$. After some straightforward calculation, we obtain $2x^2 + (y-1)^2 = 2$, the same result obtained in section 14. Doing the same calculation for the point (0,1,0), and noting that now $\alpha - \beta = \pi/2$, we obtain

$$[x(\theta), y(\theta), z(\theta)] = \left( \frac{-\sin\theta/\sqrt{2}}{A}, \frac{\cos\theta/\sqrt{2}}{A}, \frac{\cos\theta/\sqrt{2}}{A} \right), \tag{F.28}$$

which gives $x(\theta) = y(\theta)$ and the equation $2y^2 + (x-1)^2 = 2$. Both circles lie on the torus $(x^2 + y^2 + z^2 + 1)^2 = 8(x^2 + y^2)$. Each circle wraps once around the torus's hole and once around the torus's body. They intersect the torus at the characteristic *Villarceau angle.* These diagonal circles foliate the torus. The Hopf fibration $h : S^3 \to S^2$ has the following properties:

1. Each fiber is a great circle of $S^3$.
2. Fibers inside a single solid torus are pairwise disjoint and parallel. They foliate the solid torus with circles.
3. On the boundary torus, each fiber winds with slope (1, 1) in the standard meridian–longitude coordinates.
4. A fiber in one solid torus links exactly once with every fiber in the complementary solid torus, with linking number +1 or -1, depending on orientation. Fibers within the same solid torus do not link.
5. The fibers fill the torus.

These properties characterize the Hopf fibration as a fibration of $S^3$ by circles with constant linking numbers. The decomposition

$$S^3 = \{|z_1| \geq |z_2|\} \cup \{|z_1| \leq |z_2|\}, \tag{F.29}$$

splits $S^3$ into two solid tori whose common boundary is the Clifford torus.

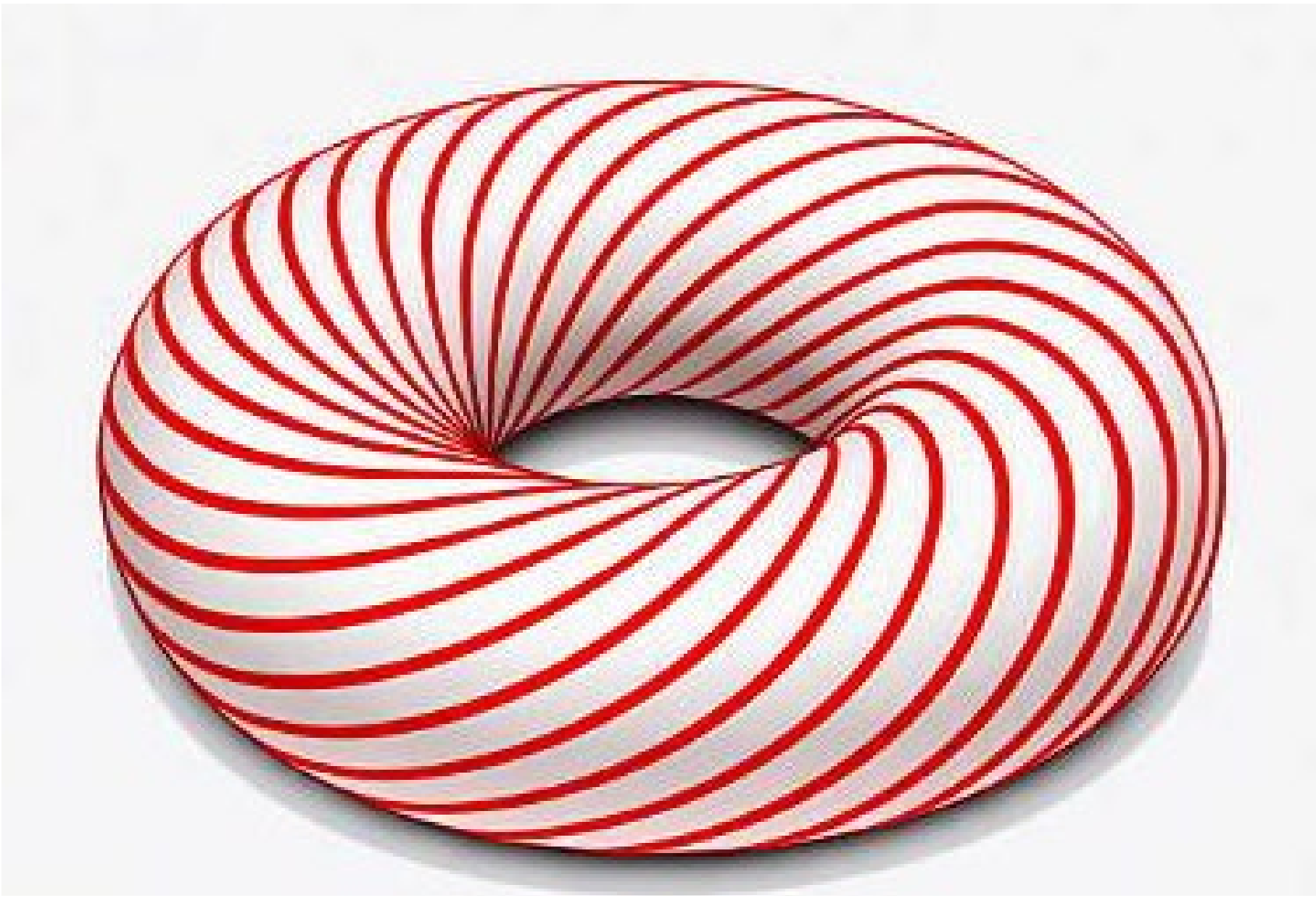

Figure F.3. The Hopf fibration.

A torus can be thought of as a square with identified edges. Each pair of parallel, identified edges becomes a generator. We draw a diagonal from one corner to its opposite corner. Then we draw a series of lines parallel to that diagonal and wrap the square into a torus. The lines become a series of circles, each circling the torus exactly once in the direction of one generator and once in the direction of the other generator. We see that the circles are linked, each passing through each other once (Fig. F.3).

As the Hopf charge increases, the possible field configurations can exhibit increasingly intricate linking and knotting. Numerical solutions of the Faddeev–Skyrme model may display toroidal, braided, or knotted structures, depending on the charge and the details of the energy-minimizing configuration. The Hopf charge constrains the topology of the field but does not uniquely determine its geometric morphology.

## Appendix G. Euler-Lagrange equation in three dimensions.

Writing the magnetization as

$$\vec{m} = (\sin\theta\cos\phi, \sin\theta\sin\phi, \cos\theta)\,, \tag{G.1}$$

the exchange energy density for the ferromagnetic Heisenberg model is given by

$$E = \frac{JM_S^2}{2}[(\nabla\theta)^2 + \sin^2\theta(\nabla\phi)^2]\,. \tag{G.2}$$

The Euler-Lagrange equations for (G.2) are

$$\frac{\partial^2\theta}{\partial x^2} + \frac{\partial^2\theta}{\partial y^2} + \frac{\partial^2\theta}{\partial z^2} - \sin\theta\cos\theta\left[\left(\frac{\partial\phi}{\partial x}\right)^2 + \left(\frac{\partial\phi}{\partial y}\right)^2 + \left(\frac{\partial\phi}{\partial z}\right)^2\right] = 0\,, \tag{G.3}$$

and

$$2\cos\theta\left[\frac{\partial\theta}{\partial x}\frac{\partial\phi}{\partial x} + \frac{\partial\theta}{\partial y}\frac{\partial\phi}{\partial y} + \frac{\partial\theta}{\partial z}\frac{\partial\phi}{\partial z}\right] + \sin\theta\left[\frac{\partial^2\phi}{\partial x^2} + \frac{\partial^2\phi}{\partial y^2} + \frac{\partial^2\phi}{\partial z^2}\right] = 0\,. \tag{G.4}$$

These equations have no analytical solutions in three dimensions, even if we assume localized solutions with finite energy. If $\theta(\vec{r})$ and $\phi(\vec{r})$ are solutions, then $\theta(\lambda\vec{r})$ and $\phi(\lambda\vec{r})$ are also solutions with energy $E(\lambda) = E(\lambda = 1)/\lambda$. Therefore, the soliton size is not fixed; it is proportional to its energy and can shrink and disappear. That does not happen in a magnetic system because other terms in the Hamiltonian are present, as shown in Eq. (291).

The LLG equations, with

$$H_{eff} = -\frac{1}{M_S}\frac{\delta E}{\delta\vec{m}}\,, \tag{G.5}$$

are

$$\frac{1}{\gamma M_S}\sin\theta\frac{\partial\phi}{\partial t} = -\sin\theta\cos\theta\left[\left(\frac{\partial\phi}{\partial x}\right)^2 + \left(\frac{\partial\phi}{\partial y}\right)^2 + \left(\frac{\partial\phi}{\partial z}\right)^2\right] + \left[\frac{\partial^2\theta}{\partial x^2} + \frac{\partial^2\theta}{\partial y^2} + \frac{\partial^2\theta}{\partial z^2}\right], \tag{G.6}$$

and

$$\frac{1}{\gamma M_S}\sin\theta\frac{\partial\theta}{\partial t} = -2\cos\theta\left[\frac{\partial\theta}{\partial x}\frac{\partial\phi}{\partial x} + \frac{\partial\theta}{\partial y}\frac{\partial\phi}{\partial y} + \frac{\partial\theta}{\partial z}\frac{\partial\phi}{\partial z}\right] - \sin\theta\left[\frac{\partial^2\phi}{\partial x^2} + \frac{\partial^2\phi}{\partial y^2} + \frac{\partial^2\phi}{\partial z^2}\right]. \tag{G.7}$$

A uniformly moving soliton with a stationary profile and uniform precession frequency is a solution of the above equations with

$$\theta(\vec{r},t)=\theta(\vec{r}-\vec{v}t)\,,\qquad \phi(\vec{r},t)=\phi(\vec{r}-\vec{v}t)+\omega t\,. \tag{G.8}$$

As noted in Section 13, the Hopf map determines the solution's topological structure, but the actual field configuration comes from solving the nonlinear Euler-Lagrange equations. The map is not, by itself, a solution to any physical field equation. It is only a *topological representative* of the homotopy class.

The nonlinear equations that define hopfions have no closed-form solutions. All known hopfions are obtained by choosing a Hopf-map-based ansatz, minimizing the nonlinear energy functional, and relaxing to a stable knotted configuration. The topology is exact; the geometry is numerical. If we simulate a Hopfion with the LLG equation, the "torus" we see is not geometric, not the stereographic projection of a Hopf torus, but a region where the magnetization wraps around in a toroidal fashion. It is a topological soliton, not a geometric embedding.